\documentclass[twocolumns,10pt,final]{IEEEtran}
\usepackage{latexsym}
\usepackage{cite}
\usepackage{amssymb}
\usepackage{bbm}
\usepackage{bm}
\usepackage[usenames,dvipsnames]{color}
\usepackage{comment}
\usepackage{dsfont}
\usepackage{amsmath, amssymb}
\DeclareMathOperator{\supp}{supp}
\usepackage[dvips]{graphics}
\usepackage{graphicx}
\usepackage[caption=false,font=footnotesize]{subfig}
\usepackage{fixltx2e}
\usepackage{mathtools}
\usepackage{amsmath}
\usepackage{psfrag}
\usepackage{lipsum}
\usepackage{ntheorem}
\usepackage{setspace}
\usepackage{algorithm}
\usepackage{algpseudocode}
\usepackage{theorem}
\usepackage{float}
\usepackage{mathtools}
 \allowdisplaybreaks

\newtheorem{definition}{Definition}

\newtheorem{theorem}{Theorem}
\newtheorem{lemma}{Lemma}
\newtheorem{remark}{Remark}

\newtheorem{corollary}[theorem]{Corollary}
\newtheorem{proposition}[theorem]{Proposition}

\newcommand\numberthis{\addtocounter{equation}{1}\tag{\theequation}}
\renewcommand{\arraystretch}{1.2} 
\catcode`~=11 
\newcommand{\urltilde}{\kern -.15em\lower .7ex\hbox{~}\kern .04em}
\catcode`~=13 

\begin{document}

\makeatletter
\newcommand{\vasti}{\bBigg@{3}}
\newcommand{\vast}{\bBigg@{4}}
\newcommand{\Vast}{\bBigg@{5}}
\makeatother
\newcommand{\be}{\begin{equation}}
\newcommand{\ee}{\end{equation}}
\newcommand{\ba}{\begin{align}}
\newcommand{\ea}{\end{align}}
\newcommand{\baa}{\begin{align*}}
\newcommand{\eaa}{\end{align*}}
\newcommand{\bea}{\begin{eqnarray}}
\newcommand{\eea}{\end{eqnarray}}
\newcommand{\beaa}{\begin{eqnarray*}}
\newcommand{\eeaa}{\end{eqnarray*}}
\newcommand{\p}[1]{\left(#1\right)}
\newcommand{\pp}[1]{\left[#1\right]}
\newcommand{\ppp}[1]{\left\{#1\right\}}
\newcommand{\ber}{$\ \mbox{Ber}$}
\newcommand\aatop[2]{\genfrac{}{}{0pt}{}{#1}{#2}}

\title{Strong Secrecy for Cooperative Broadcast Channels}

\author{Ziv Goldfeld, \emph{Student Member, IEEE}, Gerhard Kramer, \emph{Fellow, IEEE}, Haim H. Permuter, \emph{Senior Member, IEEE}, and Paul Cuff, \emph{Member, IEEE}
\thanks{Z. Goldfeld and H. H. Permuter were supported in part by the Cyber Security Research Center within the Ben-Gurion University of the Negev, in part by the European Research Council under the European Union's Seventh Framework Programme (FP7/2007-2013)/ERC grant agreement n$^\circ$337752 and in part by the Israel Science Foundation. G. Kramer was supported by an Alexander von Humboldt Professorship endowed by the German Federal Ministry of Education and Research. P. Cuff was supported in part by the National Science Foundation under Grant CCF-1350595 and CCF-1116013 and in part by the Air Force Office of Scientific Research under Grant FA9550-15-1-0180 and FA9550-12-1-0196. 
\newline This paper was presented in part at the 2015 IEEE International Symposium on Information Theory, Hong-Kong, and in part at the 2016 International Zurich Seminar on Communications, Zurich, Switzerland.
\newline Z. Goldfeld and H. H. Permuter are with the Department of Electrical and Computer Engineering, Ben-Gurion University of the Negev, Beer-Sheva, Israel (gziv@post.bgu.ac.il, haimp@bgu.ac.il). G. Kramer is with the Institute for Communications Engineering, Technical University of Munich, Munich D-80333, Germany (gerhard.kramer@tum.de). Paul Cuff is with the Department of Electrical Engineering, Princeton University, Princeton, NJ 08544 USA (e-mail: cuff@princeton.edu).}}
\maketitle


\begin{abstract}
A broadcast channel (BC) where the decoders cooperate via a one-sided link is considered. One common and two private messages are transmitted and the private message to the cooperative user should be kept secret from the cooperation-aided user. The secrecy level is measured in terms of strong secrecy, i.e., a vanishing information leakage. An inner bound on the capacity region is derived by using a channel-resolvability-based code that \emph{double-bins} the codebook of the secret message, and by using a \emph{likelihood encoder} to choose the transmitted codeword. The inner bound is shown to be tight for semi-deterministic and physically degraded BCs and the results are compared to those of the corresponding BCs without a secrecy constraint. Blackwell and Gaussian BC examples illustrate the impact of secrecy on the rate regions. Unlike the case without secrecy, where sharing information about both private messages via the cooperative link is optimal, our protocol conveys parts of the common and non-confidential messages only. This restriction reduces the transmission rates more than the usual rate loss due to secrecy requirements. An example that illustrates this loss is provided.

\end{abstract}



\section{Introduction}\label{sec_introduction}


\par
\begin{figure}[t!]
    \begin{center}
        \begin{psfrags}
            \psfragscanon
            \psfrag{I}[][][0.85]{$\mspace{-40mu}(\mspace{-1.5mu}M_0\mspace{-1.5mu},\mspace{-2mu}M_1\mspace{-1.5mu},\mspace{-2mu}M_2\mspace{-1.5mu})$}
            \psfrag{J}[][][0.9]{\ \ \ \ \ \ \ Enc $f^{(n)}$}
            \psfrag{K}[][][1]{\ \ \ $\mathbf{X}$}
            \psfrag{T}[][][0.9]{\ \ \ \ \ \ \ \ \ Channel}
            \psfrag{M}[][][1]{\ \ \ $\mathbf{Y}_1$}
            \psfrag{N}[][][1]{\ \ \ $\mathbf{Y}_2$}
            \psfrag{O}[][][0.9]{\ \ \ \ \ \ \ Dec $\phi^{(n)}_1$}
            \psfrag{P}[][][0.9]{\ \ \ \ \ \ \ Dec $\phi^{(n)}_2$}
            \psfrag{Q}[][][0.85]{\ \ \ \ \ \ \ \ \ \ \ $(\hat{M}_0^{(1)},\hat{M}_1)$}
            \psfrag{R}[][][0.85]{\ \ \ \ \ \ \ \ \ \ \ $(\hat{M}_0^{(2)},\hat{M}_2)$}
            \psfrag{X}[][][0.9]{\ \ \ \ \ \ \ \ \ $W^n_{Y_1,Y_2|X}$}
            \psfrag{V}[][][0.85]{\ \ \ \ \ \ \ \ \ \ \ \ \ \ \ \ $M_{12}=g_{12}^{(n)}(\mathbf{Y}_1)$}
            \psfrag{U}[][][0.85]{\ $M_1$}
            \hspace{5mm}\includegraphics[scale = .37]{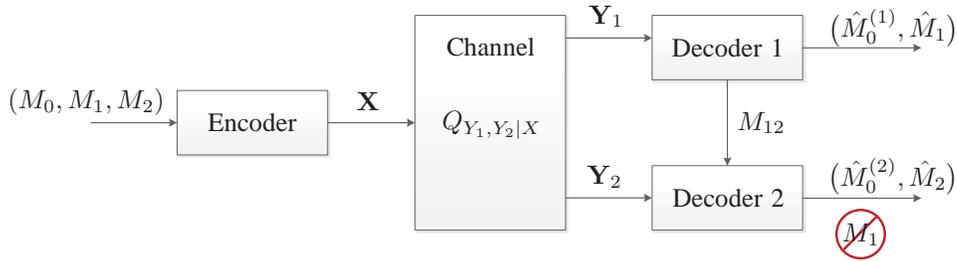}
            \caption{Cooperative BCs with one confidential message.} \label{FIG:BC_secrecy}
            \psfragscanoff
        \end{psfrags}
     \end{center}
 \end{figure}

 User cooperation and security are two essential aspects of modern communication systems. Cooperation can increase transmission rates, whereas security requirements can limit these rates. To shed light on the interaction between these two phenomena, we study broadcast channels (BCs) with one-sided decoder cooperation and one confidential message (Fig. \ref{FIG:BC_secrecy}). Cooperation is modeled as \emph{conferencing}, i.e., information exchange via a rate-limited link that extends from one receiver (referred to as the \emph{cooperative receiver}) to the other (the \emph{cooperation-aided receiver}). The cooperative receiver possesses confidential information that should be kept secret from the other user. 


Secret communication over noisy channels was modeled by Wyner who introduced the degraded wiretap channel (WTC) and derived its secrecy-capacity \cite{Wyner_Wiretap1975}. Wyner's wiretap code relied on a \emph{capacity-based} approach, i.e., the code is a union of subcodes that operate just below the capacity of the eavesdropper's channel. Csisz{\'a}r and K{\"o}rner \cite{Csiszar_Korner_BCconfidential1978} generalized Wyner's result to a general BC. Multiuser settings with secrecy have since been extensively treated in the literature. Broadcast and interference channels with two confidential messages were studied in \cite{BC_Confidential_Yates2008,Semi-det_BC_secrect_two2009,Semi-det_BC_secrect_one2009,goldfeld_leakage2015,Ulukus_Cooperative_RBC2011}. Gaussian multiple-input multiple-output (MIMO) BCs and WTCs were studied in \cite{Poor_Gaussian_MIMO_BC_Secrecy2009,Liu_Shamai_MIMOWTC2009,Poor_Shamai_Gaussian_MIMO_BC_Secrecy2010,Khitsi_MIMOWTC2010,Ulukus_Gaussian_Wiretap2011,Hassibi_MINOWTC2011}, while \cite{Ulukus_External_Eve2009,Bagherikaram_Gaussin_External_Eve2009,Piantanida_External_Eve2014} focus on BCs with an eavesdropper as an external entity from which all messages are kept secret.




\par The above papers consider the \emph{weak secrecy} metric, i.e., a vanishing information leakage \emph{rate} to the eavesdropper. Although the leakage rate vanishes asymptotically with the blocklength, the eavesdropper can decipher an increasing number of bits of the confidential message. This drawback was highlighted in \cite{Maurer_Strong_Secrecy_Chapter1994,Maurer_Wolf_Strong_Secrecy2000,Bloch_Barros_Secrecy_Book2011} (see also \cite{Csiszar_Strong_Secrecy1996}), which advocated using the \emph{information leakage} as a secrecy measure referred to as \emph{strong secrecy}. We consider strong secrecy by relying on work by Csisz{\'a}r \cite{Csiszar_Strong_Secrecy1996} and Hayashi \cite{Hayashi_Secrecy_Resolvability2006} to relate the coding mechanism for secrecy to \emph{channel-resolvability}.


\par The problem of channel resolvability, closely related to the early work of Wyner \cite{Wyner_Common_Information1975}, was formulated by Han and Verd{\'u} \cite{Han_Verdu_Resolvability1993} in terms of total variation (TV). Recently, \cite{Kramer_resolvability2013} advocated replacing the TV metric with \emph{unnormalized relative entropy}. In \cite{Cuff_Synthesis2013}, the coding mechanism for the resolvability problem was extended to various scenarios under the name \emph{soft-covering lemma}. These extensions were used to design secure communication protocols for several source coding problems under different secrecy measures \cite{Cuff_Distortion_Secrecy2014,Cuff_Henchman_Secrecy2014,Song_Cuff_Secrecy2014,Cuff_Secrecy_Coordination2014}. A \emph{resolvability-based} wiretap code associates with each message a subcode that operates just above the resolvability of the eavesdropper's channel. Using such constructions, \cite{Bloch_Resolvability_Secrecy2013} extended the results of \cite{Csiszar_Korner_BCconfidential1978} to strong secrecy for continuous random variables and channels with memory. In \cite{Kramer_EffectiveSecrecy2014} (see also \cite[Remark 2.2]{Han_WTC_Cost2014}), resolvability-based codes were used to establish the strong secrecy-capacities of the discrete and memoryless (DM) WTC and the DM-BC with confidential messages by using a metric called \emph{effective secrecy}.


\par Our inner bound on the strong secrecy-capacity region of the cooperative BC is based on a resolvability-based \emph{Marton} code. Specifically, we consider a state-dependent channel over which an encoder with non-causal access to the state sequence aims to make the conditional probability mass function (PMF) of the channel output given the state a product PMF. The resolvability code coordinates the transmitted codeword with the state sequence by means of multicoding, i.e., by associating with every message a bin that contains enough codewords to ensure joint encoding (similar to a Gelfand-Pinsker codebook). Most encoders use joint typicality tests to determine the transmitted codeword. We adopt the \emph{likelihood encoder}, recently proposed as a coding strategy for source coding problems \cite{Cuff_Song_Likelihood2014}, as our multicoding mechanism. Doing so significantly simplifies the distribution approximation analysis. We prove that the TV between the induced output PMF and the target product PMF approaches zero exponentially fast in the blocklength, which implies convergence in unnormalized relative entropy \cite[Theorem 17.3.3]{CovThom06}. 



\par Next, we construct a BC code in which the relation between the codewords corresponds to the relation between the channel states and the channel inputs in the resolvability problem. To this end we associate with every confidential message a subcode that adheres to the structure of the aforementioned resolvability code. Accordingly, the confidential message codebook is double-binned to allow joint encoding via the likelihood encoder (outer bin layer) and preserves confidentiality (inner bin layer). The bin sizes are determined by the rate constraints for the resolvability problem, which ensures strong secrecy. The inner bound induced by this coding scheme is shown to be tight for semi-deterministic (SD) and physically-degraded (PD) BCs.

\par Our protocol uses the cooperation link to convey information about the non-confidential message and the common message. Without secrecy constraints, the optimal scheme shares information on \emph{both} private messages as well as the common message \cite{Goldfeld_BC_Cooperation2014}. We show that the restricted protocol results in an additional rate loss on top of standard losses due to secrecy. To this end we compare the achievable regions induced by each cooperation strategy for a cooperative BC \emph{without secrecy}. We show that the restricted protocol does not lose rate when the BC is deterministic or PD, but it is sub-optimal in general.


\par To the best of our knowledge, we present here the first resolvability-based Marton code. This is also a first demonstration of the likelihood encoder's usefulness in the context of secrecy for channel coding problems. From a broader perspective, our resolvability result is a tool for proving strong secrecy in settings with Marton coding. As a special case, we derive the secrecy-capacity region of the SD-BC (without cooperation) where the message of the deterministic user is confidential - a new result that has merit on its own. The structure of the obtained region provides insight into the effect of secrecy on the coding strategy for BCs. A comparison between the cooperative PD-BC with and without secrecy is also given.

\par The results are visualized by considering a Blackwell BC (BW-BC) \cite{vanderMeulen_Blackwell1975,Gelfand_Blackwell1977} and a Gaussian BC. An explicit strong secrecy-achieving coding strategy for an extreme point of the BW-BC region is given. Although the BW-BC's input is ternary, to maximize the transmission rate of the confidential message only a binary subset of the input's alphabet is used. As a result, a zero-capacity channel is induced to the other user, who, therefore, cannot decode any of the secret bits. Further, we show that in the BW-BC scenario, an improved subchannel (given by the identity mapping) to the legitimate receiver does not increase the strong secrecy-capacity region.

\par This paper is organized as follows. Section \ref{SEC:preliminaries} provides preliminaries and restates some useful basic properties. In Section \ref{SEC:soft_covering} we state a resolvability lemma. Section \ref{SEC:BC} introduces the cooperative BC with one confidential message and gives an inner bound on its strong secrecy-capacity region. The secrecy-capacity regions for the SD and PD scenarios are then characterized. In Section \ref{SEC:SDBC_cooperation_effect} the effect of secrecy constraints on the optimal cooperation protocol is discussed. Section \ref{SEC:secrecy_effect} compares the capacity regions of SD- and PD-BCs with and without secrecy. Blackwell and Gaussian BCs visualise the results. Finally, proofs are provided in Section \ref{SEC:proofs}, while Section \ref{SEC:summary} summarizes the main achievements and insights of this work.

\section{Notations and Preliminary Definition}\label{SEC:preliminaries}


\subsection{Notations}

 We use the following notations. As customary $\mathbb{N}$ is the set of natural numbers (which does not include 0), while $\mathbb{R}$ denotes the reals. We further define $\mathbb{R}_+=\{x\in\mathbb{R}|x\geq 0\}$ and $\mathbb{R}_{++}=\mathbb{R}\setminus\{0\}$. Given two real numbers $a,b$, we denote by $[a\mspace{-3mu}:\mspace{-3mu}b]$ the set of integers $\big\{n\in\mathbb{N}\big| \lceil a\rceil\leq n \leq\lfloor b \rfloor\big\}$. Calligraphic letters denote sets, e.g., $\mathcal{X}$, the complement of $\mathcal{X}$ is denoted by $\mathcal{X}^c$, while $|\mathcal{X}|$ stands for its cardinality. $\mathcal{X}^n$ denoted the $n$-fold Cartesian product of $\mathcal{X}$. An element of $\mathcal{X}^n$ is denoted by $x^n=(x_1,x_2,\ldots,x_n)$; whenever the dimension $n$ is clear from the context, vectors (or sequences) are denoted by boldface letters, e.g., $\mathbf{x}$. A substring of $\mathbf{x}\in\mathcal{X}^n$ is denoted by $x_i^j=(x_i,x_{i+1},\ldots,x_j)$, for $1\leq i\leq j \leq n$; when $i=1$, the subscript is omitted. We also define $x^{n\backslash i}=(x_1,\ldots,x_{i-1},x_{i+1},\ldots,x_n)$.


Let $\big(\mathcal{X},\mathcal{F},\mathbb{P}\big)$ be a probability space, where $\mathcal{X}$ is the sample space, $\mathcal{F}$ is the $\sigma$-algebra and $\mathbb{P}$ is the probability measure. Random variables over $\big(\mathcal{X},\mathcal{F},\mathbb{P}\big)$ are denoted by uppercase letters, e.g., $X$, with conventions for random vectors similar to those for deterministic sequences. The probability of an event $\mathcal{A}\in\mathcal{F}$ is denoted by $\mathbb{P}(\mathcal{A})$, while $\mathbb{P}(\mathcal{A}\big|\mathcal{B}\mspace{2mu})$ denotes conditional probability of $\mathcal{A}$ given $\mathcal{B}$. We use $\mathds{1}_\mathcal{A}$ to denote the indicator function of $\mathcal{A}$. The set of all probability mass functions (PMFs) on a finite set $\mathcal{X}$ is denoted by $\mathcal{P}(\mathcal{X})$, i.e.,
\begin{equation}
    \mathcal{P}(\mathcal{X})=\left\{P:\mathcal{X}\to[0,1]\Bigg| \sum_{x\in\mathcal{X}}P(x)=1]\right\}.
\end{equation}
PMFs are denoted by the uppercase letters such as $P$ or $Q$, with a subscript that identifies the random variable and its possible conditioning. For example, for a discrete probability space $\big(\mathcal{X},\mathcal{F},\mathbb{P}\big)$ and two correlated random variables $X$ and $Y$ over that space, we use $P_X$, $P_{X,Y}$ and $P_{X|Y}$ to denote, respectively, the marginal PMF of $X$, the joint PMF of $(X,Y)$ and the conditional PMF of $X$ given $Y$. In particular, $P_{X|Y}$ represents the stochastic matrix whose elements are given by $P_{X|Y}(x|y)=\mathbb{P}\big(X=x|Y=y\big)$. Expressions such as $P_{X,Y}=P_XP_{Y|X}$ are to be understood as $P_{X,Y}(x,y)=P_X(x)P_{Y|X}(y|x)$, for all $(x,y)\in\mathcal{X}\times\mathcal{Y}$. Accordingly, when three random variables $X$, $Y$ and $Z$ satisfy $P_{X|Y,Z}=P_{X|Y}$, they form a Markov chain, which we denote by $X-Y-Z$. We omit subscripts if the arguments of a PMF are lowercase versions of the random variables. The support of a PMF $P$ and the expectation of a random variable $X\sim P$ are denoted by $\supp(P)$ and $\mathbb{E}_P\big[X\big]$, respectively; when the distribution of $X$ is clear from the context we write its expectation simply as $\mathbb{E}\big[X\big]$. Similarly, $H_P$ and $I_P$ denote entropy and mutual information that are calculated with respect to an underlying PMF $P$.

For a discrete measurable space $(\mathcal{X},\mathcal{F})$, a PMF $Q\in\mathcal{P}(\mathcal{X})$ gives rise to a probability measure on $(\mathcal{X},\mathcal{F})$, which we denote by $\mathbb{P}_Q$; accordingly, $\mathbb{P}_Q\big(\mathcal{A})=\sum_{x\in\mathcal{A}}Q(x)$, for every $\mathcal{A}\in\mathcal{F}$. For a sequence of random variables $X^n$, if the entries of $X^n$ are drawn in an independent and identically distributed (i.i.d.) manner according to $P_X$, then for every $\mathbf{x}\in\mathcal{X}^n$ we have $P_{X^n}(\mathbf{x})=\prod_{i=1}^nP_X(x_i)$ and we write $P_{X^n}(\mathbf{x})=P_X^n(\mathbf{x})$. Similarly, if for every $(\mathbf{x},\mathbf{y})\in\mathcal{X}^n\times\mathcal{Y}^n$ we have $P_{Y^n|X^n}(\mathbf{y}|\mathbf{x})=\prod_{i=1}^nP_{Y|X}(y_i|x_i)$, then we write $P_{Y^n|X^n}(\mathbf{y}|\mathbf{x})=P_{Y|X}^n(\mathbf{y}|\mathbf{x})$. The conditional product PMF $P_{Y|X}^n$ given a specific sequence $\mathbf{x}\in\mathcal{X}^n$ is denoted by $P_{Y|X}^n(\cdot|\mathbf{x})$.

Let $\mathcal{X}$ be a finite set. The empirical PMF $\nu_{\mathbf{x}}$ of a sequence $\mathbf{x}\in\mathcal{X}^n$ is
\begin{equation}
	\nu_{\mathbf{x}}(x)\triangleq\frac{N(x|\mathbf{x})}{n},
\end{equation}
where $N(x|\mathbf{x})=\sum_{i=1}^n\mathds{1}_{\{x_i=x\}}$. We use $\mathcal{T}_\delta^{n}(P)$ to denote the set of letter-typical sequences of length $n$ with respect to the PMF $P\in\mathcal{P}(\mathcal{X})$ and the positive number $\delta$ \cite[Chapter 3]{Massey_Applied}, i.e., we have
\begin{equation}
	\mathcal{T}_\delta^{n}(P)\mspace{-1mu}=\mspace{-1mu}\Big\{\mathbf{x}\in\mathcal{X}^n\Big|\mspace{5mu}\big|\nu_{\mathbf{x}}(x)-P(x)\big|\leq\delta P(x),\ \forall x\mspace{-1mu}\in\mspace{-1mu}\mathcal{X}\Big\}.\label{EQ:typical_set_def}
\end{equation}

\begin{figure*}[t!]
    \begin{center}
        \begin{psfrags}
            \psfragscanon
            \psfrag{A}[][][1]{$\ \ \ \ \ \ \ \ \ \ \ \ \ \ \ \ \ \ \ \ \ W\sim\mbox{Unif}\big[1:2^{n\tilde{R}}\big]$}
            \psfrag{B}[][][0.9]{\ \ \ \ \ \ \ \ \ $\begin{array}{cc}
                 & \mspace{-8mu}\mbox{Likelihood}\\
                 & \mbox{Encoder}\  \mathcal{B}_n
            \end{array}$}
            \psfrag{C}[][][0.95]{\ \ \ \ \ \ \ \ \ \ \ \ \ \ \ \ \ $\mathbf{U}\big(\mathbf{S}_0,W,I\big)$}
            \psfrag{D}[][][1]{\ \ \ \ \ \ \ \ \ \ \ \ $Q^n_{V|U,S_0,S}$}
            \psfrag{E}[][][1]{\ \ \ \ \ \ \ \ \ \ \ \ \ \ \ \ \ \ \ \ \ $\mathbf{V}\sim P_{\mathbf{V}|\mathbf{S}_0,\mathbf{S},\mathsf{B}_n=\mathcal{B}_n}$} \psfrag{F}[][][1]{$\mspace{-5mu}(\mathbf{S}_0,\mathbf{S})$}
            \psfrag{G}[][][1]{\ \ \ \ \ \ \ $Q^n_{S_0,S}$}
            \psfrag{X}[][][1]{$\ \ \ \ w=1$}
            \psfrag{Y}[][][1]{$\ \ \ \ w=2$}
            \psfrag{Z}[][][1]{$\ \ \ \ w=2^{n\tilde{R}}$}
            \psfrag{T}[][][1]{\ \ $\hdots$}
            \psfrag{U}[][][1]{\ \ \ \ \ \ \ \ \ \ \ \ \ \ \ \ \ \ \ \ \ \ \ \ \ \ \ \ \ \ \ $\mathcal{B}_n(\mathbf{s}_0)$: generated $\sim\prod Q_{U|S_0}^n(\cdot|\mathbf{s}_0)$}
            \psfrag{W}[][][1]{\ \ \ \ \ \ \ \ \ \ \ \ \ \ \ \ \ \ \ \ \ \ \ {\color{Mahogany}$2^{nR'}\ u$-codewords}}
            \psfrag{V}[][][1]{\ \ \ \ \ \ \ \ \ \ \ \ \ \ \ \ \ \ \ \ \ \ \ \ \ {\color{OliveGreen}$\mathbf{u}(\mathbf{s}_0,1,i)$: $i$ chosen by}}
            \psfrag{S}[][][1]{\ \ \ \ \ \ \ \ \ \ \ \ \ \ \ \ \ \ \ \ {\color{OliveGreen}likelihood encoder}}
            \hspace{-10mm}\includegraphics[scale = .4]{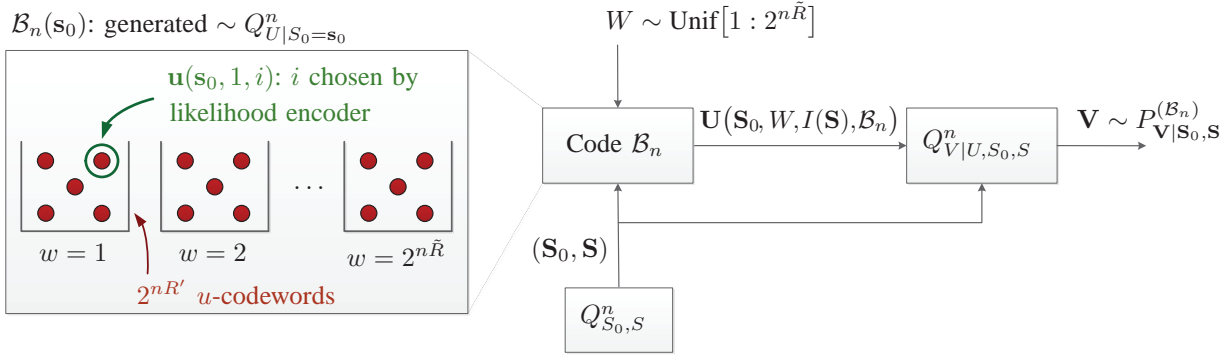}
            \caption{Coding problem for approximating $P_{\mathbf{V}|\mathbf{S}_0,\mathbf{S},\mathsf{B}_n=\mathcal{B}_n}\approx Q_{V|S_0,S}^n$ under a resolvability codebook that is superimposed on $\mathbf{s}_0\in\mathcal{S}_0^n$: For each $\mathbf{s}_0\in\mathcal{S}^n_0$, the codebook $\mathcal{B}_n(\mathbf{s}_0)$ contains $2^{n(\tilde{R}+R')}$ $u$-codewords drawn independently according to $Q_{U|S_0}^n(\cdot|\mathbf{s}_0)$. The codewords are partitioned into $2^{n\tilde{R}}$ bins, each associated with a certain $w\in\big[1:2^{n\tilde{R}}\big]$. The $u$-codeword that is fed into the channel is selected by first randomly and uniformly drawing a bin index $W$ from $\big[1:2^{n\tilde{R}}\big]$, and then drawing $I$ from $\big[1:2^{nR'}\big]$ by means of the likelihood encoder from \eqref{EQ:likelihood_definition}.}\label{FIG:soft_covering}
            \psfragscanoff
        \end{psfrags}
     \end{center}
 \end{figure*}


\subsection{Measures of Distribution Proximity}

\begin{definition}[Relative Entropy]
Let $(\mathcal{X},\mathcal{F})$ be a measurable space and let $P$ and $Q$ be two probability measures on $\mathcal{F}$, with $P\ll Q$ (i.e., $P$ is absolutely continuous with respect to $Q$). The relative entropy between $P$ and $Q$ is
\begin{equation}
	D(P||Q)=\int_\mathcal{X} dP\log\left(\frac{dP}{dQ}\right),\label{EQ:relative_entropy_def}
\end{equation}
where $\frac{dP}{dQ}$ denotes the Radon-Nikodym derivative of $P$ with respect to $Q$. If the sample space $\mathcal{X}$ is countable, \eqref{EQ:relative_entropy_def} reduces to
\begin{equation}
	D(P||Q)=\sum_{x\in\supp(P)}P(x)\log\left(\frac{P(x)}{Q(x)}\right)\label{EQ:relative_entropy_def_discrete}.
\end{equation}
\end{definition}

%
\begin{definition}[Total Variation]
	Let $(\mathcal{X},\mathcal{F})$ be a measurable and $P$ and $Q$ be two probability measures on $\mathcal{F}$. The total variation between $P$ and $Q$ is
	\begin{equation}
		||P-Q||_{\mathsf{TV}}=\sup_{\mathcal{A}\in\mathcal{F}}\big|P(\mathcal{A})-Q(\mathcal{A})\big|.\label{EQ:total_variation_def}
	\end{equation}
	If the sample space $\mathcal{X}$ is countable, \eqref{EQ:total_variation_def} reduces to
	\begin{equation}
		||P-Q||_{\mathsf{TV}}=\frac{1}{2}\sum_{x\in\mathcal{X}}\big|P(x)-Q(x)\big|.\label{EQ:total_variation_def_discrete}
	\end{equation}
\end{definition}

\begin{remark}[TV Dominates Relative Entopy]\label{REM:TV_divergence_relation}
Pinsker's inequality shows that relative entropy is larger than TV. A reverse inequality is sometimes valid. For example, if $\mathcal{X}$ is a finite set, 
$\big\{P_n\big\}_{n\in\mathbb{N}}$ is a sequence of distributions with $P_n\in\mathcal{P}(\mathcal{X}^n)$, $Q\in\mathcal{P}(\mathcal{X})$ and $P_n\ll Q^n$ for every $n\in\mathbb{N}$, then\footnote{$f(n)\in\mathcal{O}\big(g(n)\big)$ means that $f(n)\leq k\cdot g(n)$, for some $k$ independent of $n$ and sufficiently large $n$.} (see \cite[Equation (29)]{Cuff_Synthesis2013})
\begin{equation}
D(P_n||Q^n)\mspace{-2mu}\in\mspace{-2mu}\mathcal{O}\mspace{-2mu}\left(\left[n+\log\frac{1}{||P_n-Q^n||_\mathsf{TV}}\right]||P_n-Q^n||_\mathsf{TV}\mspace{-2mu}\right)\label{EQ:TV_divergence_relation}.
\end{equation}
In particular, \eqref{EQ:TV_divergence_relation} implies that an exponential decay of the TV in $n$ produces an (almost, up to a $\frac{\log n}{n}$ term) exponential decay of the relative entropy with the same exponent.
\end{remark}

\section{A Channel Resolvability Lemma for Strong Secrecy}\label{SEC:soft_covering}


\par Consider a state-dependent discrete memoryless channel (DMC) over which an encoder with non-causal access to the i.i.d. state sequence transmits a codeword (Fig. \ref{FIG:soft_covering}). Each channel state is a pair $(S_0,S)$ of random variables drawn according to $Q_{S_0,S}\in\mathcal{P}(\mathcal{S}_0\times\mathcal{S})$. The encoder superimposes its codebook on $S_0$ and then uses a \emph{likelihood encoder} with respect to $S$ to choose the channel input sequence. The structure of a subcode that is superimposed on some $\mathbf{s}_0\in\mathcal{S}_0^n$ is also illustrated in Fig. \ref{FIG:soft_covering}. The conditional PMF of the channel output given the states should approximate a conditional product distribution in terms of unnormalized relative entropy. A formal description of the setup is as follows.

Let $\mathcal{S}_0$, $\mathcal{S}$, $\mathcal{U}$ and $\mathcal{V}$ be finite sets. Fix any $Q_{S_0,S,U,V}\in\mathcal{P}(\mathcal{S}_0\times\mathcal{S}\times\mathcal{U}\times\mathcal{V})$ and let $W$ be a random variable uniformly distributed over\footnote{To simplify notation, from here on out we assume that quantities of the form $2^{nR}$, where $n\in\mathbb{N}$ and $R\in\mathbb{R}_+$, are integers. Otherwise, simple modifications of some of the subsequent expressions using floor operations are needed.} $\mathcal{W}_n=\big[1:2^{n\tilde{R}}\big]$ that is independent of $(\mathbf{S}_0,\mathbf{S})\sim Q_{S_0,S}^n$.

%
\textbf{Codebook:} For every $\mathbf{s}_0\in\mathcal{S}_0^n$, let $\mathsf{B}_n(\mathbf{s}_0)\triangleq\big\{\mathbf{U}(\mathbf{s}_0,w,i)\big\}_{(w,i)\in\mathcal{W}_n\times\mathcal{I}_n}$, where $\mathcal{I}_n=\big[1:2^{nR'}\big]$, be a collection of $2^{n(\tilde{R}+R')}$ conditionally independent random vectors of length 
$n$, each distributed according to $Q^n_{U|S_0}(\cdot|\mathbf{s}_0)$. A realization of $\mathsf{B}_n(\mathbf{s}_0)$, for $\mathbf{s}_0\in\mathcal{S}^n_0$, is denoted by $ \mathcal{B}_n(\mathbf{s}_0)\triangleq\big\{\mathbf{u}(\mathbf{s}_0,w,i)\big\}_{(w,i)\in\mathcal{W}_n\times\mathcal{I}_n}$. Each codebook $\mathcal{B}_n(\mathbf{s}_0)$ can be thought of as comprising $2^{n\tilde{R}}$ bins, each associated with a different message $w\in\mathcal{W}_n$ and contains $2^{nR'}$ $u$-codewords. We also denote $\mathsf{B}_n\triangleq\big\{\mathsf{B}_n(\mathbf{s}_0)\big\}_{\mathbf{s}_0\in\mathcal{S}_0^n}$, which is referred to as the random resolvability codebook. A possible value of $\mathsf{B}_n$ is denoted by $\mathcal{B}_n$ and we set $\mathfrak{B}_n$ as the collection of all such possible values.

The above codebook construction induces a PMF $\lambda\in\mathcal{P}(\mathfrak{B}_n)$ over the codebook ensemble. For every $\mathcal{B}_n\in\mathfrak{B}_n$, we have 
\begin{equation}
    \lambda(\mathcal{B}_n)=\prod_{\mathbf{s}_0\in\mathcal{S}_0^n}\prod_{\substack{(w,i)\\\in\mathcal{W}_n\times\mathcal{I}_n}}Q^n_{U|S_0}\big(\mathbf{u}(\mathbf{s}_0,w,i)\big|\mathbf{s}_0\big).\label{EQ:lambda_PMF}
\end{equation}
    

\textbf{Encoding and Induced PMF:} For each codebook $\mathcal{B}_n\in\mathfrak{B}_n$, consider the \emph{likelihood encoder} described by conditional PMF
\begin{equation}
\hat{P}^{(\mathcal{B}_n)}(i|w,\mathbf{s}_0,\mathbf{s})=\frac{Q_{S|U,S_0}^n\big(\mathbf{s}\big|\mathbf{u}(\mathbf{s}_0,w,i),\mathbf{s}_0\big)}{\sum\limits_{i'\in\mathcal{I}_n}Q_{S|U,S_0}^n\big(\mathbf{s}\big|\mathbf{u}(\mathbf{s}_0,w,i'),\mathbf{s}_0\big)}.\label{EQ:likelihood_definition}
\end{equation}
Upon observing $(w,\mathbf{s}_0,\mathbf{s})$, an index $i\in\mathcal{I}_n$ is drawn randomly according to \eqref{EQ:likelihood_definition}. The codeword $\mathbf{u}(\mathbf{s}_0,w,i)\in\mathcal{B}_n(\mathbf{s}_0)$ is passed through the DMC $Q_{V|U,S_0,S}^n$. For a fixed codebook $\mathcal{B}_n\in\mathfrak{B}_n$, the induced joint distribution is
\begin{align*}
P^{(\mathcal{B}_n)}(\mathbf{s}_0,\mathbf{s},w,i,\mathbf{u},\mathbf{v})&=Q_{S_0,S}^n(\mathbf{s}_0,\mathbf{s})2^{-n\tilde{R}}\hat{P}^{(\mathcal{B}_n)}(i|w,\mathbf{s}_0,\mathbf{s})\\
&\times\mspace{-3mu}\mathds{1}\mspace{-2mu}_{\big\{\mspace{-2mu}\mathbf{u}=\mathbf{u}(\mathbf{s}_0,w,i)\mspace{-2mu}\big\}}\mspace{-1mu}Q^n_{V|U,S_0,S}(\mathbf{v}|\mathbf{u},\mathbf{s}_0,\mathbf{s}).\numberthis\label{EQ:soft_proof_P_PMF}
\end{align*}
Accounting for the random codebook generation, we also set
\begin{equation}
P(\mathcal{B}_n,\mathbf{s}_0,\mathbf{s},w,i,\mathbf{u},\mathbf{v})=\lambda(\mathcal{B}_n)P^{(\mathcal{B}_n)}(\mathbf{s}_0,\mathbf{s},w,i,\mathbf{u},\mathbf{v}).\label{EQ:soft_proof_P_PMF_code}
\end{equation}

\begin{lemma}[Sufficient Conditions for Approximation]\label{LEMMA:soft_covering}
For any $Q_{S_0,S,U,V}\in\mathcal{P}(\mathcal{S}_0\times\mathcal{S}\times\mathcal{U}\times\mathcal{V})$, if $(\tilde{R},R')\in\mathbb{R}_+^2$ satisfies
\begin{subequations}
\begin{align}
R'&>I(U;S|S_0)\label{EQ:region_soft_covering1}\\
R'+\tilde{R}&>I(U;S,V|S_0),\label{EQ:region_soft_covering2}
\end{align}\label{EQ:region_soft_covering}%
\end{subequations}
then
\begin{equation}
\mathbb{E}_{\mathsf{B}_n}D\Big(P_{\mathbf{V}|\mathbf{S}_0,\mathbf{S},\mathsf{B}_n}\Big|\Big|Q_{V|S_0,S}^n\Big|Q_{S_0,S}^n\Big)\xrightarrow[n\to\infty]{}0.\label{EQ:soft_covering_approximation}
\end{equation}
\end{lemma}

The proof of Lemma \ref{LEMMA:soft_covering} (see Section \ref{SUBSEC:soft_covering_proof}) shows that the TV decays exponentially fast with the blocklength $n$. By Remark \ref{REM:TV_divergence_relation} this implies an almost exponential decay of the desired relative entropy. Another useful property is that the chosen $u$-codeword is jointly letter-typical with $(\mathbf{S}_0,\mathbf{S})$ with high probability.
\begin{lemma}[Typical with High Probability]\label{LEMMA:typicallity}
 If $(\tilde{R},R')\mspace{-2mu}\in\mathbb{R}_+^2$ satisfies \eqref{EQ:region_soft_covering}, then for any $w\in\mathcal{W}_n$ and $\epsilon>0$, we have
\begin{equation}
\mathbb{E}_{\mathsf{B}_n}\mathbb{P}_P\Big(\big(\mathbf{S}_0,\mathbf{S},\mathbf{U}(\mathbf{S}_0,w,I)\big)\notin\mathcal{T}_\epsilon^n(Q_{S_0,S,U})\Big|\mathsf{B}_n\Big)\xrightarrow[n\to\infty]{}0.\label{EQ:typicallity_proposition_tozero}
\end{equation}
\end{lemma}
The proof of Lemma \ref{LEMMA:typicallity} is given in Section \ref{SUBSEC:typicallity_proof}.


\section{Cooperative Broadcast Channels with One Confidential Message}\label{SEC:BC}


\subsection{Problem Definition}\label{SUBSEC:BC_definition}


\par The $\big(\mathcal{X},\mathcal{Y}_1,\mathcal{Y}_2,W_{Y_1,Y_2|X}:\mathcal{X}\to\mathcal{P}(\mathcal{Y}_1\times\mathcal{Y}_2)\big)$ cooperative DM-BC with one confidential message is illustrated in Fig. \ref{FIG:BC_secrecy}. The channel has one sender and two receivers. The sender uniformly chooses a triple $(m_0,m_1,m_2)$ of indices from the product set $\big[1:2^{nR_0}\big]\times\big[1:2^{nR_1}\big]\times\big[1:2^{nR_2}\big]$ and maps it to a sequence $\mathbf{x}\in\mathcal{X}^n$, which is the channel input (the mapping may be random). The sequence $\mathbf{x}$ is transmitted over a BC with transition probability $W_{Y_1,Y_2|X}:\mathcal{X}\to\mathcal{P}(\mathcal{Y}_1\times\mathcal{Y}_2)$. The output sequence $\mathbf{y}_j\in\mathcal{Y}^n_j$, where $j=1,2$, is received by decoder $j$. Decoder $j$ produces a pair of estimates $\big(\hat{m}_0^{(j)},\hat{m}_j\big)$ of $(m_0,m_j)$. Furthermore, the message $m_1$ is to be kept secret from Decoder 2 and there is a one-sided noiseless cooperation link of rate $R_{12}$ that extends from Decoder 1 to Decoder 2. By conveying a message $m_{12}\in\big[1:2^{nR_{12}}\big]$ over this link, Decoder 1 can share with Decoder 2 information about $\mathbf{y}_1$, $\big(\hat{m}_0^{(1)},\hat{m}_1\big)$, or both.

\begin{remark}[Specific Classes of BCs]
We sometimes specialize to the following classes of BCs:
\begin{itemize}
    \item \underline{Semi-Deterministic BCs:} A BC is SD if its channel transition matrix factors as $W_{Y_1,Y_2|X}=\mathds{1}_{\{Y_1=y_1(X)\}}W_{Y_2|X}$, where $y_1:\mathcal{X}\to\mathcal{Y}_1$ and $W_{Y_2|X}:\mathcal{X}\to\mathcal{P}(\mathcal{Y}_2)$.

    \item \underline{Physically-Degraded BCs:} A BC is PD if its channel transition matrix factors as $W_{Y_1,Y_2|X}=W_{Y_1|X}W_{Y_2|Y_1}$, where $W_{Y_1|X}:\mathcal{X}\to\mathcal{P}(\mathcal{Y}_1)$ and $W_{Y_2|Y_1}:\mathcal{Y}_1\to\mathcal{P}(\mathcal{Y}_2)$.

    \item \underline{Deterministic BCs:} A BC is deterministic if its channel transition matrix factors as $W_{Y_1,Y_2|X}=\mathds{1}_{\{Y_1=y_1(X)\}\cap\{Y_2=y_2(X)\}}$, where $y_j:\mathcal{X}\to\mathcal{Y}_j$, for $j=1,2$.
\end{itemize}

\end{remark}

\begin{definition}[Code]
An $(n,R_{12},R_0,R_1,R_2)$ code $c_n$ for the BC with cooperation and one confidential message has:
\begin{enumerate}
\item Four message sets $\mathcal{M}^{(n)}_{12}=\big[1:2^{nR_{12}}\big]$ and $\mathcal{M}^{(n)}_j=\big[1:2^{nR_j}\big]$, for $j=0,1,2$.
\item A stochastic encoder $f^{(n)}:\mathcal{M}^{(n)}_0\times\mathcal{M}^{(n)}_1\times\mathcal{M}^{(n)}_2\to\mathcal{P}(\mathcal{X}^n)$.
\item A decoder cooperation function $g^{(n)}_{12}: \mathcal{Y}_1^n\to \mathcal{M}^{(n)}_{12}$.
\item Two decoding functions $\phi^{(n)}_1: \mathcal{Y}_1^n\to \mathcal{M}_0\times\mathcal{M}^{(n)}_1$ and $\phi^{(n)}_2: \mathcal{M}^{(n)}_{12}\times\mathcal{Y}_2^n\to\mathcal{M}^{(n)}_0\times\mathcal{M}^{(n)}_2$.
\end{enumerate}
\end{definition}

The joint distribution induced by an $(n,R_{12},R_0,R_1,R_2)$ code $c_n$ is:
\begin{align*}
	&P^{(c_n)}\Big(m_0,m_1,m_2,\mathbf{x},\mathbf{y}_1,\mathbf{y}_2,m_{12},\big(\hat{m}_0^{(1)},\hat{m}_1\big),\big(\hat{m}_0^{(2)},\hat{m}_2\big)\Big)\\
	&=\mspace{-3mu}\left(\prod_{j=0,1,2}\frac{1}{\big|\mathcal{M}_j^{(n)}\big|}\mspace{-2mu}\right)\mspace{-3.5mu}f^{(n)}\mspace{-2mu}(\mathbf{x}|m_0,\mspace{-2mu}m_1\mspace{-1mu},\mspace{-1mu}m_2)W^n_{Y_1,Y_2|X}(\mathbf{y}_1\mspace{-1.5mu},\mspace{-2mu}\mathbf{y}_2|\mathbf{x})\\
	&\times\mathds{1}_{\big\{\mspace{-2mu}\hat{m}_{12}=g^{(n)}_{12}(\mathbf{y}_1),\big(\hat{m}^{(1)}_0\mspace{-2mu},\hat{m}_1\big)=\phi^{(n)}_1(\mathbf{y}_1),\big(\hat{m}^{(2)}_0\mspace{-2mu},\hat{m}_2\big)=\phi^{(n)}_2\mspace{-2mu}(m_{12},\mathbf{y}_2)\mspace{-2mu}\big\}}\mspace{-2mu}.\numberthis\label{EQ:induced_PMF}
\end{align*}
The performance of $c_n$ is evaluated in terms of its rate tuple $(R_{12},R_0,R_1,R_2)$, the average decoding error probability and the strong secrecy metric.

\begin{definition}[Average Error Probability] The average error probability for an $(n,R_{12},R_0,R_1,R_2)$ code $c_n$ is
\begin{equation}
P_{e}(c_n)=\mathbb{P}_{P^{(c_n)}}\left(\bigcup_{j=1,2}\bigg\{\Big(\hat{M}_0^{(j)},\hat{M}_j\Big)\neq (M_0,M_j)\bigg\}\right),\label{EQ:BC_error_prob}
\end{equation}
where $\left(\hat{M}_0^{(1)},\hat{M}_1\right)=\phi^{(n)}_1(\mathbf{Y}_1)$ and $\left(\hat{M}_0^{(2)},\hat{M}_2\right)=\phi^{(n)}_2\Big(g_{12}^{(n)}(\mathbf{Y}_1,\mathbf{Y}_2)\Big)$.
\end{definition}

\begin{definition}[Information Leakage] The information leakage at receiver 2 under an $(n,R_{12},R_0,R_1,R_2)$ code $c_n$ is
\begin{equation}
\ell(c_n)=I_{P^{(c_n)}}(M_1;M_{12},Y_2^n),\label{EQ:infromation_leakage_def}
\end{equation}
where the subscript $P^{(c_n)}$ indicates that the mutual information term is calculated with respect to the marginal PMF $P^{(c_n)}_{M_1,M_{12},\mathbf{Y}_2}$ of the induced joint distribution from \eqref{EQ:induced_PMF}.
\end{definition}



\begin{definition}[Achievability]\label{DEF:achievability} $(R_{12},R_0,R_1,R_2)\in\mathbb{R}_+^4$ is {\it{achievable}} if for any $\epsilon>0$ there exists an $(n,R_{12},R_0,R_1,R_2)$ code $c_n$, such that
\begin{subequations}
\begin{align}
&P_e(c_n)\leq\epsilon\label{EQ:error_prob}\\
&\ell(c_n)\leq \epsilon.\label{EQ:achieve_secrecy}
\end{align}\label{EQ:achiev_realibility_secrecy}%
\end{subequations}
\end{definition}

\begin{definition}[Secrecy-Capacity Region]
The strong secrecy-capacity region $\mathcal{C}_\mathsf{S}$ is the closure of the set of the achievable rates.
\end{definition}


\subsection{Strong Secrecy-Capacity Bounds and Results}\label{SUBSEC:BC_result}

\par We state an inner bound on the strong secrecy-capacity region $\mathcal{C}_\mathsf{S}$ of a cooperative BC with one confidential message.

\begin{theorem}[Inner Bound]\label{TM:inner_bound}
Let $W_{Y_1,Y_2|X}$ be a transition probability of a BC and let $\mathcal{R}_\mathsf{I}$ be the closure of the union of rate tuples $(R_{12},R_0,R_1,R_2)\in\mathbb{R}^4_+$ satisfying:
\begin{subequations}
\begin{align}
R_1&\leq I(U_1;Y_1|U_0)-I(U_1;U_2,Y_2|U_0)\label{EQ:region_inner1}\\
R_0+R_1 &\leq I(U_0,U_1;Y_1)-I(U_1;U_2,Y_2|U_0)\label{EQ:region_inner01}\\
R_0+R_2&\leq I(U_0,U_2;Y_2)+R_{12}\label{EQ:region_inner02}\\
R_0+R_1+R_2&\leq I(U_0,U_1;Y_1)+I(U_2;Y_2|U_0)\nonumber
\\&\mspace{124mu}-I(U_1;U_2,Y_2|U_0)\label{EQ:region_inner012}
\end{align}\label{EQ:region_inner}%
\end{subequations}
where the union is over all PMFs $Q_{U_0,U_1,U_2,X}\in\mathcal{P}(\mathcal{U}_0\times\mathcal{U}_1\times\mathcal{U}_2\times\mathcal{X})$, each inducing a joint distribution $Q_{U_0,U_1,U_2,X}W_{Y_1,Y_2|X}$. Then the following inclusion holds:
\begin{equation}
\mathcal{R}_\mathsf{I}\subseteq\mathcal{C}_\mathsf{S}.\label{EQ:inclusion_inner}
\end{equation}
Furthermore, $\mathcal{R}_\mathsf{I}$ is convex and one may choose $|\mathcal{U}_0|\leq|\mathcal{X}|+5$, $|\mathcal{U}_1|\leq|\mathcal{X}|$ and  $|\mathcal{U}_2|\leq|\mathcal{X}|$.
\end{theorem}


The proof of Theorem \ref{TM:inner_bound} relies on a channel-resolvability-based Marton code and is given in Section \ref{SUBSEC:inner_proof}. Two key ingredients allow us to keep $M_1$ secret while still utilizing the cooperation link to help Receiver 2. First, the cooperation strategy is modified compared to the case without secrecy that was studied in \cite{Goldfeld_BC_Cooperation2014}, where $M_{12}$ conveyed information about \emph{both} private messages as well as the common message. Here, the confidentiality of $M_1$ restricts the cooperation message from containing any information about $M_1$, and therefore, we use an $M_{12}$ that is a function of the decoded $\big(\hat{M}^{(2)}_0,\hat{M}_2)$ only. Since the protocol requires Receiver 1 to decode the information it shares with Receiver 2, this modified cooperation strategy results in a rate loss in $R_1$ when compared to \cite{Goldfeld_BC_Cooperation2014}; the loss is expressed in the first mutual information term in \eqref{EQ:region_inner1} being conditioned on $U_0$ rather than having $U_0$ next to~$U_1$.

The second ingredient is associating with each $m_1\in\mathcal{M}_1$ a resolvability-subcode that adheres to the construction for Lemmas \ref{LEMMA:soft_covering} and \ref{LEMMA:typicallity} described in Section \ref{SEC:soft_covering}. By doing so, the relations between the codewords in the Marton code correspond to those between the channel states and its input in the resolvability problem. Marton coding combines superposition coding and binning, hence the state sequences $\mathbf{S}_0$ and $\mathbf{S}$ play different roles in our resolvability setup. Reliability is established with the help of Lemma \ref{LEMMA:typicallity}, while Lemma \ref{LEMMA:soft_covering} essentially produces strong secrecy.

The inner bound from Theorem \ref{TM:inner_bound} is tight for SD- and PD-BCs, giving rise to the new strong secrecy-capacity results stated in Theorems \ref{TM:SDBC_secrecy_capacity} and \ref{TM:PDBC_secrecy_capacity}.

\begin{theorem}[SD-BC Secrecy-Capacity]\label{TM:SDBC_secrecy_capacity}
The strong secrecy-capacity region $\mathcal{C}^{(\mathsf{SD})}_\mathsf{S}$ of a cooperative SD-BC $\mathds{1}_{\{Y_1=y_1(X)\}}W_{Y_2|X}$ with one confidential message is the closure of the union of rate tuples $(R_{12},R_0,R_1,R_2)\in\mathbb{R}^4_+$ satisfying:
\begin{subequations}
\begin{align}
R_1 &\leq H(Y_1|W,V,Y_2)\label{EQ:region_SDBC11}\\
R_0+R_1 &\leq H(Y_1|W,V,Y_2)+I(W;Y_1)\label{EQ:region_SDBC12}\\
R_0+R_2 &\leq I(W,V;Y_2)+R_{12}\label{EQ:region_SDBC22}\\
R_0+R_1+R_2 &\leq H(Y_1|W,V,Y_2)+I(V;Y_2|W)+I(W;Y_1)\label{EQ:region_SDBC_sum}
\end{align}\label{EQ:region_SDBC}%
\end{subequations}
where the union is over all PMFs $Q_{W,V,Y_1,X}\in\mathcal{P}(\mathcal{W}\times\mathcal{V}\times\mathcal{Y}_1\times\mathcal{X})$ with $Y_1=y_1(X)$, each inducing a joint distribution $Q_{W,V,Y_1,X}W_{Y_2|X}$. Furthermore, $\mathcal{C}^{(\mathsf{SD})}_\mathsf{S}$ is convex and one may choose $|\mathcal{W}|\leq|\mathcal{X}|+3$ and $|\mathcal{V}|\leq|\mathcal{X}|$.
\end{theorem}

The direct part of Theorem \ref{TM:SDBC_secrecy_capacity} follows from Theorem \ref{TM:inner_bound} by setting $U_0=W$, $U_1=Y_1$ and $U_2=V$. The converse is proven in Section \ref{SUBSEC:SD_proof_converse}.

\begin{theorem}[PD-BC Secrecy-Capacity]\label{TM:PDBC_secrecy_capacity}
The strong secrecy-capacity region $\mathcal{C}^{(\mathsf{PD})}_\mathsf{S}$ of a cooperative PD-BC $W_{Y_1|X}W_{Y_2|Y_1}$ with one confidential message is the closure of the union of rate tuples $(R_{12},R_0,R_1,R_2)\in\mathbb{R}^4_+$ satisfying:
\begin{subequations}
\begin{align}
R_1 &\leq I(X;Y_1|W)-I(X;Y_2|W)\label{EQ:region_PDBC1}\\
R_0+R_2 &\leq I(W;Y_2)+R_{12}\label{EQ:region_PDBC02}\\
R_0+R_1+R_2 &\leq I(X;Y_1)-I(X;Y_2|W)\label{EQ:region_PDBC_sum}
\end{align}\label{EQ:region_PDBC}%
\end{subequations}
where the union is over all PMFs $Q_{W,X}\in\mathcal{P}(\mathcal{W}\times\mathcal{X})$, each inducing a joint distribution $Q_{W,X}W_{Y_1|X}W_{Y_2|Y_1}$. Furthermore, $\mathcal{C}^{(\mathsf{PD})}_\mathsf{S}$ is convex and one may choose $|\mathcal{W}|\leq|\mathcal{X}|+2$.
\end{theorem}

The achievability of $\mathcal{C}^{(\mathsf{PD})}_\mathsf{S}$ is a consequence of Theorem \ref{TM:inner_bound} by taking $U_0=W$, $U_1=X$ and $U_2=0$. For the converse see Section \ref{SUBSEC:PD_proof_converse}.

\begin{remark}[Converse]
We use two distinct converse proofs for Theorems \ref{TM:SDBC_secrecy_capacity} and \ref{TM:PDBC_secrecy_capacity}. In the converse of Theorem \ref{TM:SDBC_secrecy_capacity}, the bound in \eqref{EQ:region_SDBC_sum} does not involve $R_{12}$ since the auxiliary random variable $W_i$ contains $M_{12}$. With respect to this choice of $W_i$ (see \eqref{EQ:SD_converse_r1UB_final}), showing that $W-X-(Y_1,Y_2)$ forms a Markov chain relies on the SD property of the channel. For the PD-BC, however, such an  auxiliary is not feasible as it violates the Markov relation $W-X-Y_1-Y_2$ induced by the channel. To circumvent this, in the converse of Theorem \ref{TM:PDBC_secrecy_capacity} we define $W_i$ without $M_{12}$ and use the structure of the channel to keep $R_{12}$ from appearing in \eqref{EQ:region_PDBC_sum}. Specifically, this argument relies on the relation $M_{12}=g^{(n)}_{12}(\mathbf{Y}_1)$ and on $Y_2$ being a degraded version of $Y_1$ (which implies that all three messages $(M_0,M_1,M_2)$ can be reliably decoded from $\mathbf{Y}_1$ only) .
\end{remark}

\begin{remark}[Weak versus Strong Secrecy]
The results of Theorems \ref{TM:inner_bound}, \ref{TM:SDBC_secrecy_capacity} and \ref{TM:PDBC_secrecy_capacity} remain unchanged if the strong secrecy requirement (see \eqref{EQ:infromation_leakage_def} and \eqref{EQ:achieve_secrecy}) is replaced with the weak secrecy constraint. As weak secrecy refers to a vanishing normalized information leakage, to formally define the corresponding achievability, one should replace the left-hand side (LHS) of \eqref{EQ:achieve_secrecy} with $\frac{1}{n}\ell(c_n)$. To see that the results of the preceding theorems coincide under both metrics, first notice that strong secrecy implies weak secrecy (which validates the claim from Theorem \ref{TM:inner_bound}). Furthermore, the converse proofs of Theorems \ref{TM:SDBC_secrecy_capacity} and \ref{TM:PDBC_secrecy_capacity} (given in Sections \ref{SUBSEC:SD_proof_converse} and \ref{SUBSEC:PD_proof_converse}, respectively) are readily reformulated under the weak secrecy metric by replacing $\epsilon$ with $n\epsilon$ in \eqref{EQ:SD_converse_secrecy_cond_info}-\eqref{EQ:SD_converse_secrecy_cond_info2} and \eqref{EQ:PD_converse_secrecy_cond_info}-\eqref{EQ:PD_converse_secrecy_cond_info2}.
\end{remark}

\begin{remark}[Cardinality Bounds]
The cardinality bounds on the auxiliary random variables in Theorems \ref{TM:inner_bound}, \ref{TM:SDBC_secrecy_capacity} and \ref{TM:PDBC_secrecy_capacity} are established using the perturbation method \cite{Perturbation2012} and the 
Eggleston-Fenchel-Carath{\'e}odory theorem \cite[Theorem 18]{Eggleston_Convexity1958}.
\end{remark}


\section{Restricted Cooperation Scheme is Sub-Optimal Without Secrecy Constraints}\label{SEC:SDBC_cooperation_effect}

The cooperation protocol for the BC with a secret $M_1$ uses the cooperative link to convey information that is a function of the non-confidential message and the common message. Without secrecy constraints, it was shown in \cite{Goldfeld_BC_Cooperation2014} that the best cooperation strategy uses a public message that comprises parts of \emph{both} private messages as well as the common message. To understand whether the restricted protocol reduces the transmission rates beyond standard losses due to secrecy (which are discussed in Section \ref{SEC:secrecy_effect}), we compare the achievable regions induced by each scheme for the cooperative BC \emph{without secrecy}. The formal description of this BC instance (see \cite{Goldfeld_BC_Cooperation2014}) closely follows the definitions from Section \ref{SUBSEC:BC_definition} up to removing the security requirement \eqref{EQ:achieve_secrecy} from Definition \ref{DEF:achievability} of achievability. For simplicity we consider the setting without a common message, i.e., when $R_0=0$.

To isolate the (possible) rate-loss due to the restricted cooperation scheme used in this paper from other losses due to secrecy, we subsequently describe an adaptation of our coding scheme to the case where $M_1$ is not confidential. Namely, we remove the secrecy requirement on $M_1$ but still limit the cooperation protocol to share information on $M_2$ only. This results in an achievable scheme for the cooperative BC with no security requirements, and the induced achievable region is compared with the result from \cite{Goldfeld_BC_Cooperation2014}.

At first glance it might seem that even without secrecy requirements, the restricted cooperation protocol is optimal. After all, why should the cooperative receiver (Decoder 1) share information about $M_1$ with the cooperation-aided receiver (Decoder 2), which is not required to decode it? Yet, we show that this intuitive argument fails and that the restricted protocol is sub-optimal in general. For BCs in which Decoder 1 can decode more than $nR_{12}$ bits of $M_2$ (e.g., PD-BCs), both protocols achieve the same rates and $M_1$ need not be shared. However, when Decoder 1 can decode strictly less than $nR_{12}$ bits of $M_2$, then sharing $M_1$ achieves higher $R_2$ values, since now $M_1$ serves as side information for Decoder 2 in decoding $M_2$ (note that this side information is also available at the encoder).


The achievable region $\mathcal{R}_{\mathsf{NS}}$ for the cooperative BC $W_{Y_1,Y_2|X}$ without secrecy that was characterized in \cite{Goldfeld_BC_Cooperation2014} (see also \cite{Liang_Veeravalli_RBC2007,Liang_Kramer_RBC2007}) is the union over the same domain as \eqref{EQ:region_inner} of rate triples $(R_{12},R_1,R_2)\in\mathbb{R}^3_+$ satisfying:
\begin{subequations}
\begin{align}
\mspace{-8mu}R_1 &\mspace{-1.5mu}\leq\mspace{-1.5mu} I(U_0,U_1;Y_1)\label{region_nosec_SDBC1}\\
\mspace{-8mu}R_2 &\mspace{-1.5mu}\leq\mspace{-1.5mu} I(U_0,U_2;Y_2)+R_{12}\label{region_nosec_SDBC2}\\
\mspace{-8mu}R_1\mspace{-3.5mu}+\mspace{-3.5mu}R_2 &\mspace{-1.5mu}\leq\mspace{-1.5mu} I(U_0,U_1;Y_1)\mspace{-3.5mu}+\mspace{-3.5mu}I(U_2;Y_2|U_0)\mspace{-3.5mu}-\mspace{-3.5mu}I(U_1;U_2|U_0)\label{region_nosec_SDBC1+2a}\\
\mspace{-8mu}R_1\mspace{-3.5mu}+\mspace{-3.5mu}R_2 &\mspace{-1.5mu}\leq\mspace{-1.5mu} I(U_1;\mspace{-1mu}Y_1|U_0)\mspace{-3.5mu}+\mspace{-3.5mu}I(U_0,\mspace{-1.5mu}U_2;\mspace{-1mu}Y_2)\mspace{-3.5mu}-\mspace{-3.5mu}I(U_1;\mspace{-1mu}U_2|U_0\mspace{-1mu})\mspace{-3mu}+\mspace{-3mu}R_{12}.\label{region_nosec_SDBC1+2b}
\end{align}\label{region_nosec_SDBC}%
\end{subequations}
The cooperation scheme that achieves \eqref{region_nosec_SDBC} uses the pair $(M_{10},M_{20})$ (where $M_{j0}$ refers to the public part of the message $M_j$ and has rate $R_{j0}\leq R_j$, for $j=1,2$) as a public message that is decoded by both users. The public message codebook (generated by i.i.d. samples of the random variable $U_0$ in \eqref{region_nosec_SDBC}) is partitioned into $2^{nR_{12}}$ bins and is first decoded by User 1. The partitioning is defined by a mapping $m_{12}:\big[1:2^{nR_{10}}\big]\times\big[1:2^{nR_{20}}\big]\to\mathcal{M}_{12}^{(n)}$ and the bin number $m_{12}\big((\hat{M}_{10},\hat{M}_{20})\big)$ of the decoded public message is shared with User 2 over the cooperative link. This reduces the search space by a factor of $2^{nR_{12}}$. The dependence of the public message on $\hat{M}_{10}$ essentially allows User 1 to achieve rates up $I(U_0,U_1;Y_1)$.

The cooperation protocol used in this work (constructed to account for the secrecy constraint on $M_1$) removes $M_{10}$ from the public message, while keeping the rest of the protocol unchanged. The region $\tilde{\mathcal{R}}_{\mathsf{NS}}$ achieved by the restricted cooperation protocol is derived by repeating the steps in the proof of \cite[Theorem 6]{Goldfeld_BC_Cooperation2014} while setting $R_{10}=0$. One obtains that $\tilde{\mathcal{R}}_{\mathsf{NS}}$ is characterized by the same rate bounds as \eqref{region_nosec_SDBC}, up to replacing \eqref{region_nosec_SDBC1} with 
\begin{equation}
R_1 \leq I(U_1;Y_1|U_0)+\Big[I(U_2;Y_2|U_0)-I(U_1;U_2|U_0)\Big]^+\label{region_nosec_subopt_SDBC1}
\end{equation}
where $[x]^+=\max\big\{0,x\big\}$. Since $\tilde{\mathcal{R}}_{\mathsf{NS}}$ is achieved by specializing the scheme that achieves $\mathcal{R}_{\mathsf{NS}}$ (i.e., setting $R_{10}=0$ therein), we have that $\tilde{\mathcal{R}}_{\mathsf{NS}}\subseteq\mathcal{R}_{\mathsf{NS}}$.

Note that $\tilde{\mathcal{R}}_{\mathsf{NS}}=\mathcal{R}_{\mathsf{NS}}$ for any BC where setting $U_0=0$ in \eqref{region_nosec_SDBC} is optimal. In particular, we have the following proposition.
\begin{proposition}[Optimality of Restricted Protocol]\label{PROP:cooperation_schemes_same}
If a BC $W_{Y_1,Y_2|X}$ is PD or deterministic, i.e., it satisfies $W_{Y_1,Y_2|X}=W_{Y_1|X}W_{Y_2|Y_1}$ or $W_{Y_1,Y_2|X}=\mathds{1}_{\{Y_1=y_1(X)\}\cap\{Y_2=y_2(X)\}}$, respectively, then $\tilde{\mathcal{R}}_{\mathsf{NS}}=\mathcal{R}_{\mathsf{NS}}=\mathcal{C}_{\mathsf{NS}}$.
\end{proposition}

\begin{IEEEproof}
For the PD-BC, setting $U_0=W$, $U_1=X$ and $U_2=0$ into $\tilde{\mathcal{R}}_{\mathsf{NS}}$ recovers the region from \cite[Equation (17)]{Dikstein_PDBC_Cooperation2014}, which is the capacity region of the cooperative PD-BC. The capacity region of the cooperative deterministic BC (DBC) given in \cite[Corollary 12]{Goldfeld_BC_Cooperation2014} is recovered from $\tilde{\mathcal{R}}_{\mathsf{NS}}$ by taking $U_0=0$, $U_1=Y_1$ and $U_2=Y_2$.
\end{IEEEproof}


\begin{proposition}[Restricted Protocol can be Sub-Optimal]\label{PROP:cooperation_schemes_not_same}
There exist BCs $W_{Y_1,Y_2|X}$ for which $\tilde{\mathcal{R}}_{\mathsf{NS}}\subsetneq\mathcal{R}_{\mathsf{NS}}$.
\end{proposition}

The proof of Proposition \ref{PROP:cooperation_schemes_not_same} is given in Appendix \ref{APPEN:cooperation_schemes_not_same_proof}, where we construct an example for which the maximal achievable $R_1$ in both regions is the same, but the highest achievable $R_2$ while keeping $R_1$ at its maximum is strictly smaller in $\tilde{\mathcal{R}}_{\mathsf{NS}}$. 

\begin{figure}[!t]
\begin{center}
\begin{psfrags}
    \psfragscanon
    \psfrag{A}[][][1]{$X_1$}
    \psfrag{C}[][][1]{$X_2$}
    \psfrag{B}[][][1]{$\ \ \ Y_1$}
    \psfrag{D}[][][1]{$\mspace{-35mu}Y_2$}
    \psfrag{E}[][][1]{$\ \ \ \ \ \ \ \ \ W_{Y_2|X_1,X_2}$}
    \psfrag{F}[][][1]{$\ \ \ \ \ \ \ \ \ \mathrm{BSC}(0.1)$}
    \includegraphics[scale=0.5]{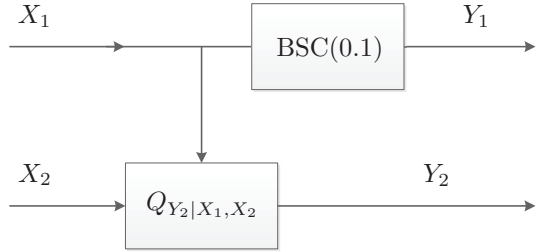}
\caption{A semi-orthogonal BC.} \label{FIG:counter_example}
\psfragscanoff
\end{psfrags}
\end{center}
\end{figure}

We start with a family of BCs as illustrated in Fig. \ref{FIG:counter_example}, where the channel input is $X=(X_1,X_2)$, the output $Y_1$ is produced by feeding $X_1$ into a binary symmetric channel (BSC) with crossover probability\footnote{The actual value of the crossover probability is of no real importance as long as it is not $0.5$.} $0.1$, while $Y_2$ is generated by the DMC $W_{Y_2|X_1,X_2}$. All alphabets are binary, i.e., $\mathcal{X}_1=\mathcal{X}_2=\mathcal{Y}_1=\mathcal{Y}_2=\big\{0,1\big\}$. The maximal achievable $R_1$ in both schemes is the capacity of the aforementioned BSC, i.e., $c\triangleq 1-H_b(0.1)$, where $H_b:[0,1]\to [0,1]$ is the binary entropy function. Setting the capacity of the cooperation link to $R_{12}=c$, we show that the highest $R_2$ such that $(R_{12},R_1,R_2)=(c,c,R_2)\in\mathcal{R}_{\mathsf{NS}}$ is lower bounded by the capacity of the state-dependent channel $W_{Y_2|X_1,X_2}$ (with $X_1$ and $X_2$ playing the roles of the state and the input, respectively) with non-causal channel state information (CSI) available at the transmitting and receiving ends. This is because $R_{12}=c$ in the permissive protocol allows Decoder 1 to share the decoded $\mathbf{X}_1$ with Decoder 2 despite its dependence on $M_1$.


The corresponding value of $R_2$ in $\tilde{\mathcal{R}}_{\mathsf{NS}}$ is then upper bounded by the capacity of the same channel but with non-causal CSI at the transmitter only (also known as a Gelfand-Pinsker (GP) channel). The cooperation link is, in fact, useless in this scenario since the entire capacity of the BSC was used to reliably convey bits of $M_1$, on which the restricted protocol prohibits exchanging information. Thus, the proof boils down to choosing $W_{Y_2|X_1,X_2}$ as a channel for which the capacity with full CSI is strictly larger than the GP capacity. The binary dirty-paper (BDP) channel \cite{Zamir_BDPC2002,Barron_BDPC2003,Toly_BDPMAC2010} qualifies and completes the proof.

\section{Effect of Secrecy on the Capacity-Region of Cooperative Broadcast Channels}\label{SEC:secrecy_effect}


The impact of the secrecy constraint on $M_1$ on the cooperation strategy and the resulting reduction of transmission rates was discussed in Section \ref{SEC:SDBC_cooperation_effect}. However, secrecy requirements affect BC codes even when no user cooperation is allowed. Thus, when considering a scenario that combines secrecy and cooperation, both these effects occur simultaneously. We highlight this by comparing the SD and PD versions of the cooperative BC to their corresponding models without secrecy. For simplicity, throughout this section we again assume BCs with private messages only, i.e., $R_0=0$.



\subsection{Semi-Deterministic Broadcast Channels}\label{SUBSEC:SDBC_comparison}


\subsubsection{Capacity Region Comparison}\label{SUBSUBSEC:SDBC_comparison_discussion} 


Consider the SD-BC without cooperation (i.e., where $R_{12}=0$) in which $M_1$ is secret. By Theorem \ref{TM:SDBC_secrecy_capacity}, the strong secrecy-capacity region of the SD-BC with one confidential message, which was an unsolved problem until this work, is as follows.


\begin{corollary}[Non-Cooperative SD-BC Secrecy-Capacity]\label{TM:SDBC_capacity_nocoop}
The strong secrecy-capacity region $\tilde{\mathcal{C}}^{(\mathsf{SD})}_\mathsf{S}$ of the SD-BC $\mathds{1}_{\{Y_1=y_1(X)\}}W_{Y_2|X}$ with one confidential message is the union of rate pairs $(R_1,R_2)\in\mathbb{R}^2_+$ satisfying:
\begin{subequations}
\begin{align}
R_1 &\leq H(Y_1|V,Y_2)\label{EQ:region_SDBC_nocoop1}\\
R_2 &\leq I(V;Y_2)\label{EQ:region_SDBC_nocoop2}
\end{align}\label{EQ:region_SDBC_nocoop}%
\end{subequations}
where the union is over all PMFs $Q_{V,Y_1,X}\in\mathcal{P}(\mathcal{V}\times\mathcal{Y}_1\times\mathcal{X})$ with $Y_1=y_1(X)$, each inducing a joint distribution $Q_{V,Y_1,X}W_{Y_2|X}$.
\end{corollary}

The region \eqref{EQ:region_SDBC_nocoop} coincides with $\mathcal{C}^{(\mathsf{SD})}_\mathsf{S}$ in \eqref{EQ:region_SDBC_sum} (where $R_{12}=R_0=0$) by noting that the bound \eqref{EQ:region_SDBC_sum} is redundant because if $Q_{W,V,Y_1,X}$ is a PMF for which \eqref{EQ:region_SDBC_sum} is active, then replacing $W$ and $V$ with $\tilde{W}=0$ and $\tilde{V}=(W,V)$ achieves a larger region. Removing \eqref{EQ:region_SDBC_sum} from $\mathcal{C}^{(\mathsf{SD})}_\mathsf{S}$ and setting $\tilde{V}=(W,V)$ recovers \eqref{EQ:region_SDBC_nocoop}.

Marton coding achieves the capacity region of the classic SD-BC \cite{GP_SemideterministicBC1980}. The capacity is the union of rate pairs $(R_1,R_2)\in\mathbb{R}^2_+$ satisfying:
\begin{subequations}
\begin{align}
R_1 &\leq H(Y_1)\label{EQ:region_SDBC_classical1}\\
R_2 &\leq I(V;Y_2)\label{EQ:region_SDBC_classical2}\\
R_1+R_2 &\leq H(Y_1|V)+I(V;Y_2)\label{EQ:region_SDBC_classical1+2}
\end{align}\label{EQ:region_SDBC_classical}%
\end{subequations}
where the union is over the same domain as in Corollary \ref{TM:SDBC_capacity_nocoop}.


\begin{figure}[t]
\begin{center}
\begin{psfrags}
    \psfragscanon
    \psfrag{A}[][][0.9]{$\ \ \ R_1$}
    \psfrag{G}[][][0.9]{$R_2$}
    \psfrag{D}[][][0.9]{$0$}
    \psfrag{E}[][][0.9]{\ $I(V;Y_2)-I(V;Y_1)$}
    \psfrag{C}[][][0.9]{$\ \ \ \ H(Y_1|V)$}
    \psfrag{F}[][][0.9]{$I(V;Y_2)$}
    \psfrag{H}[][][0.9]{$H(Y_1|V,Y_2)$}
    \psfrag{B}[][][0.9]{$\ \ \ \ H(Y_1)$}
    \psfrag{I}[][][0.9]{$\ \ \ \ I(Y_1;Y_2|V)$}
    \psfrag{X}[][][0.9]{$\ \ \ \ \ \ I(V;Y_1)$}
    \psfrag{J}[][][0.8]{\ \ \ \ \ \ \ \ Secrecy}
    \psfrag{K}[][][0.8]{\ \ \ \ \ \ \ \ \ \ \ \ No secrecy}
\includegraphics[scale=0.43]{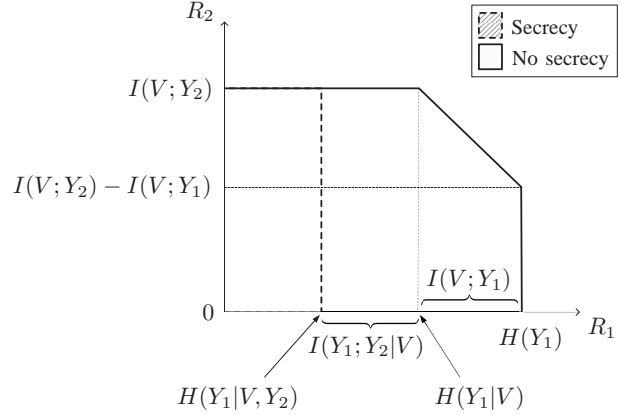}
\caption{Capacity region without secrecy vs. strong secrecy-capacity region where $M_1$ is confidential for the SD-BC (without cooperation).} \label{FIG:SDBC_region_comparison}
\psfragscanoff
\end{psfrags}
\end{center}
\end{figure}


\par The regions in \eqref{EQ:region_SDBC_nocoop} and \eqref{EQ:region_SDBC_classical} (for a fixed $Q_{W,Y_1,X}$) are depicted in Fig. \ref{FIG:SDBC_region_comparison}. When $M_1$ is secret, one can no longer operate on both corner points of Marton's region. Rather, the optimal coding scheme is the one with the lower transmission rate to the 1st user. This essentially means that the redundancy in the codebook needed for multicoding befalls solely on User 1 (whose message is to be kept secret). Consequently, a loss of $I(V;Y_1)$, which corresponds to the sizes of the bins used for joint encoding, is inflicted on $R_1$. An additional rate-loss of $I(Y_1;Y_2|V)$ in $R_1$ is caused by a second layer of binning used to conceal $M_1$ from the 2nd user. A coding scheme for the higher corner point of the region without secrecy, i.e., the point $\big(\mspace{3mu}H(Y_1)\mspace{3mu},\mspace{3mu}I(V;Y_2)-I(V;Y_1)\mspace{3mu}\big)$, is not feasible with secrecy since the larger value of $R_1$ violates the secrecy constraint. A similar effect occurs for the corresponding regions with cooperation.


\subsubsection{Blackwell BC Example}\label{SUBSUBSEC:blackwell}


\begin{figure}[t!]
    \begin{center}
        \begin{psfrags}
            \psfragscanon
            \psfrag{A}[][][1]{$0$}
            \psfrag{B}[][][1]{$1$}
            \psfrag{C}[][][1]{$2$}
            \psfrag{D}[][][1]{\ $X$}
            \psfrag{E}[][][1]{\ \ }
            \psfrag{F}[][][1]{\ \ \ $Y_1$}
            \psfrag{G}[][][1]{$0$}
            \psfrag{H}[][][1]{$1$}
            \psfrag{I}[][][1]{$0$}
            \psfrag{J}[][][1]{$1$}
            \psfrag{K}[][][1]{\ \ \ }
            \psfrag{L}[][][1]{\ \ \ $Y_2$}
            \psfrag{M}[][][1]{\ \ \ }
            \psfrag{N}[][][1]{\ \ \ $R_{12}$}
            \psfrag{X}[][][1]{\ \ \ \ \ \ \ \ \ \ \ \ \ \ \ \ \ \ }
            \psfrag{Y}[][][1]{\ \ \ \ \ \ \ \ \ \ \ \ \ \ \ \ \ \ }
            \psfrag{O}[][][1]{$2$}
            \hspace{-2mm}\subfloat[]{\includegraphics[scale = .5]{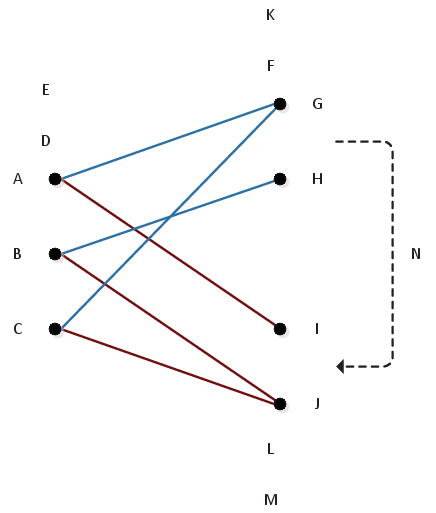}}\ \ \ \ \ \ \ \ \ \hspace{-1mm}\subfloat[]{\includegraphics[scale = .5]{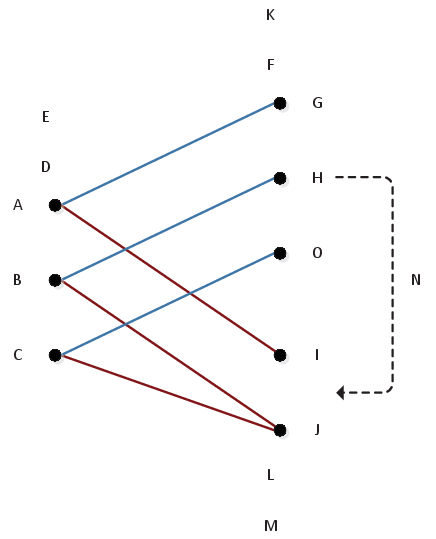}}
            \caption{(a) Cooperative Blackwell BC; (b) Cooperative Blackwell-like PD-BC.} \label{FIG:blackwell}
            \psfragscanoff
        \end{psfrags}
     \end{center}
 \end{figure}




Suppose the channel from the transmitter to receivers 1 and 2 is the BW-BC without a common message as illustrated in Fig \ref{FIG:blackwell}(a) \cite{vanderMeulen_Blackwell1975,Gelfand_Blackwell1977}. Noting that the BW-BC is deterministic, we set $R_0=0$ into the region from Theorem \ref{TM:SDBC_secrecy_capacity} to characterize the strong secrecy-capacity region of a DBC as follows.

\begin{corollary}
[DBC Secrecy-Capacity]\label{COR:DBC_secrecy_capacity}
The strong secrecy-capacity region $\mathcal{C}^{(\mathsf{D})}_\mathsf{S}$ of a cooperative DBC $\mathds{1}_{\{Y_1=y_1(X)\}\cap\{Y_2=y_2(X)\}}$ with one confidential message is the union of rate triples $(R_{12},R_1,R_2)\in\mathbb{R}^3_+$ satisfying:
\begin{subequations}
\begin{align}
R_1 &\leq H(Y_1|Y_2)\label{EQ:region_DBC1}\\
R_2 &\leq H(Y_2)+R_{12}\label{EQ:region_DBC2}\\
R_1+R_2 &\leq H(Y_1,Y_2)\label{EQ:region_DBC12}
\end{align}\label{EQ:region_DBC}%
\end{subequations}
where the union is over all input distributions $Q_X\in\mathcal{P}(\mathcal{X})$.
\end{corollary}

Corollary \ref{COR:DBC_secrecy_capacity} follows by arguments similar to those in the proof of \cite[Corollary 12]{Goldfeld_BC_Cooperation2014}. By parameterizing the input PMF $Q_X$ as
\begin{equation}
Q_X(0)=\alpha\ ,\ Q_X(1)=\beta\ ,\ Q_X(2)=1-\alpha-\beta\label{EQ:BBC_input_parameter}
\end{equation}
where $\alpha,\beta\in\mathbb{R}_+$ and $\alpha+\beta\leq1$, the strong secrecy-capacity region $\mathcal{C}^{(\mathsf{BW})}_\mathsf{S}$ of the BW-BC is the union of rate pairs $(R_1,R_2)\in\mathbb{R}^2_+$ satisfying:
\begin{subequations}
\begin{align}
R_1&\leq (1-\alpha)H_b\left(\frac{\beta}{1-\alpha}\right)\\
R_2&\leq H_b(\alpha)+R_{12}\\
R_1+R_2&\leq H_b(\alpha)+(1-\alpha)H_b\left(\frac{\beta}{1-\alpha}\right)
\end{align}\label{EQ:region_BBC}%
\end{subequations}
where the union is over all $\alpha,\beta\in\mathbb{R}_+$ with $\alpha+\beta\leq 1$.


\begin{figure}[t!]
    \begin{center}
        \begin{psfrags}
            \psfragscanon
            \psfrag{A}[][][0.9]{$R_1$ [bits/use]}
            \psfrag{B}[][][0.9]{$R_2$ [bits/use]}
            \psfrag{C}[][][0.8]{\ \ \ \ \ \ \ \ \ \ $\mspace{-4mu}R_{12}=0$}
            \psfrag{D}[][][0.8]{\ \ \ \ \ \ \ \ \ \ \ $\mspace{-2mu}R_{12}\mspace{-2mu}=\mspace{-2mu}0.2$}
            \psfrag{E}[][][0.8]{\ \ \ \ \ \ \ \ \ \ \ $\mspace{-2mu}R_{12}\mspace{-2mu}=\mspace{-2mu}0.4$}
            \psfrag{F}[][][0.8]{\ \ \ \ \ \ \ \ \ \ \ $\mspace{-2mu}R_{12}\mspace{-2mu}=\mspace{-2mu}0.6$}
            \psfrag{I}[][][0.9]{\ \ \ \ \ \ \ \ \ \ \ \ \ \ No secrecy}
            \psfrag{H}[][][0.9]{\ \ \ \ \ \ \ \ \ \ Secrecy}
            \psfrag{X}[][][0.8]{\ \ \ (b)}
            \psfrag{Y}[][][0.8]{\ \ \ (a)}
            \hspace{3mm}\includegraphics[scale = 0.52]{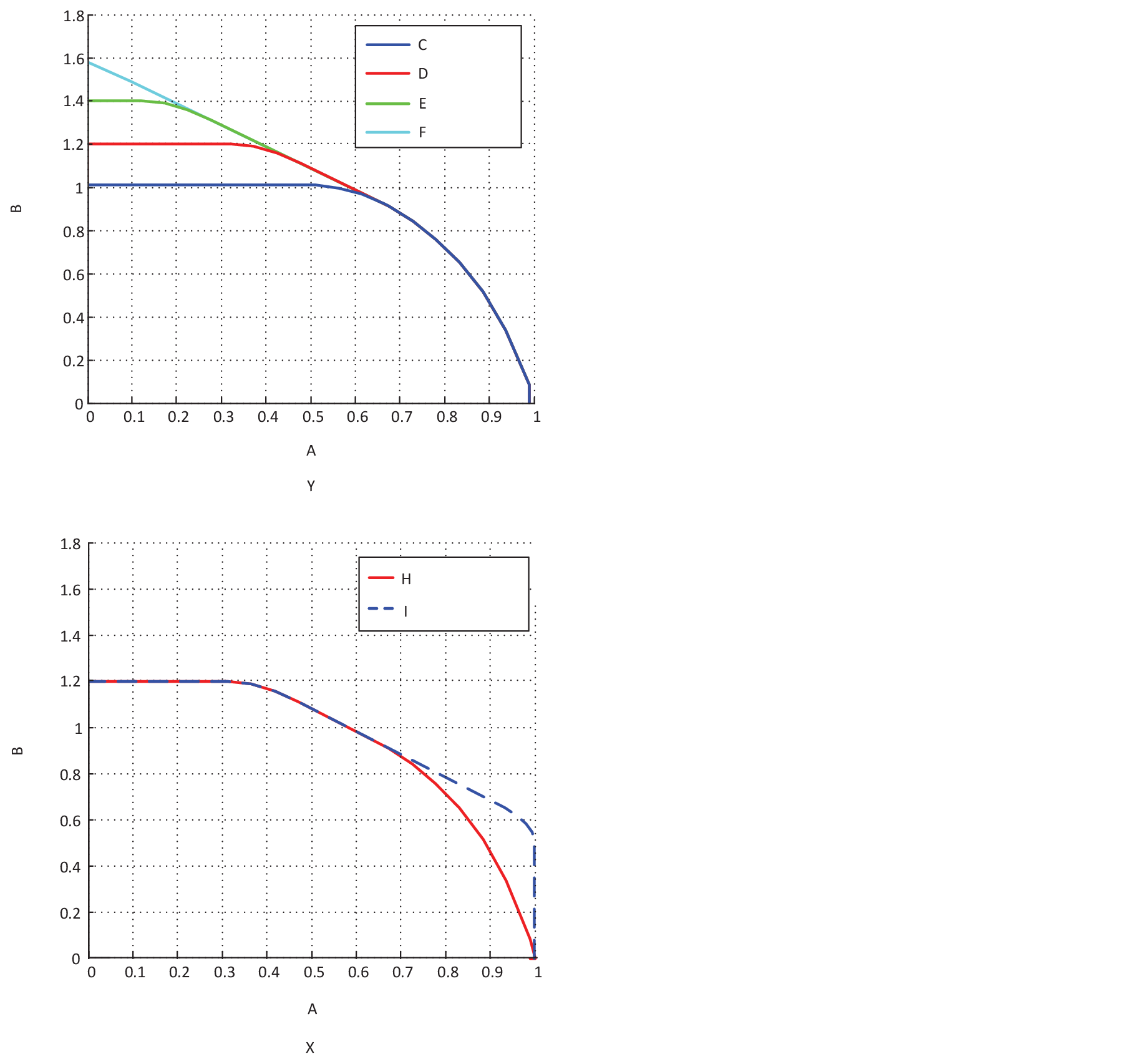}
            \caption{(a) Projection of the strong secrecy-capacity region of the cooperative BW-BC with one confidential message onto the plane $(R_1,R_2)$ for different values of $R_{12}$; (b) Cooperative BW-BC with $R_{12}=0.2$: Strong secrecy-capacity region where $M_1$ is confidential vs. Capacity region without secrecy.} \label{FIG:blackwell_capacity}
            \psfragscanoff
        \end{psfrags}
     \end{center}
 \end{figure}


The projection of $\mathcal{C}^{(\mathsf{BW})}_\mathsf{S}$ onto the plane $(R_1,R_2)$ for different values of $R_{12}$ is shown in Fig. \ref{FIG:blackwell_capacity}(a). For every $R_{12}\in\mathbb{R}_+$, the maximal achievable $R_1$ in $\mathcal{C}^{(\mathsf{BW})}_\mathsf{S}$ equals 1 [bits/use] (while the corresponding $R_2$ is zero). The rate triple $(R_{12},1,0)$ is achieved by setting $\alpha=0$ and $\beta=\frac{1}{2}$ in the bounds in \eqref{EQ:region_BBC}. These probability values provide insight into the coding strategy that maximizes the transmission rate to User 1. Namely, the encoder chooses each channel input symbol uniformly from the set $\{1,2\}\subsetneq\mathcal{X}$. By doing so, Decoder 1 effectively sees a clean binary channel (by mapping every received $Y_1=0$ to the input symbol $X=2$) with capacity 1. Decoder 2, on the other hand, sees a flat channel with zero capacity since both $X=1$ and $X=2$ are mapped to $Y_2=1$. Thus, Decoder 2 has no information about the transmitted sequence, and therefore, strong secrecy is achieved while conveying one secured bit to Decoder 1 in each channel use.

\begin{remark}[Clean Channel to User 1 Does Not Help] An improved subchannel to the legitimate user does not enlarge the strong secrecy-capacity region. We illustrate this by considering the BW-like PD-BC shown in Fig. \ref{FIG:blackwell}(b), where $\mathcal{Y}_1=\mathcal{X}$ and $Y_1=X$ ($\mathcal{Y}_2$ and the mapping from $\mathcal{X}$ to $\mathcal{Y}_2$ remain as in the BW-BC). Evaluating the strong secrecy-capacity region of the BW-like PD-BC reveals that it coincides with $\mathcal{C}^{(\mathsf{BW})}_\mathsf{S}$. This implies that the $Q_X$ that maximizes $R_1$ while keeping Decoder 2 ignorant of $M_1$ has $\alpha=0$ and $\beta=\frac{1}{2}$, which coincides with the input PMF that maximizes $R_1$ while transmitting over the classic BW-BC. Thus, to ensure secrecy over the BW-like PD-BC, the encoder overlooks the improved channel to Decoder 1 and ends up not using the symbol $X=0$.
\end{remark}






The effect of secrecy on the capacity region of a cooperative BC is illustrated by comparing to the BW-BC (Fig. \ref{FIG:blackwell}(a)) without a secrecy constraint. Using the characterization of the capacity region of a cooperative DBC given in \cite[Corollary 12]{Goldfeld_BC_Cooperation2014} and the parametrization in \eqref{EQ:BBC_input_parameter}, the capacity region $\mathcal{C}^{(\mathsf{BW})}_{\mathsf{NS}}$ of the cooperative BW-BC is the union of rate triples $(R_{12},R_1,R_2)\in\mathbb{R}^3_+$ satisfying:
\begin{subequations}
\begin{align}
R_1&\leq H_b(\alpha+\beta)\\
R_2&\leq H_b(\alpha)+R_{12}\\
R_1+R_2&\leq H_b(\alpha)+(1-\alpha)H_b\left(\frac{\beta}{1-\alpha}\right)
\end{align}\label{region_blackwell_no_sec}%
\end{subequations}
where the union is over all $\alpha,\beta\in\mathbb{R}_+$ with $\alpha+\beta\leq 1$.

Fig. \ref{FIG:blackwell_capacity}(b) compares the regions with and without secrecy. The dashed red line represents the capacity region for the case without secrecy while the blue line depicts the region where $M_1$ is confidential. Evidently, $\mathcal{C}^{(\mathsf{BW})}_{\mathsf{NS}}$ is strictly larger than $\mathcal{C}^{(\mathsf{BW})}_\mathsf{S}$. Note that up to approximately $R_1\approx 0.6597\triangleq R_1^{(\mathrm{Th})}$, the two regions coincide. Thus, as long as $R_1\leq R_1^{(\mathrm{Th})}$, concealing $M_1$ is achieved without any rate loss in $R_2$. When $R_1>R_1^{(\mathrm{Th})}$, on the other hand, an increased confidential message rate leads to a reduced $R_2$ value compared to the case without secrecy. Further, if \emph{no secrecy constraint} is imposed on $M_1$, one can transmit it at its maximal rate of $R_1=1$ and still have a positive value of $R_2$ (up to approximately $0.5148$). When $M_1$ is confidential then $R_1=1$ is achievable only if $R_2=0$.

\subsection{Physically Degraded BCs}\label{SUBSEC:PDBC_comparison}

\subsubsection{Capacity Region Comparison}\label{SUBSUBSEC:PDBC_comparison_discussion}


\begin{figure}[t]
\begin{center}
\begin{psfrags}
    \psfragscanon
    \psfrag{A}[][][0.9]{$\ \ \ R_1$}
    \psfrag{G}[][][0.9]{$R_2$}
    \psfrag{D}[][][0.9]{$0$}
    \psfrag{J}[][][0.8]{\ \ \ \ \ \ \ \ Secrecy}
    \psfrag{K}[][][0.8]{\ \ \ \ \ \ \ \ \ \ \ \ No secrecy}
    \psfrag{O}[][][0.9]{$\ \ \ \ \ \ \ \ \ \ \ \ I(W;Y_1)$}
    \psfrag{M}[][][0.9]{$\ \ \ \ \ \ \ \ \ \ I(X;Y_1|W)$}
    \psfrag{P}[][][0.9]{$\mspace{-65mu}I(W;Y_2)+R_{12}$}
    \psfrag{R}[][][0.9]{$\mspace{-20mu}I(X;Y_1|W)-I(W;Y_2)+R_{12}$}
    \psfrag{S}[][][0.9]{$\ \ \ \ \ \ \ \ I(X;Y_2|W)$}
\includegraphics[scale=0.45]{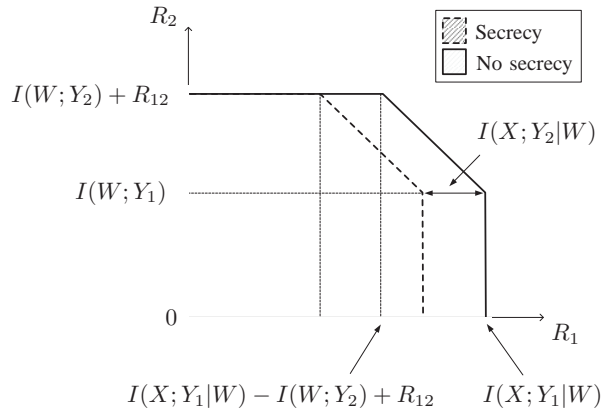}
\caption{Capacity region without secrecy vs. strong secrecy-capacity region where $M_1$ is confidential for the cooperative PD-BC.} \label{FIG:PDBC_region_comparison}
\psfragscanoff
\end{psfrags}
\end{center}
\end{figure}


When the BC is PD, the reduction in $R_1$ is due to the extra layer of bins in the codebook of $M_1$ only, while the modified cooperation scheme results in no loss (in accordance with Proposition \ref{PROP:cooperation_schemes_same}). To see this, consider the capacity region $\mathcal{C}^{(\mathsf{PD})}_\mathsf{NS}$ of cooperative PD-BC without a secrecy constraint on $M_1$ (see \cite{Dikstein_PDBC_Cooperation2014} and \cite{DaboraServetto06BC}), which is the union over the same domain as \eqref{EQ:region_PDBC} of rate triples $(R_{12},R_1,R_2)\in\mathbb{R}^3_+$ satisfying:
\begin{subequations}
\begin{align}
R_1 &\leq I(X;Y_1|W)\label{EQ:region_PDBC_classical1}\\
R_2 &\leq I(W;Y_2)+R_{12}\label{EQ:region_PDBC_classical2}\\
R_1+R_2 &\leq I(X;Y_1).\label{EQ:region_PDBC_classical1+2}
\end{align}\label{EQ:region_PDBC_classical}%
\end{subequations}

In contrast to the SD case, the only impact of the secrecy requirement on the capacity region is expressed in a rate-loss of $I(X;Y_2|W)$ in $R_1$ (see \eqref{EQ:region_PDBC1} in comparison to \eqref{EQ:region_PDBC_classical1}) that is due to the extra layer of bins needed for secrecy. Otherwise, the optimal code construction (and the optimal cooperation protocol) for both problems is the same. The similarity is because, whether $M_1$ is secret or not, its codebook is superimposed on the codebook of $M_2$, and decoding $M_2$ as part of the cooperation protocol comes without cost by the degraded property of the channel. Thus, for a fixed $Q_{W,X}$, if $(R_{12},R_1,R_2)\in\mathcal{C}^{(\mathsf{PD})}_\mathsf{NS}$ then $\Big(R_{12},\big[R_1-I(X;Y_2|W)\big]^+,R_2\Big)\in\mathcal{C}^{(\mathsf{PD})}_\mathsf{S}$, and vice versa. This relation is illustrated in Fig. \ref{FIG:PDBC_region_comparison} for some fixed value of $R_{12}$ and under the assumption that $I(W;Y_2)+R_{12}>I(W;Y_1)$.  


\subsubsection{Gaussian BC Example}\label{SUBSUBSEC:Gaussian}

\begin{figure}[t!]
    \begin{center}
        \begin{psfrags}
            \psfragscanon
            \psfrag{D}[][][1]{\ $X$}
            \psfrag{F}[][][1]{\ $Y_1$}
            \psfrag{L}[][][1]{\ $Y_2$}
            \psfrag{A}[][][1]{\ $Z_1\sim\mathcal{N}(0,\mathrm{N}_1)$}
            \psfrag{B}[][][1]{\ $Z_2\sim\mathcal{N}(0,\mathrm{N}_2)$}
            \psfrag{C}[][][1]{\ \ \ \ \ \ \ \ \ \ \ \ \ \ \ \ \ \ \ \ \ $Z_2\sim\mathcal{N}(0,\mathrm{N}_2-\mathrm{N}_1)$}
            \psfrag{N}[][][1]{\ \ \ $R_{12}$}
            \psfrag{X}[][][1]{\ }
            \includegraphics[scale = .85]{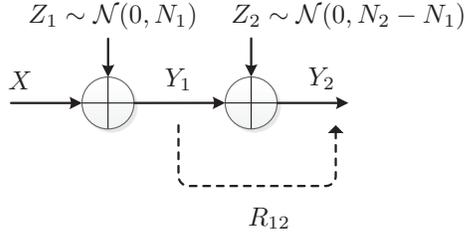}
            \caption{Cooperative Gaussian PD-BC.} \label{FIG:Gaussian}
            \psfragscanoff
        \end{psfrags}
     \end{center}
 \end{figure}


Consider next the cooperative Gaussian PD-BC (without a common message) shown in Fig. \ref{FIG:Gaussian}, where for every time instance $i\in[1:n]$, we have
\begin{subequations}
\begin{align}
Y_{1,i}&=X_i+Z_{1,i},\\
Y_{2,i}&=X_i+Z_{1,i}+Z_{2,i}
\end{align}
\end{subequations}
and $\big\{Z_{1,i}\big\}_{i=1}^n$ and $\big\{Z_{2,i}\big\}_{i=1}^n$ are mutually independent sequences of i.i.d. Gaussian random variables with $Z_{1,i}\sim\mathcal{N}(0,\mathrm{N}_1)$, $Z_{2,i}\sim\mathcal{N}(0,\mathrm{N}_2-\mathrm{N}_1)$ and $\mathrm{N}_2>\mathrm{N}_1$, for $i\in[1:n]$. The channel input is subject to an average power constraint
\begin{equation}
\frac{1}{n}\sum_{i=1}^n\mathbb{E}\big[X_i^2\big]\leq \mathrm{P}.
\end{equation}

By using continuous alphabets with an input power constraint adaptation of Theorem \ref{TM:PDBC_secrecy_capacity} we characterize the strong secrecy-capacity region $\mathcal{C}^{(\mathsf{G})}_\mathsf{S}$ of the cooperative Gaussian PD-BC with one confidential message as the union of rate triples $(R_{12},R_1,R_2)\in\mathbb{R}^3_+$ satisfying:
\begin{subequations}
\begin{align}
R_1&\leq \frac{1}{2}\log\left(1+\frac{\alpha \mathrm{P}}{\mathrm{N}_1}\right)-\frac{1}{2}\log\left(1+\frac{\alpha \mathrm{P}}{\mathrm{N}_2}\right)\\
R_2&\leq \frac{1}{2}\log\left(1+\frac{\bar{\alpha} \mathrm{P}}{\alpha \mathrm{P}+\mathrm{N}_2}\right)+R_{12}\\
R_1+R_2&\leq \frac{1}{2}\log\left(1+\frac{\mathrm{P}}{\mathrm{N}_1}\right)-\frac{1}{2}\log\left(1+\frac{\alpha \mathrm{P}}{\mathrm{N}_2}\right)
\end{align}\label{EQ:region_Gaussian_BC}%
\end{subequations}
where the union is over all $\alpha\in[0,1]$.

The achievability of \eqref{EQ:region_Gaussian_BC} follows from Theorem \ref{TM:PDBC_secrecy_capacity} with the following choice of
random variables:
\begin{equation}
W\sim\mathcal{N}(0,\alpha P)\ ,\ \tilde{W}\sim\mathcal{N}(0,\bar{\alpha} P)\ ,\ X=W+\tilde{W}
\end{equation}
where $W$ and $\tilde{W}$ are independent. The optimality of Gaussian inputs is proven in Appendix \ref{APPEN:Gaussian_proof}.


\begin{figure}[t!]
    \begin{center}
        \begin{psfrags}
            \psfragscanon
            \psfrag{A}[][][0.9]{$R_1$ [bits/use]}
            \psfrag{B}[][][0.9]{$R_2$ [bits/use]}
            \psfrag{C}[][][0.8]{\ \ \ \ \ \ $\mspace{-7mu}R_{12}=0$}
            \psfrag{D}[][][0.8]{\ \ \ \ \ \ \ $\mspace{-4mu}R_{12}\mspace{-2mu}=\mspace{-2mu}0.2$}
            \psfrag{E}[][][0.8]{\ \ \ \ \ \ \ $\mspace{-4mu}R_{12}\mspace{-2mu}=\mspace{-2mu}0.4$}
            \psfrag{F}[][][0.8]{\ \ \ \ \ \ \ $\mspace{-4mu}R_{12}\mspace{-2mu}=\mspace{-2mu}0.6$}
            \psfrag{I}[][][0.8]{\ \ \ \ \ \ \ \ \ No secrecy}
            \psfrag{H}[][][0.8]{\ \ \ \ \ Secrecy}
            \psfrag{X}[][][0.9]{\ \ \ \ \ $\alpha=0$}
            \psfrag{Y}[][][0.8]{\ \ \ \ \ \ \ \ \ \ \ \ \ $\frac{1}{2}\log\left(1+\frac{\alpha P}{\mathrm{N}_2}\right)$}
            \psfrag{Z}[][][0.9]{\ \ \ \ \ $\alpha=1$}
            \subfloat[]{\includegraphics[scale = .47]{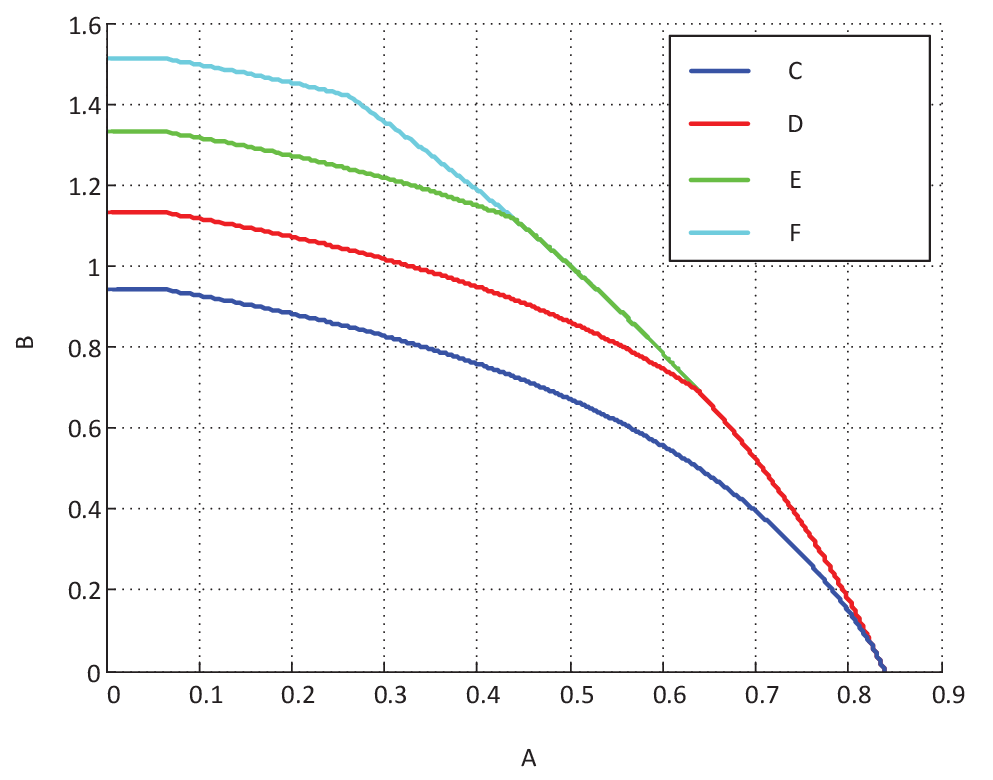}}\\
            \subfloat[]{\includegraphics[scale = .47]{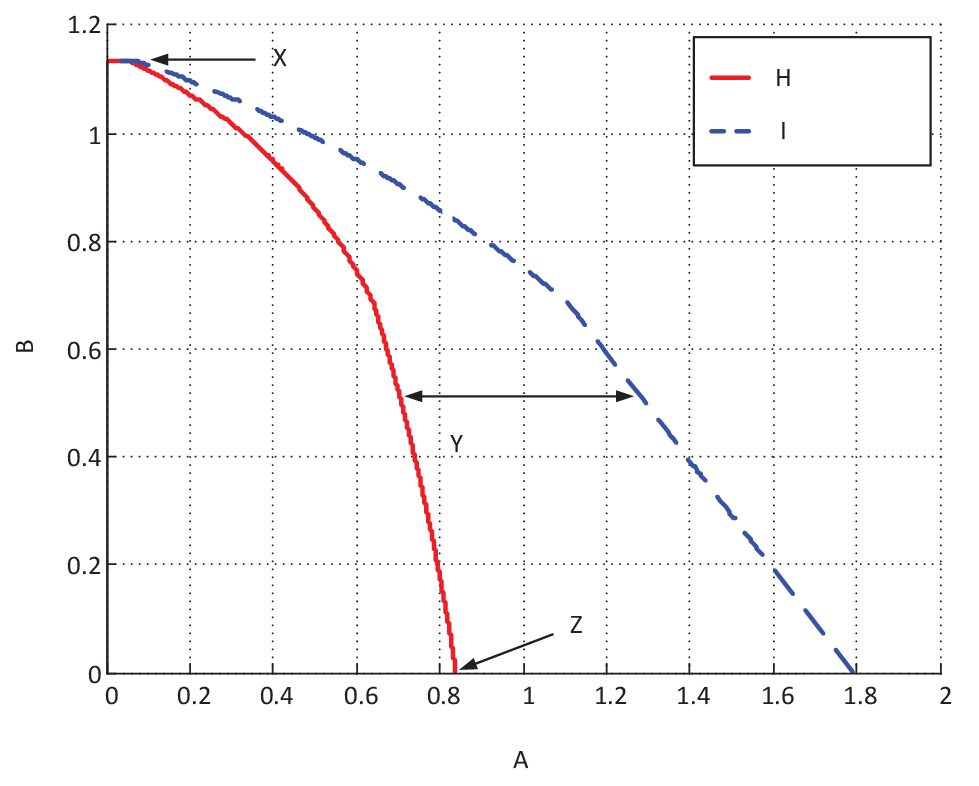}}
            \caption{(a) Projection of the strong secrecy-capacity region of the cooperative Gaussian BC with one confidential message onto the plane $(R_1,R_2)$ for different values of $R_{12}$; (b) Cooperative Gaussian BC with $R_{12}=0.2$: Strong secrecy-capacity region where $M_1$ is confidential vs. capacity region without secrecy.} \label{FIG:Gaussian_comparison}
            \psfragscanoff
        \end{psfrags}
     \end{center}
 \end{figure}


Setting $\mathrm{P}=11$, $\mathrm{N}_1=1$ and $\mathrm{N}_2=4$, Fig. \ref{FIG:Gaussian_comparison}(a) shows the strong secrecy-capacity region of the cooperative Gaussian BC for different $R_{12}$ values, while Fig. \ref{FIG:Gaussian_comparison}(b) compares the optimal rate regions when a secrecy constraint on $M_1$ is and is not present. The red line in both figures coincide and represent the secrecy-capacity region when $R_{12}=0.2$. The dashed blue line in Fig \ref{FIG:Gaussian_comparison}(b) shows the capacity region $\mathcal{C}^{(\mathsf{G})}_\mathsf{NS}$ of the cooperative Gaussian BC without secrecy constraints, which is given by the union over all $\alpha\in[0,1]$ of rate triples $(R_{12},R_1,R_2)\in\mathbb{R}^3_+$ satisfying:
\begin{subequations}
\begin{align}
R_1&\leq \frac{1}{2}\log\left(1+\frac{\alpha \mathrm{P}}{\mathrm{N}_1}\right)\\
R_2&\leq \frac{1}{2}\log\left(1+\frac{\bar{\alpha} \mathrm{P}}{\alpha \mathrm{P}+\mathrm{N}_2}\right)+R_{12}\\
R_1+R_2&\leq \frac{1}{2}\log\left(1+\frac{\mathrm{P}}{\mathrm{N}_1}\right)
\end{align}\label{EQ:region_NS_Gaussian_BC}%
\end{subequations}
The derivation of \eqref{EQ:region_NS_Gaussian_BC} relies on \cite[Equation (17)]{Dikstein_PDBC_Cooperation2014} and uses standard arguments for proving the optimality of Gaussian inputs.

By the structure of the rate bounds in \eqref{EQ:region_Gaussian_BC} and \eqref{EQ:region_NS_Gaussian_BC}, for every fixed $\alpha\in[0,1]$, if $(R_{12},R_1,R_2)\in\mathcal{C}^{(\mathsf{G})}_\mathsf{NS}$, we have
\begin{equation}
\left(R_{12},R_1-\frac{1}{2}\log\left(1+\frac{\alpha \mathrm{P}}{\mathrm{N}_2}\right),R_2\right)\in\mathcal{C}^{(\mathsf{G})}_\mathsf{S}.
\end{equation}
This agrees with the discussion in Section \ref{SUBSUBSEC:PDBC_comparison_discussion} as $I(X;Y_2|W)=\frac{1}{2}\log\left(1+\frac{\alpha \mathrm{P}}{\mathrm{N}_2}\right)$.


\section{Proofs}\label{SEC:proofs}


\subsection{Proof of Lemma \ref{LEMMA:soft_covering}}\label{SUBSEC:soft_covering_proof}

Recall that the factorization in \eqref{EQ:soft_proof_P_PMF_code} implies that $P_{\mathbf{S}_0,\mathbf{S},W,I,\mathbf{U},\mathbf{V}|\mathsf{B}_n=\mathcal{B}_n}=P^{(\mathcal{B}_n)}_{\mathbf{S}_0,\mathbf{S},W,I,\mathbf{U},\mathbf{V}}$, where $\mathcal{B}_n\in\mathfrak{B}_n$ and the RHS is given in \eqref{EQ:soft_proof_P_PMF}. Throughout this proof we use $P^{(\mathcal{B}_n)}_{\mathbf{S}_0,\mathbf{S},W,I,\mathbf{U},\mathbf{V}}$ when the codebook $\mathcal{B}_n\in\mathfrak{B}_n$ is fixed, and prefer $P_{\mathbf{S}_0,\mathbf{S},W,I,\mathbf{U},\mathbf{V}|\mathsf{B}_n}$ when the codebook is random. Furthermore, on account of the factorization in \eqref{EQ:soft_proof_P_PMF} we have $P^{(\mathcal{B}_n)}_{\mathbf{S}_0,\mathbf{S}}=Q_{S_0,S}^n$, for each $\mathcal{B}_n\in\mathfrak{B}_n$. Therefore, to establish Lemma \ref{LEMMA:soft_covering} we show that 
\begin{equation}
\mathbb{E}_{\mathsf{B}_n}D\\\Big(P_{\mathbf{S}_0,\mathbf{S},\mathbf{V}|\mathsf{B}_n}\\\Big|\\\Big|Q_{S_0,S,V}^n\\\Big)\xrightarrow[n\to\infty]{}0.\label{EQ:soft_proof_divergence_goto0}
\end{equation}

\begin{lemma}[Absolute Continuity]\label{LEMMA:absolute_continues}
For any $\mathcal{B}_n\in\mathfrak{B}_n$, we have $P^{(\mathcal{B}_n)}_{\mathbf{S}_0,\mathbf{S},\mathbf{V}}\ll Q_{S_0,S,V}^n$, i.e., $P^{(\mathcal{B}_n)}_{\mathbf{S}_0,\mathbf{S},\mathbf{V}}$ is absolutely continues with respect to $Q_{S_0,S,V}^n$.
\end{lemma}

The proof of Lemma \ref{LEMMA:absolute_continues} is relegated to Appendix \ref{APPEN:lemma_absolute_proof}. Combining this with Remark \ref{REM:TV_divergence_relation}, a sufficient condition for \eqref{EQ:soft_proof_divergence_goto0} is that
\begin{equation}
\mathbb{E}_{\mathsf{B}_n}\Big|\Big|P_{\mathbf{S}_0,\mathbf{S},\mathbf{V}|\mathsf{B}_n}-Q_{S_0,S,V}^n\Big|\Big|\xrightarrow[n\to\infty]{}0 \label{EQ:soft_proof_TV_goto0}
\end{equation}
at an exponential rate. 

To evaluate the TV in \eqref{EQ:soft_proof_TV_goto0}, for any $\mathcal{B}_n\in\mathfrak{B}_n$, define the \emph{ideal} PMF on $\mathcal{S}_0^n\times\mathcal{S}^n\times\mathcal{W}_n\times\mathcal{I}_n\times\mathcal{U}^n\times\mathcal{V}^n$ as
\begin{subequations}
\begin{align*}
&\Gamma^{(\mathcal{B}_n)}(\mathbf{s}_0,w,i,\mathbf{u},\mathbf{s},\mathbf{v})\\
&=Q_{S_0}^n(\mathbf{s}_0)2^{-n(\tilde{R}+R')}\mathds{1}_{\big\{\mathbf{u}=\mathbf{u}(\mathbf{s}_0,w,i)\big\}}Q^n_{S,V|U,S_0}(\mathbf{s},\mathbf{v}|\mathbf{u},\mathbf{s}_0)\numberthis\label{EQ:soft_proof_Gamma_PMF_codebook}
\end{align*}
and further set 
\begin{equation}
\Gamma(\mathcal{B}_n,\mathbf{s}_0,w,i,\mathbf{u},\mathbf{s},\mathbf{v})=\lambda(\mathcal{B}_n)\Gamma^{(\mathcal{B}_n)}(\mathbf{s}_0,w,i,\mathbf{u},\mathbf{s},\mathbf{v}).\label{EQ:soft_proof_Gamma_PMF}
\end{equation}
\end{subequations}
Note that $\Gamma$ describes an encoding process where the choice of the $u$-codeword from a certain bin is uniform, as opposed to $P$ in \eqref{EQ:soft_proof_P_PMF} that uses a likelihood encoder. Furthermore, the structure of $\Gamma$ implies that the sequence $\mathbf{s}$ is generated by feeding $\mathbf{s}_0$ and the chosen $u$-codeword into the DMC $Q^n_{S|U,S_0}$. 

Using the TV triangle inequality, we upper bound the LHS of \eqref{EQ:soft_proof_TV_goto0} by
\begin{align*}
\mathbb{E}&_{\mathsf{B}_n}\Big|\Big|P_{\mathbf{S}_0,\mathbf{S},\mathbf{V}|\mathsf{B}_n}-Q_{S_0,S,V}^n\Big|\Big|_{\mathsf{TV}}\\
&\begin{multlined}[b][.4\textwidth]
\leq\mathbb{E}_{\mathsf{B}_n}\Big|\Big|P_{\mathbf{S}_0,\mathbf{S},\mathbf{V}|\mathsf{B}_n}-\Gamma_{\mathbf{S}_0,\mathbf{S},\mathbf{V}|\mathsf{B}_n}\Big|\Big|_{\mathsf{TV}}\\+\mathbb{E}_{\mathsf{B}_n}\Big|\Big|\Gamma_{\mathbf{S}_0,\mathbf{S},\mathbf{V}|\mathsf{B}_n}-Q_{S_0,S,V}^n\Big|\Big|_{\mathsf{TV}}
\end{multlined}.\numberthis\label{eq:soft_proof_TV_UB}
\end{align*}
By \cite[Corollary VII.5]{Cuff_Synthesis2013}, the second expected TV on the RHS of \eqref{eq:soft_proof_TV_UB} decays exponentially fast as $n\to\infty$ if 
\begin{equation}
\tilde{R}+R'>I(U;S,V|S_0).
\end{equation}

For the first term in \eqref{eq:soft_proof_TV_UB}, we use the following relations between $\Gamma$ and $P$. For every $\mathcal{B}_n\in\mathfrak{B}_n$, we have
\begin{subequations}
\begin{align}
&\Gamma^{(\mathcal{B}_n)}_{I|W,\mathbf{S}_0,\mathbf{S}}=\hat{P}^{(\mathcal{B}_n)}_{I|W,\mathbf{S}_0,\mathbf{S}}=P^{(\mathcal{B}_n)}_{I|W,\mathbf{S}_0,\mathbf{S}}\label{EQ:soft_proof_Gamma_P_relations1}\\
&\Gamma^{(\mathcal{B}_n)}_{\mathbf{U}|I,W,\mathbf{S}_0,\mathbf{S}}=\mathds{1}_{\big\{\mathbf{U}=\mathbf{u}(\mathbf{S}_0,W,I)\big\}}=P^{(\mathcal{B}_n)}_{\mathbf{U}|I,W,\mathbf{S}_0,\mathbf{S}}\label{EQ:soft_proof_Gamma_P_relations2}\\
&\Gamma^{(\mathcal{B}_n)}_{\mathbf{V}|\mathbf{U},I,W,\mathbf{S}_0,\mathbf{S}}=Q_{V|U,S_0,S}^n=P^{(\mathcal{B}_n)}_{\mathbf{V}|\mathbf{U},I,W,\mathbf{S}_0,\mathbf{S}}.\label{EQ:soft_proof_Gamma_P_relations3}
\end{align}\label{EQ:soft_proof_Gamma_P_relations}%
\end{subequations}
While \eqref{EQ:soft_proof_Gamma_P_relations2}-\eqref{EQ:soft_proof_Gamma_P_relations3} follow directly from \eqref{EQ:soft_proof_P_PMF} and \eqref{EQ:soft_proof_Gamma_PMF}, the justification for \eqref{EQ:soft_proof_Gamma_P_relations1} is that for every $(\mathcal{B}_n,\mathbf{s}_0,\mathbf{s},w,i)\in\mathfrak{B}_n\times\mathcal{S}^n_0\times\mathcal{S}^n\times\mathcal{W}_n\times\mathcal{I}_n$, we have
\begin{align*}
&\Gamma^{(\mathcal{B}_n)}(i|w,\mathbf{s}_0,\mathbf{s})\\
&=\frac{\Gamma^{(\mathcal{B}_n)}(\mathbf{s}_0,w,i,\mathbf{s})}{\Gamma^{(\mathcal{B}_n)}(\mathbf{s}_0,w,\mathbf{s})}\\                                                 &=\frac{\sum_{\mathbf{u}}Q_{S_0}^n(\mathbf{s}_0)2^{-n(\tilde{R}+R')}\mathds{1}_{\big\{\mathbf{u}=\mathbf{u}(\mathbf{s}_0,w,i)\big\}}Q^n_{S|U,S_0}(\mathbf{s}|\mathbf{u},\mathbf{s}_0)}{\sum_{\mathbf{u},i'}Q_{S_0}^n(\mathbf{s}_0)2^{-n(\tilde{R}+R')}\mathds{1}_{\big\{\mathbf{u}=\mathbf{u}(\mathbf{s}_0,w,i')\big\}}Q^n_{S|U,S_0}(\mathbf{s}|\mathbf{u},\mathbf{s}_0)}\\
&=\frac{Q^n_{S|U,S_0}\big(\mathbf{s}\big|\mathbf{u}(\mathbf{s}_0,w,i),\mathbf{s}_0\big)}{\sum_{i'}Q^n_{S|U,S_0}\big(\mathbf{s}\big|\mathbf{u}(\mathbf{s}_0,w,i'),\mathbf{s}_0\big)}\\
&\stackrel{(a)}=\hat{P}^{(\mathcal{B}_n)}(i|w,\mathbf{s}_0,\mathbf{s})\numberthis
\end{align*}
where (a) follows from \eqref{EQ:likelihood_definition}. The relations in \eqref{EQ:soft_proof_Gamma_P_relations} yield
\begin{align*}
&\mathbb{E}_{\mathsf{B}_n}\Big|\Big|P_{\mathbf{S}_0,\mathbf{S},\mathbf{V}|\mathsf{B}_n}-\Gamma_{\mathbf{S}_0,\mathbf{S},\mathbf{V}|\mathsf{B}_n}\Big|\Big|_{\mathsf{TV}}\\
&\leq\mathbb{E}_{\mathsf{B}_n}\Big|\Big|P_{\mathbf{S}_0,\mathbf{S},W,I,\mathbf{U},\mathbf{V}|\mathsf{B}_n}-\Gamma_{\mathbf{S}_0,\mathbf{S},W,I,\mathbf{U},\mathbf{V}|\mathsf{B}_n}\Big|\Big|_{\mathsf{TV}}\\
&\stackrel{(a)}=\mathbb{E}_{\mathsf{B}_n}\Big|\Big|P_{\mathbf{S}_0,\mathbf{S},I,\mathbf{U},\mathbf{V}|W=1,\mathsf{B}_n}-\Gamma_{\mathbf{S}_0,\mathbf{S},I,\mathbf{U},\mathbf{V}|W=1,\mathsf{B}_n}\Big|\Big|_{\mathsf{TV}}\\
&\stackrel{(b)}=\mathbb{E}_{\mathsf{B}_n}\Big|\Big|Q_{S_0,S}^n-\Gamma_{\mathbf{S}_0,\mathbf{S}|W=1,\mathsf{B}_n}\Big|\Big|_{\mathsf{TV}}\numberthis\label{eq:soft_proof_2ndTV_UB}
\end{align*}
where:\\
(a) is because $\Gamma^{(\mathcal{B}_n)}(w)=P^{(\mathcal{B}_n)}(w)=2^{-n\tilde{R}}$, for every $w\in\mathcal{W}_n$ and $\mathcal{B}_n\in\mathfrak{B}_n$, the independence of $\mathsf{B}_n$ and $W$, and the symmetry of the codebook construction with respect to $W$;\\
(b) is by \eqref{EQ:soft_proof_Gamma_P_relations} and because $P^{(\mathcal{B}_n)}_{\mathbf{S}_0,\mathbf{S}}=Q_{S_0,S}^n$ for every $\mathcal{B}_n\in\mathfrak{B}_n$.

Invoking \cite[Corollary VII.5]{Cuff_Synthesis2013} once more yields
\begin{equation}
\mathbb{E}_{\mathsf{B}_n}\Big|\Big|Q_{S_0,S}^n-\Gamma^{(\mathsf{B}_n)}_{\mathbf{S}_0,\mathbf{S}|W=1}\Big|\Big|_{\mathsf{TV}}\xrightarrow[n\to\infty]{}0\label{EQ:soft_proof_TVtoZero2}
\end{equation}
exponentially fast, as long as
\begin{equation}
R'>I(U;S|S_0).\label{EQ:soft_proof_R'_bound}
\end{equation}
This implies that there exists $\gamma>0$ such that
\begin{equation}
\mathbb{E}_{\mathsf{B}_n}\Big|\Big|P_{\mathbf{S}_0,\mathbf{S},\mathbf{V}|\mathsf{B}_n}-Q_{S_0,S,V}^n\Big|\Big|_{\mathsf{TV}}\leq e^{-n\gamma}.\label{eq:soft_proof_TVtozero}
\end{equation}
%


\subsection{Proof of Lemma \ref{LEMMA:typicallity}}\label{SUBSEC:typicallity_proof}

The proof uses the following property of the TV (see, e.g., \cite[Property 1]{Song_Cuff_Secrecy2014}): Let $\mu,\nu$ be two probability measures on a measurable space $(\mathcal{X},\mathcal{F})$ and $g:\mathcal{X}\to\mathbb{R}$ be a measurable function
bounded by $b\in\mathbb{R}$. We then have
\begin{equation}
\big|\mathbb{E}_\mu g-\mathbb{E}_\nu g\big|\leq b\cdot\big|\big|\mu-\nu\big|\big|_{TV}\label{EQ:TV_property}
\end{equation}

Fix $\epsilon>0$ and consider the $\Gamma$ PMF defined in \eqref{EQ:soft_proof_Gamma_PMF}. With respect to the random experiment described by $\Gamma$, we have
\begin{equation}
\mathbb{E}_{\mathsf{B}_n}\mathbb{P}_{\Gamma}\Big(\big(\mathbf{S}_0,\mathbf{S},\mathbf{U}(\mathbf{S}_0,w,I)\big)\notin\mathcal{T}_\epsilon^{n}(Q_{S_0,S,U})\Big|\mathsf{B}_n\Big)\xrightarrow[n\to\infty]{}0\label{EQ:typicallity_gamma_tozero}
\end{equation}
because $\mathbf{U}(\mathbf{S}_0,w,i)\sim Q_{U|S_0}^n$, for every $i\in\mathcal{I}_n$, and $\mathbf{S}$ is obtained by feeding $(\mathbf{S}_0,\mathbf{U}(\mathbf{S}_0,w,i)\big)$ into the DMC $Q_{S|U,S_0}^n$. Thus, \eqref{EQ:typicallity_gamma_tozero} holds by the weak law of large numbers (WLLN). Further, basic properties of the TV and the analysis in Section \ref{SUBSEC:soft_covering_proof} (see \eqref{eq:soft_proof_2ndTV_UB}) imply
\begin{align*}
&\mathbb{E}_{\mathsf{B}_n}\Big|\Big|P_{\mathbf{S}_0,\mathbf{S},\mathbf{U}|\mathsf{B}_n}-\Gamma_{\mathbf{S}_0,\mathbf{S},\mathbf{U}|\mathsf{B}_n}\Big|\Big|_{\mathsf{TV}}\\
&\leq\mathbb{E}_{\mathsf{B}_n}\Big|\Big|P_{\mathbf{S}_0 ,\mathbf{S},W,I,\mathbf{U},\mathbf{V}|\mathsf{B}_n}-\Gamma_{\mathbf{S}_0,\mathbf{S},W,I,\mathbf{U},\mathbf{V}|\mathsf{B}_n}\Big|\Big|_{\mathsf{TV}}\xrightarrow[n\to\infty]{}0.\numberthis\label{EQ:typicallity_TV_tozero}
\end{align*}

Now, let $g_n:\mathcal{S}_0^n\times\mathcal{S}^n\times\mathcal{U}^n\to\mathbb{R}$ be defined by $g_n(\mathbf{s}_0,\mathbf{s},\mathbf{u})\triangleq \mathds{1}_{\big\{(\mathbf{s}_0,\mathbf{s},\mathbf{u})\notin\mathcal{T}_\epsilon^{n}(Q_{S_0,S,U})\big\}}$ and consider
\begin{align*}
\mathbb{E}_{\mathsf{B}_n}&\mathbb{P}_P\Big(\big(\mathbf{S}_0,\mathbf{S},\mathbf{U}(\mathbf{S}_0,w,I)\big)\notin\mathcal{T}_\epsilon^{n}(Q_{S_0,S,U})\Big|\mathsf{B}_n\Big)\\
&=\mathbb{E}_{\mathsf{B}_n}\mathbb{E}_P\Big[g_n\big(\mathbf{S}_0,\mathbf{S},\mathbf{U}(\mathbf{S}_0,w,I)\big)\Big| \mathsf{B}_n\Big]\\
&\begin{multlined}[b][.43\textwidth]\leq\mathbb{E}_{\mathsf{B}_n}\mathbb{E}_{\Gamma}\Big[g_n\big(\mathbf{S}_0,\mathbf{S},\mathbf{U}(\mathbf{S}_0,w,I)\big)\Big|\mathsf{B}_n\Big]\\+\mathbb{E}_{\mathsf{B}_n}\bigg|\mathbb{E}_P\Big[g_n\big(\mathbf{S}_0,\mathbf{S},\mathbf{U}(\mathbf{S}_0,w,I)\big)\Big|\mathsf{B}_n\Big]\\-\mathbb{E}_{\Gamma}\Big[g_n\big(\mathbf{S}_0,\mathbf{S},\mathbf{U}(\mathbf{S}_0,w,I)\big)\Big|\mathsf{B}_n\Big]\bigg|\end{multlined}\\
&\begin{multlined}[b][.43\textwidth]\stackrel{(a)}\leq\mathbb{E}_{\mathsf{B}_n}\mathbb{P}_{\Gamma}\Big(\big(\mathbf{S}_0,\mathbf{S},\mathbf{U}(\mathbf{S}_0,w,I)\big)\notin\mathcal{T}_\epsilon^{n}(Q_{S_0,S,U})\Big|\mathsf{B}_n\Big)\\+\mathbb{E}_{\mathsf{B}_n}\Big|\Big|P_{\mathbf{S}_0,\mathbf{S},\mathbf{U}|\mathsf{B}_n}-\Gamma_{\mathbf{S}_0,\mathbf{S},\mathbf{U}|\mathsf{B}_n}\Big|\Big|_{\mathsf{TV}}\end{multlined}\numberthis\label{EQ:Typicallity_probability_UB}
\end{align*}
where (a) uses \eqref{EQ:TV_property} and $g_n$ being bounded by $b=1$, for any $n\in\mathbb{N}$. By \eqref{EQ:typicallity_gamma_tozero}-\eqref{EQ:typicallity_TV_tozero}, the RHS of \eqref{EQ:Typicallity_probability_UB} approaches 0 as $n\to\infty$.


\subsection{Proof of Theorem \ref{TM:inner_bound}}\label{SUBSEC:inner_proof}

Fix $n\in\mathbb{N}$, $\epsilon,\delta>0$, a PMF $Q_{U_0,U_1,U_2,X}\in\mathcal{P}(\mathcal{U}_0\times\mathcal{U}_1\times\mathcal{U}_2\times\mathcal{X})$ and denote $Q_{U_0,U_1,U_2,X,Y_1,Y_2}\triangleq Q_{U_0,U_1,U_2,X}W_{Y_1,Y_2|X}$. In the following we omit the blocklength $n$ from our notations of the involved sets of indices, e.g., we write $\mathcal{M}_0$ instead of $\mathcal{M}_0^{(n)}$, etc. Furthermore, we assume that quantities of the form $2^{nR}$, where $n\in\mathbb{N}$ and $R\in\mathbb{R}_+$, are integers.

\par\textbf{Message Splitting:} Split each $m_2\in\mathcal{M}_2$ into two sub-messages denoted by $(m_{20},m_{22})$. The pair $m_p\triangleq(m_0,m_{20})$ is referred to as a \emph{public message} and is to be decoded by both receivers, while $m_1$ and $m_{22}$, that serve as \emph{private messages}, are to be decoded by receiver 1 and receiver 2, respectively. The cooperation protocol will use the link to convey information about the decoded $m_p$ from receiver 1 to receiver 2. The rates associated with $m_{20}$ and $m_{22}$ are denoted by $R_{20}$ and $R_{22}$, while the corresponding alphabets are $\mathcal{M}_{20}$ and $\mathcal{M}_{22}$, respectively. Furthermore, we use $R_p\triangleq R_0+R_{20}$ and $\mathcal{M}_p\triangleq\mathcal{M}_0\times\mathcal{M}_{20}$. Since $|\mathcal{M}_p|=2^{nR_p}$, with some abuse of notation, we also use $\mathcal{M}_p=\big[1:2^{nR_p}\big]$. The partial rates $R_{20}$ and $R_{22}$ satisfy
\begin{equation}
R_2=R_{20}+R_{22}.\label{EQ:achiev_partial_rates_sum}
\end{equation}
With respect to the above, the random variable $M_2$ is split into two independent random variables $M_{20}$ and $M_{22}$ that are uniform over $\mathcal{M}_{20}$ and $\mathcal{M}_{22}$, respectively. The random variable  $M_p\triangleq(M_0,M_{20})$ is uniformly distributed over $\mathcal{M}_p$. Moreover, let $W$ be a random variable uniformly distributed over $\mathcal{W}=\big[1:2^{n\tilde{R}}\big]$ and independent of $(M_0,M_1,M_2)$ (which implies its independence of $(M_p,M_1,M_{22})$).

\par\textbf{Cooperation Protocol Preliminaries:} Fix a partitioning\footnote{The partitioning may be preformed in any prescribed manner and it is not part of the random coding experiment.} of $\mathcal{M}_p$ into $2^{nR_{12}}$ equal-sized subsets (referred to as ``bins'') $\mathcal{B}_n(m_{12})$, where $m_{12}\in\mathcal{M}_{12}$. Let $\hat{m}_{12}:\mathcal{M}_p\to\mathcal{M}_{12}$ be the function that associates with each public message $m_p\in\mathcal{M}_p$ its bin index $\hat{m}_{12}(m_p)$, i.e., $m_p\in\mathcal{B}_n\big(\hat{m}_{12}(m_p)\big)$, for each $m_p\in\mathcal{M}_p$. 

\textbf{Codebook $\bm{\mathcal{C}_n}$:} Let $\mathsf{C}^{(n)}_0\triangleq\big\{\mathbf{U}_0(m_p)\big\}_{m_p\in\mathcal{M}_p}$ be a random public message codebook that comprises $2^{nR_p}$ i.i.d. random vectors $\mathbf{U}_0(m_p)$, each distributed according to $Q_{U_0}^n$. A realization of $\mathsf{C}^{(n)}_0$ is denoted by $\mathcal{C}^{(n)}_0\triangleq\big\{\mathbf{u}_0(m_p)\big\}_{m_p\in\mathcal{M}_p}$.

Fix a public message codebook $\mathcal{C}^{(n)}_0$. For every $m_p\in\mathcal{M}_p$, let $\mathsf{C}^{(n)}_1(m_p)\triangleq\big\{\mathbf{U}_1(m_p,m_1,w,i)\big\}_{(m_1,w,i)\in\mathcal{M}_1\times\mathcal{W}\times\mathcal{I}}$, where $\mathcal{I}\triangleq\big[1:2^{nR'}\big]$, be a random codebook of confidential messages to User 1, consisting of conditionally independent random vectors each distributed according to $Q^n_{U_1|U_0}\big(\cdot\big|\mathbf{u}_0(m_p)\big)$. A realization of $\mathsf{C}^{(n)}_1(m_p)$ is denoted by $\mathcal{C}^{(n)}_1(m_p)\triangleq\big\{\mathbf{u}_1(m_p,m_1,w,i)\big\}_{(m_1,w,i)\in\mathcal{M}_1\times\mathcal{W}\times\mathcal{I}}$. Based on this labeling, each $\mathsf{C}^{(n)}_1(m_p)$, $m_p\in\mathcal{M}_p$, can be thought of as having a $u_1$-bin associated with every pair $(m_1,w)\in\mathcal{M}_1\times\mathcal{W}_n$, each containing $2^{nR_1'}$ $u_1$-codewords.

Next, for each $m_p\in\mathcal{M}_p$, the corresponding random codebook of private message 2 is $\mathsf{C}^{(n)}_2(m_p)\triangleq\big\{\mathbf{U}_2(m_p,m_{22})\big\}_{m_{22}\in\mathcal{M}_{22}}$, and comprises $2^{nR_{22}}$ conditionally independent random vectors distributed according to $Q^n_{U_2|U_0}\big(\cdot\big|\mathbf{u}_0(m_p)\big)$. We use $\mathcal{C}^{(n)}_2(m_p)\triangleq\big\{\mathbf{u}_2(m_p,m_{22})\big\}_{m_{22}\in\mathcal{M}_{22}}$ to denote a possible outcome of $\mathsf{C}^{(n)}_2(m_p)$.

For $j=1,2$, we denote $\mathsf{C}^{(n)}_j\triangleq\Big\{\mathsf{C}^{(n)}_j(m_p)\Big\}_{m_p\in\mathcal{M}_p}$, and its realization by $\mathsf{C}^{(n)}_j$. A random codebook is denoted by $\mathsf{C}_n=\Big\{\mathsf{C}^{(n)}_0,\mathsf{C}^{(n)}_1,\mathsf{C}^{(n)}_2\Big\}$, while
$\mathcal{C}_n=\Big\{\mathcal{C}^{(n)}_0,\mathcal{C}^{(n)}_1,\mathcal{C}^{(n)}_2\Big\}$ denotes a fixed codebook (a possible realization of $\mathsf{C}_n$). Denoting the set of all possible values of $\mathsf{C}_n$ by $\mathfrak{C}_n$, the above codebook construction induces a PMF $\mu\in\mathcal{P}(\mathfrak{C}_n)$ over the codebook ensemble. For every $\mathcal{C}_n\in\mathfrak{C}_n$, we have \eqref{EQ:codebook_probability} from the top of the next page.

\begin{figure*}[!t]
\begin{equation}
\mu(\mathcal{C}_n)=\prod_{m_p\in\mathcal{M}_p}\mspace{-6mu}Q^n_{U_0}\big(\mathbf{u}_0(m_p)\big)\mspace{-10mu}\prod_{\substack{(m^{(1)}_p,m_1,w,i)\\\in\mathcal{M}_p\times\mathcal{M}_1\times\mathcal{W}\times\mathcal{I}}}\mspace{-34mu}Q^n_{U_1|U_0}\Big(\mathbf{u}_1\big(m_p^{(1)},m_1,w,i\big)\Big|\mathbf{u}_0\big(m_p^{(1)}\big)\Big)\prod_{\substack{(m_p^{(2)},m_{22})\\\in\mathcal{M}_p\times\mathcal{M}_{22}}}\mspace{-14mu}Q^n_{U_2|U_0}\Big(\mathbf{u}_2\big(m_p^{(2)},m_{22}\big)\Big|\mathbf{u}_0\big(m_p^{(2)}\big)\Big)\label{EQ:codebook_probability}
\end{equation}
\hrulefill
\setcounter{equation}{58}
\begin{align*}
&P^{(\mathcal{C}_n)}\Big(m_p,m_1,m_{22},w,m_{12},\mathbf{u}_0,\mathbf{u}_2,i,\mathbf{u}_1,\mathbf{x},\mathbf{y}_1,\mathbf{y}_2,\big(\hat{m}_0^{(1)},\hat{m}_1\big),\big(\hat{m}_0^{(2)},\hat{m}_2\big)\Big)\\&
=2^{-n(R_p+R_1+R_{22}+\tilde{R})}\mathds{1}_{\big\{m_{12}=\hat{m}_{12}(m_p),\mathbf{u}_0=\mathbf{u}_0(m_p),\mathbf{u}_2=\mathbf{u}_2(m_p,m_{22})\big\}}P^{(\mathcal{C}_n)}_\mathsf{LE}\big(i\big|w,\mathbf{u}_0(m_p),\mathbf{u}_2(m_p,m_{22})\big)\mathds{1}_{\big\{\mathbf{u}_1=\mathbf{u}_1(m_p,m_1,w,i)\big\}}\\&\times Q_{X|U_0,U_1,U_2}^n(\mathbf{x}|\mathbf{u}_0,\mathbf{u}_1,\mathbf{u}_2)Q_{Y_1,Y_2|X}^n(\mathbf{y}_1,\mathbf{y}_2|\mathbf{x})\mathds{1}_{\big\{\big(\hat{m}_0^{(1)},\hat{m}_1\big)=\phi_1^{(\mathcal{C}_n)}(\mathbf{y}_1),\big(\hat{m}_0^{(2)},\hat{m}_2\big)=\phi_2^{(\mathcal{C}_n)}(m_{12},\mathbf{y}_2)\big\}}\numberthis\label{EQ:achie_proof_induced_PMF}
\end{align*}
\hrulefill
\begin{align*}
P\Big(m_p,m_1,m_{22},w,m_{12},\mathbf{u}_0&,\mathbf{u}_2,i,\mathbf{u}_1,\mathbf{x},\mathbf{y}_1,\mathbf{y}_2,\big(\hat{m}_0^{(1)},\hat{m}_1\big),\big(\hat{m}_0^{(2)},\hat{m}_2\big)\Big)\\&=\mu(\mathcal{C}_n)P^{(\mathcal{C}_n)}\Big(m_p,m_1,m_{22},w,m_{12},\mathbf{u}_0,\mathbf{u}_2,i,\mathbf{u}_1,\mathbf{x},\mathbf{y}_1,\mathbf{y}_2,\big(\hat{m}_0^{(1)},\hat{m}_1\big),\big(\hat{m}_0^{(2)},\hat{m}_2\big)\Big)\numberthis\label{EQ:achiev_induced_PMF}
\end{align*}
\hrulefill
\end{figure*}
\setcounter{equation}{55}

For a fixed codebook $\mathcal{C}_n\in\mathfrak{C}_n$ we next describe its associated encoding function $f^{(\mathcal{C}_n)}$, cooperation function $g_{12}^{(\mathcal{C}_n)}$ and decoding functions $\phi^{(\mathcal{C}_n)}_j$, for $j=1,2$.


\textbf{Encoder $\bm{f^{(\mathcal{C}_n)}}$:} To transmit a triple $(m_0,m_1,m_2)\in\mathcal{M}_0\times\mathcal{M}_1\times\mathcal{M}_2$, the encoder transforms it into the triple $(m_p,m_1,m_{22})\in\mathcal{M}_p\times\mathcal{M}_1\times\mathcal{M}_{22}$, and draws $W$ uniformly over $\mathcal{W}$; denote the realization of $W$ by $w\in\mathcal{W}$. Given $(m_p,m_1,m_{22},w)$, an index $i\in\mathcal{I}$ is then randomly selected by the likelihood encoder according to
\begin{align*}
&P^{(\mathcal{C}_n)}_\mathsf{LE}\big(i\big|w,\mathbf{u}_0(m_p),\mathbf{u}_2(m_p,m_{22})\big)\\
&=\frac{Q_{U_2|U_1,U_0}^n\big(\mathbf{u}_2(m_p,m_{22})\big|\mathbf{u}_1(m_p,m_1,w,i),\mathbf{u}_0(m_p)\big)}{\sum\limits_{i'\in\mathcal{I}}Q_{U_2|U_1,U_0}^n\big(\mathbf{u}_2(m_p,m_{22})\big|\mathbf{u}_1(m_p,m_1,w,i'),\mathbf{u}_0(m_p)\big)}.\numberthis\label{EQ:likelihood_encoder_BC}
\end{align*}
The structure of $P^{(\mathcal{C}_n)}_\mathsf{LE}$ adheres to the setup of Lemmas \ref{LEMMA:soft_covering}-\ref{LEMMA:typicallity} from Section \ref{SEC:soft_covering} and, in particular, to the stochastic choice of indices therein as described in \eqref{EQ:likelihood_definition}.

Denoting by $i\in\mathcal{I}$ the index selected by $P^{(\mathcal{C}_n)}_\mathsf{LE}$,~the channel input sequence is then randomly generated according to the conditional product distribution $Q^n_{X|U_0,U_1,U_2}\big(\cdot\big|\mathbf{u}_0(m_p),\mathbf{u}_1(m_p,m_1,w,i),\mathbf{u}_2(m_p,m_{22})\big)$.


\par\textbf{Decoding and Cooperation:} For a fixed codebook $\mathcal{C}_n\in\mathfrak{C}_n$, we define the following:
\begin{itemize}
    \item \textbf{Decoder $\bm{\phi_1^{(\mathcal{C}_n)}}$:} Searches for a unique triple $(\hat{m}_p,\hat{m}_1,\hat{w})\in\mathcal{M}_p\times\mathcal{M}_1\times\mathcal{W}$, for which there exists an index $\hat{i}\in\mathcal{I}$ such that
\begin{equation}
\Big(\mathbf{u}_0(\hat{m}_p),\mathbf{u}_1(\hat{m}_p,\hat{m}_1,\hat{w},\hat{i}),\mathbf{y}_1\Big)\in\mathcal{T}_\epsilon^n(Q_{U_0,U_1,Y_1}).
\end{equation}
If such a unique triple is found set $\phi_1^{(\mathcal{B}_n)}(\mathbf{y}_1)=(\hat{m}_0,\hat{m}_1)$, where $\hat{m}_0$ is taken from $\hat{m}_p=(\hat{m}_0,\hat{m}_{22})$; otherwise, set $\phi_1^{(\mathcal{C}_n)}(\mathbf{y}_1)=(1,1)$.

\item \textbf{Cooperation $\bm{g_{12}^{(\mathcal{C}_n)}}$:} Having $(\hat{m}_p,\hat{m}_1,\hat{w},\hat{i})$, Decoder 1 conveys the bin number of $\hat{m}_p$, i.e., $\hat{m}_{12}(\hat{m}_p)\in\mathcal{M}_{12}$, to Decoder 2 via the cooperation link. That is, $g_{12}^{(\mathcal{C}_n)}(\mathbf{y}_1)=\hat{m}_{12}(\hat{m}_p)$.

\item \textbf{Decoder $\bm{\phi_2^{(\mathcal{C}_n)}}$:} Upon observing $\big(\hat{m}_{12}(\hat{m}_p),\mathbf{y}_2\big)$, Decoder 2 searches for a unique pair $(\hat{\hat{m}}_p,\hat{\hat{m}}_{22})\in\mathcal{M}_p\times\mathcal{M}_{22}$, such that
\begin{equation}
\Big(\mathbf{u}_0(\hat{\hat{m}}_p),\mathbf{u}_2(\hat{\hat{m}}_p,\hat{\hat{m}}_{22}),\mathbf{y}_2\Big)\in\mathcal{T}_\epsilon^n(Q_{U_0,U_2,Y_2})
\end{equation}
where $\hat{\hat{m}}_p\in\mathcal{B}_n\big(\hat{m}_{12}(\hat{m}_p)\big)$. If such a unique pair is found, set
$\phi_2^{(\mathcal{C}_n)}\big(\hat{m}_{12}(\hat{m}_p),\mathbf{y}_2\big)=\big(\hat{\hat{m}}_0,\hat{\hat{m}}_2\big)$, where $\hat{\hat{m}}_2=(\hat{\hat{m}}_{20},\hat{\hat{m}}_{22})$ in which $\hat{\hat{m}}_0$ and $\hat{\hat{m}}_{20}$ are specified by $\hat{\hat{m}}_p=\big(\hat{\hat{m}}_0,\hat{\hat{m}}_{20}\big)$; otherwise, set $\phi_2^{(\mathcal{B}_n)}\big(\hat{m}_{12}(\hat{m}_p),\mathbf{y}_2\big)=(1,1)$.
\end{itemize}


\textbf{Induced Code and Joint Distribution:} The tuple $\left(f^{(\mathcal{C}_n)},g_{12}^{(\mathcal{C}_n)},\phi^{(\mathcal{C}_n)}_1,\phi^{(\mathcal{C}_n)}_2\right)$ defined with respect to the codebook $\mathcal{C}_n\in\mathfrak{C}_n$ constitutes an $(n,R_{12},R_0,R_1,R_2)$ code $c_n$ for the cooperative BC. Thus, for every codebook $\mathcal{C}_n\in\mathfrak{C}_n$, the induced joint distribution is given in \eqref{EQ:achie_proof_induced_PMF} at the top of this page, where the random variables $\mathbf{U}_0$, $\mathbf{U}_1$ and $\mathbf{U}_2$ are the chosen codewords at the conclusion of the encoding process (from which the input $\mathbf{X}$ to the BC is generated).

Taking the random codebook generation into account, we also set \eqref{EQ:achiev_induced_PMF} from the top of this page, where $\mu\in\mathcal{P}(\mathfrak{C}_n)$ is described in \eqref{EQ:codebook_probability}. The PMF $P$ induces a probability measure $\mathbb{P}\triangleq\mathbb{P}_P$, with respect to which the subsequent analysis is preformed. Specifically, all the mutli-letter information measures in the sequel are taken with respect to $P$ from \eqref{EQ:achiev_induced_PMF}, while single-letter information terms are always calculated with respect to $Q_{U_0,U_1,U_2,X,Y_1,Y_2}$.

\begin{figure*}[!t]
\setcounter{equation}{63}
\begin{align*}
D&\Big(P^{(\mathcal{B}_n)}_{\mathbf{Y}_2|M_p,M_1,M_{22},\mathbf{U}_0,\mathbf{U}_2}\Big|\Big|P^{(\mathcal{B}_n)}_{\mathbf{Y}_2|M_p,M_{22},\mathbf{U}_0,\mathbf{U}_2}\Big|P^{(\mathcal{B}_n)}_{M_p,M_1,M_{22},\mathbf{U}_0,\mathbf{U}_2}\Big)\\
      &\leq D\Big(P^{(\mathcal{B}_n)}_{\mathbf{Y}_2|M_p,M_1,M_{22},\mathbf{U}_0,\mathbf{U}_2}\Big|\Big|Q_{Y_2|U_0,U_2}^n\Big|P^{(\mathcal{B}_n)}_{M_p,M_1,M_{22},\mathbf{U}_0,\mathbf{U}_2}\Big)-D\Big(P^{(\mathcal{B}_n)}_{\mathbf{Y}_2|M_p,M_{22},\mathbf{U}_0,\mathbf{U}_2}\Big|\Big|Q_{Y_2|U_0,U_2}^n\Big|P^{(\mathcal{B}_n)}_{M_p,M_{22},\mathbf{U}_0,\mathbf{U}_2}\Big)\numberthis\label{EQ:auxiliary_leakage_UB}
\end{align*}
\hrulefill
\begin{align*}
\mathbb{E}_{\mathsf{C}_n}&D\Big(P_{\mathbf{Y}_2|M_p,M_1,M_{22},\mathbf{U}_0,\mathbf{U}_2,\mathsf{C}_n}\Big|\Big|Q_{Y_2|U_0,U_2}^n\Big|P_{M_p,M_1,M_{22},\mathbf{U}_0,\mathbf{U}_2,\mathsf{C}_n}\Big)\\
          &\begin{multlined}[b][.87\textwidth]=\mathbb{E}_{\mathsf{C}_n}\Bigg[\sum_{m_p,m_1,m_{22},\mathbf{u}_0,\mathbf{u}_2}\mspace{-20mu}2^{-n(R_p+R_1+R_{22})}\mathds{1}_{\big\{\big(\mathbf{U}_0(m_p),\mathbf{U}_2(m_p,m_{22})\big)=(\mathbf{u}_0,\mathbf{u}_2)\big\}}\\
          \times D\Big(P_{\mathbf{Y}_2|M_p=m_p,M_1=m_1,M_{22}=m_{22},\mathbf{U}_0=\mathbf{u}_0,\mathbf{U}_2=\mathbf{u}_2,\mathsf{C}_n}\Big|\Big|Q^n_{Y_2|U_0,U_2}(\cdot|\mathbf{u}_0,\mathbf{u}_2)\Big)\Bigg]\end{multlined}\\
          &\stackrel{(a)}=\sum_{\mathbf{u}_0,\mathbf{u}_2}\mathbb{E}_{\mathsf{C}_n}\bigg[\mathds{1}_{\big\{\big(\mathbf{U}_0(1),\mathbf{U}_2(1,1)\big)=(\mathbf{u}_0,\mathbf{u}_2)\big\}}D\Big(P_{\mathbf{Y}_2|M_p=1,M_1=1,M_{22}=1,\mathbf{U}_0=\mathbf{u}_0,\mathbf{U}_2=\mathbf{u}_2,\mathsf{C}_n}\Big|\Big|Q^n_{Y_2|U_0,U_2}(\cdot|\mathbf{u}_0,\mathbf{u}_2)\Big)\bigg]\\
          &\stackrel{(b)}=\sum_{\mathbf{u}_0,\mathbf{u}_2}\mathbb{E}_{\mathsf{C}^{(n)}_{0,2}}\Bigg[\mathds{1}_{\big\{\big(\mathbf{U}_0(1),\mathbf{U}_2(1,1)\big)=(\mathbf{u}_0,\mathbf{u}_2)\big\}}\mathbb{E}_{\mathsf{C}^{(n)}_1\big|\mathsf{C}^{(n)}_{0,2}}\bigg[D\Big(P_{\mathbf{Y}_2|M_p=1,M_1=1,M_{22}=1,\mathbf{U}_0=\mathbf{u}_0,\mathbf{U}_2=\mathbf{u}_2,\mathsf{C}_n}\Big|\Big|Q^n_{Y_2|U_0,U_2}(\cdot|\mathbf{u}_0,\mathbf{u}_2)\Big)\bigg]\Bigg]\numberthis\label{EQ:expected_divergence_UB1}
\end{align*}
\hrulefill
\end{figure*}

\setcounter{equation}{60}

\setcounter{equation}{60}

\textbf{Expected Average Error Probability Analysis:} By virtue of Lemma \ref{LEMMA:typicallity} we first show that under the proper rate constraints, the above encoding process results in $u_0$-, $u_1$- and $u_2$-sequences that are jointly typical. Having that, the rest of the analysis goes through via classic joint typicality arguments. The details of the analysis are relegated to Appendix \ref{APPEN:error_analysis}, where it is shown that
\begin{equation}
\mathbb{E}P_{\mathsf{e}}(\mathsf{C}_n)\leq \eta(n,\delta,\delta'),\label{EQ:error_prob_decays}
\end{equation}
where $\delta'\in(0,\delta)$ and $\lim_{n\to\infty}\eta(n,\delta,\delta')=0$ for all $0<\delta'<\delta$, if
\begin{subequations}
\begin{align}
R'&>I(U_1;U_2|U_0)\label{EQ:achiev_rb1}\\
R'+\tilde{R}&>I(U_1;U_2,Y_2|U_0)\label{EQ:achiev_rb6}\\
R_1+\tilde{R}+R' &< I(U_1;Y_1|U_0)-\tau_\delta\label{EQ:achiev_rb2}\\
R_p+R_1+\tilde{R}+R' &< I(U_0,U_1;Y_1)-\tau_\delta\label{EQ:achiev_rb3}\\
R_{22}&< I(U_2;Y_2|U_0)-\tau_\delta\label{EQ:achiev_rb4}\\
R_p+R_{22}-
R_{12}&< I(U_0,U_2;Y_2)-\tau_\delta.\label{EQ:achiev_rb5}
\end{align}\label{EQ:achiev_rb}%
\end{subequations}with $\tau_\delta\to 0$ as $\delta\to 0$ and $\tau_{\delta'}\to 0$ as $\delta'\to 0$. To clarify, the $\delta'$ that appears in the upper bound on the expected error probability from \eqref{EQ:error_prob_decays} is a consequence of the Conditional Typicality Lemma \cite[Section 2.5]{ElGamal2011}. Namely, the lemma considers conditioning on sequences that are jointly letter-typical with respect to a slightly smaller gap than the original $\delta$. This smaller gap is $\delta'$.


\textbf{Security Analysis:} As in the proof of Lemma \ref{LEMMA:soft_covering} from Section \ref{SUBSEC:soft_covering_proof}, throughout this proof we use $P^{(\mathcal{C}_n)}$ when the codebook $\mathcal{C}_n\in\mathfrak{C}_n$ is fixed, and  $P_{\cdot|\mathsf{C}_n}$ when the codebook is random (see \eqref{EQ:achie_proof_induced_PMF}-\eqref{EQ:achiev_induced_PMF}). Fix a codebook $\mathcal{C}_n\in\mathfrak{C}_n$ and let $I_{\mathcal{C}_n}$ denote the a mutual information taken with respect to $P^{(\mathcal{C}_n)}$. Consider the following upper bound on the information leakage.
\begin{align*}
&I_{\mathcal{C}_n}(M_1;M_{12},\mathbf{Y}_2)\\
    &\leq I_{\mathcal{C}_n}(M_1;M_{12},M_p,M_{22},\mathbf{Y}_2)\\
    &\stackrel{(a)}=I_{\mathcal{C}_n}\big(M_1;\mathbf{Y}_2|M_p,M_{22},\mathbf{U}_0,\mathbf{U}_2\big)\\
    &\stackrel{(b)}\leq D\Big(P^{(\mathcal{C}_n)}_{\mathbf{Y}_2|M_p,M_1,M_{22},\mathbf{U}_0,\mathbf{U}_2}\Big|\Big|Q_{Y_2|U_0,U_2}^n\Big|P^{(\mathcal{C}_n)}_{M_p,M_1,M_{22},\mathbf{U}_0,\mathbf{U}_2}\Big)\numberthis\label{EQ:secrecy_divergence_UB1}
\end{align*}
where:\\
(a) is because $M_1$ is independent of $(M_p,M_{22})$, and since $M_{12}=\hat{m}_{12}(M_p)$, $\mathbf{U}_0=\mathbf{u}_0(M_p)$ and $\mathbf{U}_2=\mathbf{u}_2(M_p,M_{22})$ are defined by $(M_p,M_{22})$;\\
(b) follows by the relative entropy chain rule and because for every $\mathcal{C}_n\in\mathfrak{C}_n$, the definition of relative entropy gives \eqref{EQ:auxiliary_leakage_UB} from the top of this page.

Taking the expectation of the RHS of \eqref{EQ:secrecy_divergence_UB1} over the ensemble of codebooks, we get \eqref{EQ:expected_divergence_UB1} from the top of this page, where (a) uses the symmetry of the codebook with respect to the messages, while (b) is the law of total expectation and (conditioning the inner expectation on $\mathsf{C}^{(n)}_{0,2}\triangleq\left\{\mathsf{C}^{(n)}_0,\mathsf{C}^{(n)}_2\right\}$).

\setcounter{equation}{65}

Next, we adjust the RHS of \eqref{EQ:expected_divergence_UB1} so that it corresponds to the setup of Lemma \ref{LEMMA:soft_covering}. To this end, note that when $\mathcal{C}_n\in\mathfrak{C}_n$ is fixed, $P^{(\mathcal{C}_n)}_{\mathbf{Y}_2|M_p=1,M_1=1,M_{22}=1,\mathbf{U}_0=\mathbf{u}_0,\mathbf{U}_2=\mathbf{u}_2}$ is well-defined only if $\mathbf{u}_0=\mathbf{u}_0(1)$ and $\mathbf{u}_2=\mathbf{u}_2(1)$. For any other $\mathbf{u}_0$ and $\mathbf{u}_2$, we may set this conditional distribution as any arbitrary PMF on $\mathcal{Y}_2^n$, since this does not affect the joint distribution from \eqref{EQ:achie_proof_induced_PMF}. Accordingly, if $\mathbf{u}_0\neq\mathbf{u}_0(1)$ or $\mathbf{u}_2\neq\mathbf{u}_2(1,1)$, we define
\begin{equation}
    P^{(\mathcal{C}_n)}_{\mathbf{Y}_2|M_p=1,M_1=1,M_{22}=1,\mathbf{U}_0=\mathbf{u}_0,\mathbf{U}_2=\mathbf{u}_2}=Q^n_{Y_2|U_0,U_2}\big(\mspace{-1.5mu}\cdot\mspace{-1.5mu}\big|\mathbf{u}_0,\mathbf{u}_2\big).\label{EQ:cond_PMF_bad}
\end{equation}
Having this, note that for any $(\mathbf{u}_0,\mathbf{u}_2)\in\mathcal{U}_0^n\times\mathcal{U}_2^n$ and a fixed $\mathsf{C}^{(n)}_{0,2}=\mathcal{C}^{(n)}_{0,2}\triangleq\left\{\mathcal{C}^{(n)}_0,\mathcal{C}^{(n)}_2\right\}$, we have \eqref{EQ:expected_divergence_UB2}~from the top of the next page. In the derivation of \eqref{EQ:expected_divergence_UB2} (a) follows from \eqref{EQ:cond_PMF_bad} and because conditioned on $\mathbf{U}_0(1)$ and $\mathbf{U}_2(1,1)$, $P_{\mathbf{Y}_2|M_p=1,M_1=1,M_{22}=1,\mathbf{U}_0=\mathbf{u}_0,\mathbf{U}_2=\mathbf{u}_2,\mathsf{C}_n}$ is independent of all the other codewords in $\mathsf{C}_{0,2}$. Furthermore, $P_{\mathbf{Y}_2|M_p=1,M_1=1,M_{22}=1,\mathbf{U}_0=\mathbf{u}_0,\mathbf{U}_2=\mathbf{u}_2,\mathsf{C}_n}$ is actually a function of the codebook $\mathsf{C}^{(n)}_1(1)$, rather than the entire collection $\mathsf{C}_n$.

\begin{figure*}[!t]
\begin{align*}
   &\mathbb{E}_{\mathsf{C}^{(n)}_1\big|\mathsf{C}^{(n)}_{0,2}=\mathcal{C}^{(n)}_{0,2}}\bigg[D\Big(P_{\mathbf{Y}_2|M_p=1,M_1=1,M_{22}=1,\mathbf{U}_0=\mathbf{u}_0,\mathbf{U}_2=\mathbf{u}_2,\mathsf{C}_n}\Big|\Big|Q^n_{Y_2|U_0,U_1}(\cdot|\mathbf{u}_0,\mathbf{u}_2)\Big)\bigg]\\
   &\begin{multlined}[b][.87\textwidth]=\mathbb{E}_{\mathsf{C}^{(n)}_1\big|\mathsf{C}^{(n)}_{0,2}=\mathcal{C}^{(n)}_{0,2}}\bigg[\mathds{1}_{\big\{\big(\mathbf{u}_0(1),\mathbf{u}_2(1,1)\big)=(\mathbf{u}_0,\mathbf{u}_2)\big\}}D\Big(P_{\mathbf{Y}_2|M_p=1,M_1=1,M_{22}=1,\mathbf{U}_0=\mathbf{u}_0,\mathbf{U}_2=\mathbf{u}_2,\mathsf{C}_n}\Big|\Big|Q^n_{Y_2|U_0,U_1}(\cdot|\mathbf{u}_0,\mathbf{u}_2)\Big)\\
   +\mathds{1}_{\big\{\big(\mathbf{u}_0(1),\mathbf{u}_2(1,1)\big)\neq(\mathbf{u}_0,\mathbf{u}_2)\big\}}D\Big(P_{\mathbf{Y}_2|M_p=1,M_1=1,M_{22}=1,\mathbf{U}_0=\mathbf{u}_0,\mathbf{U}_2=\mathbf{u}_2,\mathsf{C}_n}\Big|\Big|Q^n_{Y_2|U_0,U_1}(\cdot|\mathbf{u}_0,\mathbf{u}_2)\Big)\bigg]\end{multlined}\\
   &\begin{multlined}[b][.87\textwidth]\stackrel{(a)}=\mathbb{E}_{\mathsf{C}^{(n)}_1\big|\mathbf{U}_0(1)=\mathbf{u}_0(1),\mathbf{U}_2(1,1)=\mathbf{u}_2(1,1)}\bigg[\mathds{1}_{\big\{\big(\mathbf{u}_0(1),\mathbf{u}_2(1,1)\big)=(\mathbf{u}_0,\mathbf{u}_2)\big\}}\\\times D\Big(P_{\mathbf{Y}_2|M_p=1,M_1=1,M_{22}=1,\mathbf{U}_0=\mathbf{u}_0,\mathbf{U}_2=\mathbf{u}_2,\mathsf{C}_1^{(n)}(1)}\Big|\Big|Q^n_{Y_2|U_0,U_1}(\cdot|\mathbf{u}_0,\mathbf{u}_2)\Big)\bigg]\end{multlined}\numberthis\label{EQ:expected_divergence_UB2}
\end{align*}
\hrulefill
\end{figure*}

Some further definitions are required in order to rigorously justify the application of Lemma \ref{LEMMA:soft_covering}. For each $\mathbf{u}_0\in\mathcal{U}_0^n$, let $\tilde{\mathsf{C}}_n(\mathbf{u}_0)\triangleq\big\{\tilde{\mathbf{U}}_1(\mathbf{u}_0,w,i)\big\}_{(w,i)\in\mathcal{W}\times\mathcal{I}}$, be a collection of i.i.d. random vectors of length $n$, each distributed according to $Q^n_{U_1|U_0}(\cdot|\mathbf{u}_0)$. The collection $\tilde{\mathsf{C}}_n\triangleq\left\{\tilde{\mathsf{C}}_n(\mathbf{u}_0)\right\}_{\mathbf{u}_0\in\mathcal{U}_0^n}$ is~independent of $\mathsf{C}_n$ and is distributed according to 
\begin{equation}
    \tilde{\lambda}(\tilde{\mathcal{C}}_n)=\prod_{\mathbf{u}_0\in\mathcal{U}_0^n}\prod_{\substack{(w,i)\\\in\mathcal{W}\times\mathcal{I}}}Q^n_{U_1|U_0}\big(\tilde{\mathbf{u}}_1(\mathbf{u}_0,w,i)\big|\mathbf{u}_0\big),\label{EQ:tilde_lambda_PMF}
\end{equation}
where, as before, $\tilde{\mathcal{C}}_n(\mathbf{u}_0)\triangleq\big\{\tilde{\mathbf{u}}_1(\mathbf{u}_0,w,i)\big\}_{(w,i)\in\mathcal{W}\times\mathcal{I}}$ stands for a realization of $\tilde{\mathsf{C}}_n(\mathbf{u}_0)$. For each $(\mathbf{u}_0,\mathbf{u}_2)\in\mathcal{U}_0^n\times\mathcal{U}_2^n$ and a corresponding $\tilde{\mathcal{C}}_n(\mathbf{u}_0)$, define a conditional PMF
\begin{align*}
&\tilde{P}^{(\tilde{\mathcal{C}}_n)}(w,i,\tilde{\mathbf{u}}_1,\mathbf{y}_2|\mathbf{u}_0,\mathbf{u}_2)\\
&\begin{multlined}[b][.42\textwidth]=2^{-n\tilde{R}}\tilde{P}^{(\tilde{\mathcal{C}}_n)}(i|w,\mathbf{u}_0,\mathbf{u}_2)\mathds{1}_{\big\{\tilde{\mathbf{u}}_1=\tilde{\mathbf{u}}_1(\mathbf{u}_0,w,i)\big\}}\\\times Q^n_{Y_2|U_0,U_1,U_2}(\mathbf{y}_2|\mathbf{u}_0,\tilde{\mathbf{u}}_1,\mathbf{u}_2)\end{multlined},\numberthis\label{EQ:achiev_proof_aux_PMF}
\end{align*}
where $\tilde{P}^{(\tilde{\mathcal{C}}_n)}(i|w,\mathbf{u}_0,\mathbf{u}_2)$ is defined exactly like $\hat{P}^{(\mathcal{B}_n)}(i|w,\mathbf{s}_0,\mathbf{s})$ from \eqref{EQ:likelihood_definition}, up to renaming $\mathbf{s}_0$, $\mathbf{s}$, $\mathbf{u}$ and $\mathcal{B}_n$ therein to $\mathbf{u}_0$, $\mathbf{u}_2$, $\tilde{\mathbf{u}}_1$ and $\tilde{\mathcal{C}}_n$, respectively. Also define 
\begin{equation}
\tilde{P}(\tilde{\mathcal{C}}_n,w,i,\tilde{\mathbf{u}}_1,\mathbf{y}_2|\mathbf{u}_0,\mathbf{u}_2)=\tilde{\lambda}(\tilde{\mathcal{C}}_n)\tilde{P}^{(\tilde{\mathcal{C}}_n)}(w,i,\tilde{\mathbf{u}}_1,\mathbf{y}_2|\mathbf{u}_0,\mathbf{u}_2).\label{EQ:achiev_proof_aux_PMF_code}
\end{equation}

For any $(\mathbf{u}_0,\mathbf{u}_2)\in\mathcal{U}_0^n\times\mathcal{U}_2^n$, the RHS of \eqref{EQ:expected_divergence_UB2} is further upper bounded by
\begin{equation}
\mathbb{E}_{\tilde{\mathsf{C}}_n}D\Big(\tilde{P}_{\mathbf{Y}_2|\mathbf{U}_0=\mathbf{u}_0,\mathbf{U}_2=\mathbf{u}_2,\tilde{\mathsf{C}}_n}\Big|\Big|Q^n_{Y_2|U_0,U_1}(\cdot|\mathbf{u}_0,\mathbf{u}_2)\Big).\label{EQ:expected_divergence_UB3}
\end{equation}
This follows by removing the indicator function and because when $\mathbf{u}_0(1)=\tilde{\mathbf{u}}_0$ and $\mathcal{C}_1^{(n)}(1)=\tilde{\mathcal{C}}_n(\tilde{\mathbf{u}}_0)$, the distributions $P_{\mathbf{Y}_2|M_p=1,M_1=1,M_{22}=1,\mathbf{U}_0=\mathbf{u}_0,\mathbf{U}_2=\mathbf{u}_2,\mathsf{C}_1^{(n)}(1)=\mathcal{C}_1^{(n)}(1)}$ and $\tilde{P}_{\mathbf{Y}_2|\mathbf{U}_0=\tilde{\mathbf{u}}_0,\mathbf{U}_2=\mathbf{u}_2,\tilde{\mathsf{C}}_n(\tilde{\mathbf{u}}_0)=\tilde{\mathcal{C}}_n(\tilde{\mathbf{u}}_0)}$ are equal as PMFs on $\mathcal{Y}_2^n$. Since \eqref{EQ:expected_divergence_UB3} falls within the framework of Lemma \ref{LEMMA:soft_covering} we can make this expectation arbitrarily small provided that \eqref{EQ:achiev_rb1}-\eqref{EQ:achiev_rb6} hold.


Inserting \eqref{EQ:expected_divergence_UB1}, \eqref{EQ:expected_divergence_UB2} and \eqref{EQ:expected_divergence_UB3} back into \eqref{EQ:secrecy_divergence_UB1}, yields
\begin{align*}
&\mathbb{E}_{\mathsf{C}_n}\ell(\mathsf{C}_n)\\
    &I(M_1;M_{12},\mathbf{Y}_2|\mathsf{C}_n)\\
    &\begin{multlined}[b][.47\textwidth]\leq\sum_{\mathbf{u}_0,\mathbf{u}_2}\mathbb{E}_{\mathsf{C}_{0,2}}\mathds{1}_{\big\{\big(\mathbf{U}_0(1),\mathbf{U}_2(1,1)\big)=(\mathbf{u}_0,\mathbf{u}_2)\big\}}\\\times\mathbb{E}_{\tilde{\mathsf{C}}_n}D\Big(\tilde{P}_{\mathbf{Y}_2|\mathbf{U}_0=\mathbf{u}_0,\mathbf{U}_2=\mathbf{u}_2,\tilde{\mathsf{C}}_n}\Big|\Big|Q^n_{Y_2|U_0,U_1}(\cdot|\mathbf{u}_0,\mathbf{u}_2)\Big)\end{multlined}\\
    &\begin{multlined}[b][.47\textwidth]\stackrel{(a)}=\mathbb{E}_{\tilde{\mathsf{C}}_n}\Bigg[\sum_{\mathbf{u}_0,\mathbf{u}_2}Q_{U_0,U_2}^n(\mathbf{u}_0,\mathbf{u}_2)\\\times D\Big(\tilde{P}_{\mathbf{Y}_2|\mathbf{U}_0=\mathbf{u}_0,\mathbf{U}_2=\mathbf{u}_2,\tilde{\mathsf{C}}_n}\Big|\Big|Q^n_{Y_2|U_0,U_1}(\cdot|\mathbf{u}_0,\mathbf{u}_2)\Big)\Bigg]\end{multlined}\\
    &=\mathbb{E}_{\tilde{\mathsf{C}}_n}D\Big(\tilde{P}_{\mathbf{Y}_2|\mathbf{U}_0,\mathbf{U}_2,\tilde{\mathsf{C}}_n}\Big|\Big|Q_{Y_2|U_0,U_2}^n\Big|Q_{U_0,U_2}^n\Big)\numberthis\label{EQ:expected_divergence_final}
\end{align*}
where (a) is since $Q_{U_0,U_2}$ is the coding PMF, which gives $\mathbb{P}_\mu\Big(\mathbf{U}_0(1)=\mathbf{u}_0,\mathbf{U}_2(1,1)=\mathbf{u}_2\Big)=Q_{U_0,U_2}^n(\mathbf{u}_0,\mathbf{u}_2)$. Invoking Lemma \ref{LEMMA:soft_covering} on the RHS of \eqref{EQ:expected_divergence_final}, while viewing $Q_{Y_2|U_0,U_1,U_2}$ as a state-dependent DMC from $\mathcal{U}_1$ to $\mathcal{Y}_2$ with state space $\mathcal{U}_0\times\mathcal{U}_2$, we see that \eqref{EQ:achiev_rb1}-\eqref{EQ:achiev_rb6} give
\begin{equation}
\mathbb{E}_{\tilde{\mathsf{C}}_n}D\Big(\tilde{P}_{\mathbf{Y}_2|\mathbf{U}_0,\mathbf{U}_2,\tilde{\mathsf{C}}_n}\Big|\Big|Q_{Y_2|U_0,U_2}^n\Big|Q_{U_0,U_2}^n\Big)\xrightarrow[n\to\infty]{}0\label{EQ:expected_divergence_to_zero}.
\end{equation}

The Selection Lemma \cite[Lemma 5]{Goldfeld_WTCII_semantic2015} (see also \cite[Lemma 2.2]{Bloch_Barros_Secrecy_Book2011}) applied to the sequence of random variables $\big\{\mathsf{C}_n\big\}_{n\in\mathbb{N}}$ and the functions $P_e$ and $\ell$ implies the existence of a sequence of codebooks $\big\{\mathcal{C}_n\big\}_{n\in\mathbb{N}}$, each giving rise to a code $c_n$ such that $P_e(c_n)\leq\epsilon$ and $\ell(c_n)\leq\epsilon$, for $n$ sufficiently large. Finally, we apply Fourier-Motzkin elimination (FME) on \eqref{EQ:achiev_rb} while using \eqref{EQ:achiev_partial_rates_sum} and the non-negativity of the involved terms, to eliminate $R_{20}$, $R'$ and $\tilde{R}$. Since the above linear inequalities have constant coefficients, the FME can be performed by a computer program, e.g., by the FME-IT algorithm \cite{FMEIT_newsletter2015}. This produces the rate bounds from \eqref{EQ:region_inner} with small subtracted terms such as $\tau_\delta$. Since $\delta>0$ and $\delta'\in(0,\delta)$ can be chosen arbitrarily small (which shrinks $\tau_\delta$), this concludes the proof of Theorem \ref{TM:inner_bound}.

{\def\arraystretch{2}
\begin{table*}[!t]
\begin{center}
\caption{Correspondence between the coding scheme for the cooperative BC and the setup of the resolvability Lemma~\ref{LEMMA:soft_covering}}\label{TABLE:BC_vs_resolvability}
\begin{tabular}{|c||c|c|}
\hline
& \textbf{Cooperative BC Code} & \textbf{Resolvability Lemma}\\ \hline\hline
\textbf{State-dependent DMC} & $Q_{Y_2|U_0,U_1,U_2}$ & $Q_{V|U,S_0,S}$ \\ \hline 
\textbf{Channel states} & $(\mathbf{U}_0,\mathbf{U}_2)$  & $(\mathbf{S}_0,\mathbf{S})$ \\ \hline
\textbf{Channel input} & $\mathbf{U}_1$  & $\mathbf{U}$ \\ \hline
\textbf{Resolvability codebook} & $\big\{\mathbf{U}_1(m_p,m_1,w,i)\big\}_{(w,i)=(1,1)}^{(2^{n\tilde{R}},2^{nR'})}$, & $\big\{\mathbf{U}(\mathbf{s}_0,w,i)\big\}_{(w,i)=(1,1)}^{(2^{n\tilde{R}},2^{nR'})}$ \\
& for each $(m_p,m_1)\in\mathcal{M}_p\times\mathcal{M}_1$ & \\ \hline
\textbf{Codebook generation} & $ \sim Q^n_{U_1|U_0}\big(\cdot\big|\mathbf{u}(m_p)\big)$ & $\sim Q^n_{U|S_0}(\cdot|\mathbf{s}_0)$ \\ \hline
\textbf{Likelihood encoder} & $P^{(\mathcal{C}_n)}_\mathsf{LE}(i|w,\mathbf{u}_0,\mathbf{u}_2)$ from \eqref{EQ:likelihood_encoder_BC} - & $\hat{P}^{(\mathcal{B}_n)}(i|w,\mathbf{s}_0,\mathbf{s})$ from \eqref{EQ:likelihood_definition} - \\
& Correlates $(\mathbf{U}_0,\mathbf{U}_1)$ with $\mathbf{U}_2$ & Correlates $(\mathbf{S}_0,\mathbf{U})$ with $\mathbf{S}_2$\\ \hline
\textbf{Rate bounds} & $R'>I(U_1;U_2|U_0)$ & $R'>I(U;S|S_0)$\\
& $R'+\tilde{R}>I(U_1;U_2,Y_2|U_0)$ & $R'+\tilde{R}>I(U;S,V|S_0)$\\ \hline
\textbf{Implied asymptotic} & $I(M_1;M_{12},\mathbf{Y}_2|\mathsf{C}_n)\to 0$ & $\mathbb{E}_{\mathsf{B}_n}D\Big(P_{\mathbf{V}|\mathbf{S}_0,\mathbf{S},\mathsf{B}_n}\Big|\Big|Q_{V|S_0,S}^n\Big|Q_{S_0,S}^n\Big)\to 0$ \\
 \textbf{behaviour} & as $n\to\infty$ & as $n\to\infty$\\
\hline
\end{tabular}
\end{center}
\hrulefill
\end{table*}}

\begin{remark}[BC Code and Resolvability Lemma Analogy]
Lemma \ref{LEMMA:soft_covering} is key in the security analysis of the proposed coding scheme. In the following, we relate the cooperative BC code construction and the setup of our resolvability lemma. Having \eqref{EQ:secrecy_divergence_UB1}, the main idea is to adjust the relative entropy on the RHS so that it corresponds to the lemma. This is done by viewing the $u_0$- and the $u_2$-codewords from the BC codebook as a pair of states of the subchannel $Q_{Y_2|U_0,U_1,U_2}$ to Decoder 2, where the $u_1$-codewords plays the role of the channel's input. The validity of this analogy stems from the structure of the BC codebook, where for each $(m_p,m_1)\in\mathcal{M}_p\times\mathcal{M}_1$, the set $\big\{\mathbf{U}_1(m_p,m_1,w,i)\big\}_{(w,i)\in\mathcal{W}\times\mathcal{I}}$ forms a resolvability codebook just like in Lemma \ref{LEMMA:soft_covering}. This resolvability codebook is superimposed on $\mathbf{U}_0(m_p)$, while the transmitted $u_1$-codeword is correlated with $\mathbf{U}_2(m_p,m_{22})$ by means of the likelihood encoder \eqref{EQ:likelihood_encoder_BC}. The correspondence between the coding scheme presented in this section and the setup of Lemma \ref{LEMMA:soft_covering} is summarized in Table \ref{TABLE:BC_vs_resolvability}.

The main challenge in applying the resolvability for the BC code is accounting for the relative entropy from the RHS of \eqref{EQ:secrecy_divergence_UB1} being conditioned on the induced joint distribution of $\mathbf{U}_0$ and $\mathbf{U}_2$, while the lemma conditions it on a product distribution. However, as the derivation between Equation \eqref{EQ:secrecy_divergence_UB1}-\eqref{EQ:expected_divergence_to_zero} shows, under the expectation over the ensemble of codebooks, the induced distribution in the conditioning can be converted to the product PMf $Q_{U_0,U_2}^n$ (according to which the codebooks $\mathbf{U}_0$ and $\mathbf{U}_2$ are drawn).
\end{remark}

\begin{remark}[Comparison to the Scheme without Secrecy]
The main differences between the coding schemes for the cooperative BC with one confidential message and the same channel without secrecy \cite{Goldfeld_BC_Cooperation2014} are threefold. First, a randomizer $W$ is used in the secrecy-achieving scheme. Second, the cooperation message $M_{12}$ depends on $M_{20}$ rather than on the pair $(M_{10},M_{20})$ ($M_{10}$ refers to the public part of the message $M_1$). Note that conveying an $M_{12}$ that holds any part of $M_1$ (in the form of its public part $M_{10}$) violates the secrecy requirement. Finally, a prefix channel $Q_{X|U_0,U_1,U_2}$ is used to optimize randomness and, in turn, to conceal $M_1$ from the 2nd receiver. In the non-secret scenario $Q_{X|U_0,U_1,U_2}$ can be replaced with a deterministic function.
\end{remark}


\subsection{Converse Proof for Theorem \ref{TM:SDBC_secrecy_capacity}}\label{SUBSEC:SD_proof_converse}

\par We show that if a rate tuple $(R_{12},R_0,R_1,R_2)$ is achievable, then there exists a PMF $Q_{W,V,Y_1,X}\in\mathcal{P}(\mathcal{W}\times\mathcal{V}\times\mathcal{Y}_1\times\mathcal{X})$ with $Y_1=y_1(X)$, such that the inequalities in \eqref{EQ:region_SDBC} are satisfied with respect to the joint distribution $Q_{W,V,Y_1,X}W_{Y_2|X}$. Fix an achievable tuple $(R_{12},R_0,R_1,R_2)$, an $\epsilon>0$, and let $c_n$ be the corresponding $(n,R_{12},R_0,R_1,R_2)$ code for some sufficiently large $n\in\mathbb{N}$ such that \eqref{EQ:achiev_realibility_secrecy} holds. All subsequent multi-letter information measures are calculated with respect to the PMF induced by $c_n$ from \eqref{EQ:induced_PMF}, with the SD-BC $W^n_{Y_1,Y_2|X}(\mathbf{y}_1,\mathbf{y}_2|\mathbf{x})=\mathds{1}_{\bigcap_{i=1}^n\big\{y_{1,i}=y_1(x_i)\big\}}W_{Y_2|X}^n(\mathbf{y}_2|\mathbf{x})$. By Fano's inequality we have
\begin{subequations}
\begin{align}
H(M_0,M_1|Y_1^n)&\leq 1+n\epsilon (R_0+R_1)\triangleq n\epsilon_n^{(1)}\label{EQ:SD_converse_Fano1}\\
H(M_0,M_2|M_{12},Y_2^n)&\leq 1+n\epsilon (R_0+R_2)\triangleq n\epsilon_n^{(2)}.\label{EQ:SD_converse_Fano2}
\end{align}
Define
\begin{equation}
\epsilon_n=\max\big\{\epsilon_n^{(1)},\epsilon_n^{(2)}\big\}.\label{EQ:SD_converse_Fano_max}
\end{equation}\label{EQ:SD_converse_Fano}%
\end{subequations}
Moreover, \eqref{EQ:achieve_secrecy} implies
\begin{align*}
\epsilon&\geq I(M_1;M_{12},Y_2^n)\\
   &=I(M_1;M_0,M_2,M_{12},Y_2^n)-I(M_1;M_0,M_2|M_{12},Y_2^n)\\
   &\stackrel{(a)}\geq I(M_1;M_{12},Y_2^n|M_0,M_2)-H(M_0,M_2|M_{12},Y_2^n)\\
   &\stackrel{(b)}\geq I(M_1;M_{12},Y_2^n|M_0,M_2)-n\epsilon_n\numberthis\label{EQ:SD_converse_secrecy_cond_info}
\end{align*}
where (a) uses the independence of $M_1$ and $(M_0,M_2)$ and the non-negativity of entropy, while (b) follows from \eqref{EQ:SD_converse_Fano}. Thus,
\begin{equation}
I(M_1;M_{12},Y_2^n|M_0,M_2)\leq\epsilon+n\epsilon_n.\label{EQ:SD_converse_secrecy_cond_info2}
\end{equation}
It follows that
\begin{align*}
  nR_1&=H(M_1)\\
      &\stackrel{(a)}=H(M_1|M_{12},M_0,M_2)+I(M_1;M_{12}|M_0,M_2)\\
      &\begin{multlined}[b][.43\textwidth]\stackrel{(b)}\leq I(M_1;Y_1^n|M_{12},M_0,M_2)+I(M_1;M_{12}|M_0,M_2)\\-I(M_1;M_{12},Y_2^n|M_0,M_2)+n\delta^{(1)}_n\end{multlined}\\
      &\begin{multlined}[b][.45\textwidth]\stackrel{(c)}=\sum_{i=1}^n\Big[I(M_1;Y_1^i,Y_{2,i+1}^n|M_{12},M_0,M_2)\\-I(M_1;Y_1^{i-1},Y_{2,i}^n|M_{12},M_0,M_2)\Big]+n\delta^{(1)}_n\end{multlined}\\
      &\begin{multlined}[b][.45\textwidth]=\sum_{i=1}^n\Big[I(M_1;Y_{1,i}|M_{12},M_0,M_2,Y_1^{i-1},Y_{2,i+1}^n)\\-I(M_1;Y_{2,i}|M_{12},M_0,M_2,Y_1^{i-1},Y_{2,i+1}^n)\Big]+n\delta^{(1)}_n\end{multlined}\\ &\begin{multlined}[b][.45\textwidth]\stackrel{(d)}=\sum_{i=1}^n\Big[H(Y_{1,i}|M_2,W_i)-H(Y_{1,i}|M_1,M_2,W_i)\\-I(M_1;Y_{2,i}|M_2,W_i)\Big]+n\delta^{(1)}_n\end{multlined}\\
      &\begin{multlined}[b][.45\textwidth]\leq\sum_{i=1}^n\Big[H(Y_{1,i}|M_2,W_i)-I(Y_{1,i};Y_{2,i}|M_1,M_2,W_i)\\-I(M_1;Y_{2,i}|M_2,W_i)\Big]+n\delta^{(1)}_n\end{multlined}\\
      &\begin{multlined}[b][.45\textwidth]=\sum_{i=1}^n\Big[H(Y_{1,i}|M_2,W_i)\\-I(M_1,Y_{1,i};Y_{2,i}|M_1,M_2,W_i)\Big]+n\delta^{(1)}_n\end{multlined}\\
      &\leq\sum_{i=1}^nH(Y_{1,i}|M_2,W_i,Y_{2,i})+n\delta^{(1)}_n\numberthis\label{EQ:SD_converse_r1UB_final}
\end{align*}
where:\\
  (a) is because $M_1$ is independent $(M_0,M_2)$;\\
  (b) follows from \eqref{EQ:SD_converse_Fano}-\eqref{EQ:SD_converse_secrecy_cond_info} and by denoting $\delta^{(1)}_n=2\epsilon_n+\frac{\epsilon}{n}$;\\
  (c) is a telescoping identity \cite[Equations (9) and (11)]{Kramer_telescopic2011};\\
  (d) defines $W_i=(M_{12},M_0,Y_1^{i-1},Y_{2,i+1}^n)$.

\par The common message rate $R_0$ satisfies
\begin{subequations}
\begin{align*}
nR_0&=H(M_0)\\
    &\stackrel{(a)}\leq I(M_0;Y_1^n)+n\epsilon_n\numberthis\label{EQ:SD_converse_0UB_final1a}\\
    &=\sum_{i=1}^nI(M_0;Y_{1,i}|Y_1^{i-1})+n\epsilon_n\\
    &\leq\sum_{i=1}^nI(M_0,Y_1^{i-1};Y_{1,i})+n\epsilon_n\\
    &\stackrel{(b)}\leq\sum_{i=1}^nI(W_i;Y_{1,i})+n\epsilon_n\numberthis\label{EQ:SD_converse_0UB_final1b}
\end{align*}
\end{subequations}
where (a) uses \eqref{EQ:SD_converse_Fano} and (b) follows by the definition of $W_i$. Combining \eqref{EQ:SD_converse_r1UB_final} with \eqref{EQ:SD_converse_0UB_final1b} yields
\begin{equation}
n(R_0+R_1)\leq\sum_{i=1}^n\Big[H(Y_{1,i}|M_2,W_i,Y_{2,i})+I(W_i;Y_{1,i})\Big]+n\delta^{(2)}_n\numberthis\label{EQ:SD_converse_r0r1UB_final}
\end{equation}
where $\delta^{(2)}_n=\delta^{(1)}_n+\epsilon_n$.

For the sum $R_0+R_2$, we have
\begin{align*}
n(R_0&+R_2)\\
    &= H(M_0,M_2)\\
    &\stackrel{(a)}\leq I(M_0,M_2;M_{12},Y_2^n)+n\epsilon_n\\
    &= I(M_0,M_2;Y_2^n|M_{12})+I(M_0,M_2;M_{12})+n\epsilon_n\\
    &\stackrel{(b)}\leq I(M_0,M_2;Y_2^n|M_{12})+nR_{12}+n\epsilon_n\\
    &= \sum_{i=1}^n I(M_0,M_2;Y_{2,i}|M_{12},Y_{2,i+1}^n)+nR_{12}+n\epsilon_n\\
    &\stackrel{(c)}\leq \sum_{i=1}^n I(M_2,W_i;Y_{2,i})+nR_{12}+n\epsilon_n\numberthis\label{EQ:SD_converse_r2UB_final}
\end{align*}
where:\\
(a) uses \eqref{EQ:SD_converse_Fano};\\
(b) is by the non-negativity of entropy and since a uniform distribution maximizes entropy;\\
(c) follows from the definition of $W_i$ and because conditioning cannot increase entropy.

\par To bound $R_0+R_1+R_2$, we begin by writing
\begin{align*}
n(R_0\mspace{-2.5mu}+\mspace{-2.5mu}R_1\mspace{-2.5mu}+\mspace{-2.5mu}R_2)&\mspace{-2mu}=\mspace{-2mu} H(M_0,M_1,M_2)\\
              &=H(M_1|M_0,M_2)\mspace{-2mu}+\mspace{-2mu}H(M_2|M_0)\mspace{-2mu}+\mspace{-2mu}H(M_0).\numberthis\label{EQ:SD_converse_UBsum}
\end{align*}
Consider now
\begin{align*}
&H(M_2|M_0)\\
    &\stackrel{(a)}\leq I(M_2;Y_2^n|M_{12},M_0)+I(M_2;M_{12}|M_0)+n\epsilon_n\\
    &\begin{multlined}[b][.45\textwidth]\stackrel{(b)}= \sum_{i=1}^n \Big[ I(M_2;Y_{2,i}^n|M_{12},M_0,Y_1^{i-1})\\-I(M_2;Y_{2,i+1}^n|M_{12},M_0,Y_1^i)\Big]+I(M_2;M_{12}|M_0)+n\epsilon_n\end{multlined}\\
    &\begin{multlined}[b][.45\textwidth]\stackrel{(c)}= \sum_{i=1}^n \Big[ I(M_2;Y_{2,i+1}^n|M_{12},M_0,Y_1^{i-1})\\+I(M_2;Y_{2,i}|W_i)-I(M_2;Y_{1,i},Y_{2,i+1}^n|M_{12},M_0,Y_1^{i-1})\\+I(M_2;Y_{1,i}|M_{12},M_0,Y_1^{i-1})\Big]+I(M_2;M_{12}|M_0)+n\epsilon_n
    \end{multlined}\\
    &\begin{multlined}[b][.477\textwidth]\stackrel{(d)}= \sum_{i=1}^n \Big[ I(M_2;Y_{2,i}|W_i)-I(M_2;Y_{1,i}|W_i)\Big]\\+I(M_2;Y_1^n|M_0)+n\epsilon_n\end{multlined}\numberthis\label{EQ:SD_converse_r2UB3}
\end{align*}
where:\\
  (a) uses \eqref{EQ:SD_converse_Fano} and the mutual information chain rule;\\
  (b) is a telescoping identity;\\
  (c) follows from the definition of $W_i$;\\
  (d) is due to the mutual information chain rule and the definition of $W_i$ (second term), and because $M_{12}$ is defined by $Y_1^n$ (third term).

\par Combining \eqref{EQ:SD_converse_0UB_final1a} with \eqref{EQ:SD_converse_r2UB3}, yields
\begin{align*}
&n(R_0+R_2)\\
        &\begin{multlined}[b][.45\textwidth]\leq \sum_{i=1}^n \Big[ I(M_2;Y_{2,i}|W_i)-I(M_2;Y_{1,i}|W_i)\Big]\\+I(M_0,M_2;Y_1^n)+2n\epsilon_n\end{multlined}\\
        &\begin{multlined}[b][.45\textwidth]\stackrel{(a)}\leq \sum_{i=1}^n \Big[ I(M_2;Y_{2,i}|W_i)-I(M_2;Y_{1,i}|W_i)+H(Y_{1,i})\\-H(Y_{1,i}|M_0,M_2,Y_1^{i-1})\Big]+2n\epsilon_n\end{multlined}\\
        &\begin{multlined}[b][.45\textwidth]\stackrel{(b)}\leq \sum_{i=1}^n \Big[ I(M_2;Y_{2,i}|W_i)+I(W_i;Y_{1,i})\\-I(M_{12},Y_{2,i+1}^n;Y_{1,i}|M_0,M_2,Y_1^{i-1})\Big]+2n\epsilon_n\end{multlined}\\
        &\stackrel{(c)}\leq \sum_{i=1}^n \Big[ I(M_2;Y_{2,i}|W_i)+I(W_i;Y_{1,i})\Big]+2n\epsilon_n\numberthis\label{EQ:SD_converse_r0r2UB}
\end{align*}
where:\\
(a) is because conditioning cannot increase entropy;\\
(b) uses the definition of $W_i$;\\
(c) is by the non-negativity of mutual information.

By inserting \eqref{EQ:SD_converse_r1UB_final} and \eqref{EQ:SD_converse_r0r2UB} into \eqref{EQ:SD_converse_UBsum}, we bound the sum of rates as
\begin{align*}
n(R_0+R_1+R_2)&\leq\sum_{i=1}^n\Big[H(Y_{1,i}|M_2,W_i,Y_{2,i})\\&+I(M_2;Y_{2,i}|W_i)+I(W_i;Y_{1,i})\Big]+n\delta^{(3)}_n\numberthis\label{EQ:SD_converse_r0r1r2UB_final}
\end{align*}
where $\delta^{(3)}_n=\delta^{(1)}_n+2\epsilon_n$.

The bounds in \eqref{EQ:SD_converse_r1UB_final}, \eqref{EQ:SD_converse_r0r1UB_final}, \eqref{EQ:SD_converse_r2UB_final} and   \eqref{EQ:SD_converse_r0r2UB}  are rewritten by introducing a time-sharing random variable $T$ that is uniformly distributed over the set $[1:n]$ and is independent of $(M_0,M_1,M_2,X^n,Y_1^n,Y_2^n)$. For instance, \eqref{EQ:SD_converse_r1UB_final} is rewritten as
\begin{align*}
  R_1&\leq\frac{1}{n}\sum_{t=1}^n H(Y_{1,t}|M_2,W_t,Y_{2,t})+\delta^{(1)}_n\\
     &=\sum_{t=1}^n\mathbb{P}\big(T=t\big) H(Y_{1,T}|M_2,W_T,Y_{2,T},T=t)+\delta^{(1)}_n\\
     &= H(Y_{1,T}|M_2,W_T,Y_{2,T},T)+\delta^{(1)}_n\numberthis\label{EQ:SD_1UBT_final1}
\end{align*}
Denote $W\triangleq (W_T,T)$, $V\triangleq (M_2,W)$, $X\triangleq X_T$, $Y_1\triangleq Y_{1,T}$ and $Y_2\triangleq Y_{2,T}$. This results in the bounds \eqref{EQ:region_SDBC} with small added terms such as $\epsilon_n$ and $\delta_n^{(1)}$. For large $n$, we can make these terms approach 0. The converse is completed by showing the PMF of $(W,V,X,Y_1,Y_2)$ factors as $Q_{W,V,Y_1,X}W_{Y_2|X}$ and satisfies $Y_1=y_1(X)$. As the functional relation between $Y_1$ and $X$ is straightforward, it remains to be shown that
\begin{equation}
(W,V,Y_1)-X-Y_2\label{EQ:SD_converse_Markov}
\end{equation}
forms a Markov chain. This is proven in Appendix \ref{APPEN:SD_Markov_Proof}.


\subsection{Converse Proof for Theorem \ref{TM:PDBC_secrecy_capacity}}\label{SUBSEC:PD_proof_converse}

\par We show that given an achievable rate tuple $(R_{12},R_0,R_1,R_2)$, there exists a PMF $Q_{W,X}\in\mathcal{P}(\mathcal{W}\times\mathcal{X})$ for which \eqref{EQ:region_PDBC} holds with respect to the joint distribution $Q_{W,X}W_{Y_1|X}W_{Y_2|Y_1}$. Let be $(R_{12},R_0,R_1,R_2)$ an achievable tuple and fix $\epsilon>0$. Let $c_n$ be the corresponding $(n,R_{12},R_0,R_1,R_2)$ code for some sufficiently large $n\in\mathbb{N}$ such that \eqref{EQ:achiev_realibility_secrecy} holds. The induced joint distribution is again given by \eqref{EQ:induced_PMF}, but now the transition matrix is of a PD-BC, i.e., $W^n_{Y_1,Y_2|X}(\mathbf{y}_1,\mathbf{y}_2|\mathbf{x})=W_{Y_1|X}^n(\mathbf{y}_1|\mathbf{x})W_{Y_2|Y_1}^n(\mathbf{y}_2|\mathbf{y}_1)$. Fano's inequality gives
\begin{subequations}
\begin{align}
H(M_0,M_1|Y_1^n)&\leq1\mspace{-2mu}+\mspace{-2mu}n\epsilon (R_0\mspace{-2mu}+\mspace{-2mu}R_1)\triangleq n\kappa_n^{(1)}\label{EQ:PD_converse_Fano1}\\
H(M_0,M_2|M_{12},Y_2^n)&\leq 1\mspace{-2mu}+\mspace{-2mu}n\epsilon (R_0\mspace{-2mu}+\mspace{-2mu}R_2)\triangleq n\kappa_n^{(2)}\label{EQ:PD_converse_Fano2}\\
H(M_0,M_1,M_2|Y_1^n,Y_2^n)&\leq 1\mspace{-2mu}+\mspace{-2mu}n\epsilon (R_0\mspace{-2mu}+\mspace{-2mu}R_1\mspace{-2mu}+\mspace{-2mu}R_2)\triangleq n\kappa_n^{(3)}\label{EQ:PD_converse_Fano3}
\end{align}
and we set
\begin{equation}
\kappa_n=\max\big\{\kappa_n^{(1)},\kappa_n^{(2)},\kappa_n^{(3)}\big\}=\kappa_n^{(3)}.\label{EQ:PD_converse_Fano_max}
\end{equation}\label{EQ:PD_converse_Fano}%
\end{subequations}
Further, by the strong secrecy constraint \eqref{EQ:achieve_secrecy}, we have
\begin{align*}
\epsilon&\geq I(M_1;M_{12},Y_2^n)\\&=I(M_1;M_0,M_2,M_{12},Y_2^n)-I(M_1;M_0,M_2|M_{12},Y_2^n)\\
                   &\stackrel{(a)}\geq I(M_1;M_{12},Y_2^n|M_0,M_2)-H(M_0,M_2|M_{12},Y_2^n)\\
                   &\stackrel{(b)}\geq I(M_1;Y_2^n|M_0,M_2)-n\kappa_n\numberthis\label{EQ:PD_converse_secrecy_cond_info}
\end{align*}
where (a) uses the independence of $M_1$ and $(M_0,M_2)$ and the non-negativity of entropy, while (b) is by \eqref{EQ:PD_converse_Fano} and since conditioning cannot increase entropy. This yields
\begin{equation}
I(M_1;Y_2^n|M_0,M_2)\leq\epsilon+n\kappa_n.\label{EQ:PD_converse_secrecy_cond_info2}
\end{equation}

We bound
\begin{subequations}
\begin{align*}
  nR_1&=H(M_1)\\
      &\stackrel{(a)}=H(M_1|M_0,M_2)\\
      &\stackrel{(b)}\leq I(M_1;Y_1^n|M_0,M_2)-I(M_1;Y_2^n|M_0,M_2)+n\eta_n\\
      &\begin{multlined}[b][.42\textwidth]\stackrel{(c)}=\sum_{i=1}^n\Big[I(M_1;Y_1^i,Y_{2,i+1}^n|M_0,M_2)\\-I(M_1;Y_1^{i-1},Y_{2,i}^n|M_0,M_2)\Big]+n\eta_n\end{multlined}\\
      &\stackrel{(d)}=\sum_{i=1}^n\Big[I(M_1;Y_{1,i}|W_i)-I(M_1;Y_{2,i}|W_i)\Big]+n\eta_n\numberthis\label{EQ:PD_converse_r1UB_temp}\\
      &\stackrel{(e)}=\sum_{i=1}^nI(M_1;Y_{1,i}|W_i,Y_{2,i})+n\eta_n\\
      &\stackrel{(f)}\leq \sum_{i=1}^nI(X_i;Y_{1,i}|W_i,Y_{2,i})+n\eta_n\\
      &\stackrel{(g)}\leq \sum_{i=1}^n\Big[I(X_i;Y_{1,i}|W_i)\mspace{-2mu}-\mspace{-2mu}I(X_i;Y_{2,i}|W_i)\Big]\mspace{-2mu}+\mspace{-2mu}n\eta_n\numberthis\label{EQ:PD_converse_r1UB_final}
\end{align*}
\end{subequations}
where:\\
  (a) uses the independence of $M_1$ and $(M_0,M_2)$;\\
  (b) is by virtue of \eqref{EQ:PD_converse_Fano}-\eqref{EQ:PD_converse_secrecy_cond_info} and by denoting $\eta_n=2\kappa_n+\frac{\epsilon}{n}$;\\
  (c) is a telescoping identity;\\
  (d) follows by defining $W_i\triangleq (M_0,M_2,Y_1^{i-1},Y_{2,i+1}^n)$;\\
  (e) and (g) rely on the mutual information chain rule and the PD property of the channel, which implies that $(M_1,X_i)-(W_i,Y_{1,i})-Y_{2,i}$ forms a Markov chain for all $i\in[1:n]$;\\
  (f) follows since $M_1-(W_i,X_i,Y_{1,i})-Y_{2,i}$ forms a Markov chain.

\par Next, we have
\begin{align*}
  n(R_0+R_2)&= H(M_0,M_2)\\
            &\stackrel{(a)}\leq I(M_0,M_2;M_{12},Y_2^n)+n\kappa_n\\
            &\stackrel{(b)}\leq I(M_0,M_2;Y_2^n)+nR_{12}+n\kappa_n\\
            &= \sum_{i=1}^n I(M_0,M_2;Y_{2,i}|Y_{2,i+1}^n)+nR_{12}+n\kappa_n\\
            &\stackrel{(c)}\leq \sum_{i=1}^n I(W_i;Y_{2,i})+nR_{12}+n\kappa_n\numberthis\label{EQ:PD_converse_r2UB_final}
\end{align*}
where:\\
(a) is by \eqref{EQ:PD_converse_Fano};\\
(b) is because entropy is non-negative and is maximized by the uniform distribution;\\
(c) follows from the definition of $W_i$ and because conditioning cannot increase entropy.

\par Finally, consider
\begin{align*}
&n(R_0+R_1+R_2)\\
        &=H(M_0,M_1,M_2)\\
        &\stackrel{(a)}\leq I(M_0,M_1,M_2;Y_1^n,Y_2^n)-I(M_1;Y_2^n|M_0,M_2)+n\eta_n\\
        &\stackrel{(b)}= I(M_0,M_1,M_2;Y_1^n)-I(M_1;Y_2^n|M_0,M_2)+n\eta_n\\
        &\begin{multlined}[b][.47\textwidth]\stackrel{(c)}=\sum_{i=1}^n\Big[I(M_0,M_1,M_2,Y_{2,i+1}^n;Y_{1,i}|Y_1^{i-1})\\-I(Y_{2,i+1}^n;Y_{1,i}|M_0,M_1,M_2,Y_1^{i-1})\\-I(M_1;Y_{2,i}|M_0,M_2,Y_{2,i+1}^n)\Big]+n\eta_n\end{multlined}\\
        &\begin{multlined}[b][.47\textwidth]\stackrel{(d)}=\sum_{i=1}^n\Big[I(M_0,M_1,M_2,Y_{2,i+1}^n;Y_{1,i}|Y_1^{i-1})\\-I(Y_1^{i-1};Y_{2,i}|M_0,M_1,M_2,Y_{2,i+1}^n)\\-I(M_1;Y_{2,i}|M_0,M_2,Y_{2,i+1}^n)\Big]+n\eta_n\end{multlined}\\
        &\begin{multlined}[b][.47\textwidth]\leq \sum_{i=1}^n \Big[I(M_0,M_1,M_2,Y_1^{i-1},Y_{2,i+1}^n;Y_{1,i})\\-I(M_1,Y_1^{i-1};Y_{2,i}|M_0,M_2,Y_{2,i+1}^n)\Big]+n\eta_n\end{multlined}\\
        &\stackrel{(e)}\leq \sum_{i=1}^n \Big[ I(W_i;\mspace{-1mu}Y_{1,i})\mspace{-3mu}+\mspace{-3mu}I(\mspace{-1mu}M_1;\mspace{-1.5mu}Y_{1,i}|W_i)\mspace{-3mu}-\mspace{-3mu}I(\mspace{-1mu}M_1;\mspace{-1.5mu}Y_{2,i}|W_i)\Big]\mspace{-4mu}+\mspace{-3mu}n\eta_n\\
        &\stackrel{(f)}\leq \sum_{i=1}^n \Big[ I(W_i;Y_{1,i})\mspace{-3mu}+\mspace{-3mu}I(X_i;Y_{1,i}|W_i)\mspace{-3mu}-\mspace{-3mu}I(X_i;Y_{2,i}|W_i)\Big]\mspace{-4mu}+\mspace{-3mu}n\eta_n\\
        &\stackrel{(g)}= \sum_{i=1}^n \Big[ I(X_i;Y_{1,i})-I(X_i;Y_{2,i}|W_i)\Big]+n\eta_n\numberthis\label{EQ:bc_converse_sumUB}
\end{align*}
where:\\
  (a) uses \eqref{EQ:PD_converse_Fano} and the definition of $\eta_n$;\\
  (b) is because $(M_0,M_1,M_2)-Y_1^n-Y_2^n$ forms a Markov chain, which is induced by the PD degraded and memoryless property of the channel;\\
  (c) is the mutual information chain rule;\\
  (d) uses the Csisz{\'a}r sum identity (see, e.g., \cite[Equation (3)]{Kramer_telescopic2011});\\
  (e) follows from the definitions of $W_i$ and because conditioning cannot increase entropy;\\
  (f) is by repeating steps \eqref{EQ:PD_converse_r1UB_temp}-\eqref{EQ:PD_converse_r1UB_final};\\
  (g ) is by the mutual information chain rule and because $W_i-X_i-Y_{1,i}$ forms a Markov chain (see Appendix \ref{APPEN:PD_Markov_Proof} for the proof).

By time-sharing arguments similar to those presented in Section \ref{SUBSEC:SD_proof_converse}, and by denoting $W\triangleq (W_T,T)$, $X\triangleq X_T$, $Y_1\triangleq Y_{1,T}$ and $Y_2\triangleq Y_{2,T}$, we obtain the bounds of \eqref{EQ:region_PDBC} with the small added terms $\kappa_n$ and $\eta_n$, which approach 0 as $n\to\infty$. In Appendix \ref{APPEN:PD_Markov_Proof} we show that the chain
\begin{equation}
W-X-Y_1-Y_2\label{EQ:PD_converse_Markov}
\end{equation}
is Markov, which establishes the converse.


\section{Summary and Concluding Remarks}\label{SEC:summary}

\par We considered cooperative BCs with one common and two private messages, where the private message to the cooperative user is confidential. An inner bound on the strong secrecy-capacity region was established by deriving a channel resolvability lemma and using it as a building block for the BC code. A resolvability-based Marton code for the BC with a double-binning of the confidential message codebook was constructed, and the resolvability lemma was invoked to achieve strong secrecy. The cooperation protocol used the link from Decoder 1 to Decoder 2 to share information on a portion of the non-confidential message and the common message only. Removing the secrecy constraint on $M_1$ allows a more flexible cooperation scheme that in general achieves strictly higher transmission rates \cite{Goldfeld_BC_Cooperation2014}. The inner bound was shown to be tight for the SD and PD cases. Two separate converse proofs were used because the structure of the joint PMFs describing the regions seem to require distinct choices of auxiliary random variable. 

\par The secrecy results were compared to those of the corresponding BCs without secrecy constraints, and the impact of secrecy on the capacity regions was highlighted. Cooperative Blackwell and Gaussian BCs illustrated the results. An explicit coding scheme that achieves strong secrecy while maximizing the transmission rate of the confidential message over the BW-BC was given. Further, it was shown that the strong secrecy-capacity region of the BW-BC remains unchanged even if the subchannel to the legitimate user is noiseless.


\appendices


\section{Proof of Proposition \ref{PROP:cooperation_schemes_not_same}}\label{APPEN:cooperation_schemes_not_same_proof}

Let $\mathcal{X}_1=\mathcal{X}_2=\mathcal{Y}_1=\mathcal{Y}_2=\{0,1\}$. Consider the BC $W_{Y_1|X_1}W_{Y_2|X_1,X_2}$ from Fig. \ref{FIG:counter_example}, where $W_{Y_1|X_1}$ is a BSC with transition probability $0.1$ and $W_{Y_2|X_1,X_2}$ is an arbitrary channel from $\{0,1\}^2$ to $\{0,1\}$ to be specified later.

For simplicity of notation we relabel $U_0=W$, $U_1=U$ and $U_2=V$ in $\mathcal{R}_{\mathsf{NS}}$, which becomes the union of rate triples $(R_{12},R_1,R_2)\in\mathbb{R}^3_+$ satisfying:
\begin{subequations}
\begin{align}
R_1 &\leq I(W,U;Y_1)\label{EQ:NSSD_region_BC1}\\
R_2 &\leq I(W,V;Y_2)+R_{12}\label{EQ:NSSD_region_BC2}\\
R_1\mspace{-3mu}+\mspace{-3mu}R_2 &\leq I(U;Y_1|W)+I(V;Y_2|W)-I(U;V|W)\nonumber\\&\mspace{57
mu}+\min\Big\{I(W;Y_1),I(W;Y_2)+R_{12}\Big\}\label{EQ:NSSD_region_BC1+2}
\end{align}\label{EQ:NSSD_region_BC}%
\end{subequations}
where the union is over all PMFs $Q_{W,U,V,X_1,X_2}\in\mathcal{P}(\mathcal{W}\times\mathcal{V}\times\mathcal{V}\times\mathcal{X}_1\times\mathcal{X}_2)$, each inducing a joint distribution $Q_{W,U,V,X_1,X_2,Y_1,Y_2}\triangleq Q_{W,U,V,X_1,X_2}W_{Y_1|X_1}W_{Y_2|X_1,X_2}$. Setting $U_0=W$, $U_1=U$ and $U_2=V$ into $\tilde{\mathcal{R}}_{\mathsf{NS}}$, gives a region described by the same rate bounds as \eqref{EQ:NSSD_region_BC}, up to replacing \eqref{EQ:NSSD_region_BC1} with
\begin{equation}
R_1 \leq I(U;Y_1|W)+\Big[I(V;Y_2|W)-I(U;V|W)\Big]^+.\label{EQ:NSSD_region_nosec_subopt_SDBC1}
\end{equation}
We outer bound $\tilde{\mathcal{R}}_{\mathsf{NS}}$ by loosening \eqref{EQ:NSSD_region_nosec_subopt_SDBC1} to
\begin{equation}
R_1 \leq I(U;Y_1|W).\label{EQ:NSSD_region_subopt_SDBC1_last}
\end{equation}
Let $\tilde{\mathcal{O}}_{\mathsf{NS}}$ denote the obtained outer bound on $\tilde{\mathcal{R}}_{\mathsf{NS}}$. We show that under the considered example $\tilde{\mathcal{O}}_{\mathsf{NS}}\subsetneq \mathcal{R}_{\mathsf{NS}}$. 

For any $r\in\mathbb{R}_+$, let
\begin{subequations}
\begin{align}
\mathcal{R}_{\mathsf{NS}}(r)&\triangleq\Big\{(R_1,R_2)\in\mathbb{R}_+^2\Big|(r,R_1,R_2)\in\mathcal{R}_{\mathsf{NS}}\Big\}\\
\tilde{\mathcal{O}}_{\mathsf{NS}}(r)&\triangleq\Big\{(R_1,R_2)\in\mathbb{R}_+^2\Big|(r,R_1,R_2)\in\tilde{\mathcal{O}}_{\mathsf{NS}}\Big\}
\end{align}
\end{subequations}
be the projections of $\mathcal{R}_{\mathsf{NS}}$ and $\tilde{\mathcal{O}}_{\mathsf{NS}}$ on the $(R_1,R_2)$ plane for $R_{12}=r$. Let $c=1-H_b(0.1)$, where $H_b:[0,1]\to [0,1]$ is the binary entropy function, and note that $R_1=c$ is the maximal achievable rate of $M_1$ in both $\mathcal{C}_{\mathsf{NS}}(c)$ and $\tilde{\mathcal{O}}_{\mathsf{NS}}(c)$. Define the supremum of all achievable $R_2$ that preserve $R_1=c$ in each region by
\begin{subequations}
\begin{align}
R_2^\star&\triangleq\sup\Big\{R_2\in\mathbb{R}_+\Big|(c,R_2)\in\mathcal{R}_{\mathsf{NS}}(c)\Big\}\\
\tilde{R}_2^\star&\triangleq\sup\Big\{R_2\in\mathbb{R}_+\Big|(c,R_2)\in\tilde{\mathcal{O}}_{\mathsf{NS}}(c)\Big\}.
\end{align}
\end{subequations}
We next evaluate $R_2^\star$ and $\tilde{R}_2^\star$, and then choose $W_{Y_2|X_1,X_2}$ for which $R_2^\star>\tilde{R}_2^\star$. 

For $\mathcal{R}_{\mathsf{NS}}(c)$, setting $W=X_1\sim\ber\left(\frac{1}{2}\right)$ achieves $R_1=c$:
\begin{equation}
R_1=I(W,U;Y_1)\stackrel{(a)}=I(X_1;Y_1)=c
\end{equation}
where (a) follows because $U-X_1-Y_1$ forms a Markov chain. Consequently, for $R_2^\star$ we have
\begin{align*}
R_2^\star&\stackrel{(a)}=\sup_{\substack{Q_{U,V,X_2|X_1}:\\(U,V)-(X_1,X_2)-Y_2}}\mspace{-10mu}\min\left\{\mspace{-5mu}\begin{array}{l}
     I(X_1,V;Y_2)+c,\\
     I(X_1,V;Y_2)-I(U;V|X_1)
\end{array}\mspace{-5mu}\right\}\\
         &\stackrel{(b)}\geq\sup_{\substack{Q_{V,X_2|X_1}:\\V-(X_1,X_2)-Y_2}} I(V;Y_2|X_1)\numberthis\label{EQ:maximal_R2_good_scheme_LB}
\end{align*}
where (a) uses the structure of $\mathcal{R}_{\mathsf{NS}}$ from \eqref{EQ:NSSD_region_BC} and the relations $R_{12}=I(X_1;Y_1)=c$ and $W=X_1$, while (b) is by setting $U=X_1$ and due to the non-negativity of mutual information.

For $\tilde{\mathcal{O}}_{\mathsf{NS}}(c)$, first note that $R_1$ is upper bounded by $c$ since
\begin{equation}
I(U;Y_1|W)\stackrel{(a)}\leq I(W,U;Y_1)\stackrel{(b)}\leq I(X_1;Y_1)\stackrel{(c)}\leq c.\label{EQ:maximal_R1_bad_scheme}
\end{equation}
However, $R_1=c$ is also achievable: (a) becomes an inequality if and only if $Y_1$ is independent of $W$; (b) is an equality if and only if $X_1-(W,U)-Y_1$ forms a Markov chain (this step also uses the Markov relation $(W,U)-X_1-Y_1$; (c) holds with equality if and only if $X_1\sim\ber\left(\frac{1}{2}\right)$. 

Now, since $Y_1$ and $X_1$ are connected by a BSC, the independence of $Y_1$ and $W$ implies that $X_1$ and $W$ are also independent. To see this observe that the independence of $Y_1$ and $W$ means that
\begin{equation}
Q_{Y_1|W}(0|w)=Q_{Y_1|W}(0|w'),\quad\forall(w,w')\in\mathcal{W}^2,\label{EQ:Y1W_independence_example}
\end{equation}
and assume by contradiction that a similar relation does not hold for $X_1$ and $W$. Namely, assume that there exists a pair $(w,w')\in\mathcal{W}^2$, such that
\begin{equation}
Q_{X_1|W}(0|w)\neq Q_{X_1|W}(0|w').
\end{equation}
Denote $Q_{X_1|W}(0|w)=\alpha$ and $Q_{X_1|W}(0|w')=\alpha'$, where $\alpha,\alpha'\in[0,1]$ and $\alpha\neq\alpha'$. Consider the following:
\begin{align*}
Q&_{Y_1|W}(0|w)\\
    &\stackrel{(a)}=Q_{X_1|W}(0|w)Q_{Y_1|X_1}(0|0)+Q_{X_1|W}(1|w)Q_{Y_1|X_1}(0|1)\\
    &=0.9\alpha+0.1(1-\alpha)\\
    &=0.1+0.8\alpha.\numberthis\label{EQ:Y1W_independence_case1}
\end{align*}
By repeating similar steps for $Q_{Y_1|W}(0|w')$, we get 
\begin{equation}
Q_{Y_1|W}(0|w')=0.1+0.8\alpha'.\label{EQ:Y1W_independence_case2}
\end{equation}
Combining \eqref{EQ:Y1W_independence_case1}-\eqref{EQ:Y1W_independence_case2} with \eqref{EQ:Y1W_independence_example} gives that $\alpha=\alpha'$, which is a contradiction. Therefore $X_1$ and $W$ must be independent. 

Furthermore, recall that from the equality in step (b) of \eqref{EQ:maximal_R1_bad_scheme}, the chain $X_1-(W,U)-Y_1$ is Markov, i.e., 
\begin{align*}
Q_{X_1,Y_1|W,U}(x_1&,y_1|w,u)\\
&=Q_{X_1|W,U}(x_1|w,u)Q_{Y_1|W,U}(y_1|w,u)\numberthis\label{EQ:compare_chen_factor1}
\end{align*}
for all $(w,u,x_1,y_1)\in\mathcal{W}\times\mathcal{U}\times\mathcal{X}_1\times\mathcal{Y}_1$. Since $(W,U)-X_1-Y_1$ is also a Markov chain, we have that $Q_{X_1,Y_1|W,U}$ also factors as
\begin{equation}
Q_{X_1,Y_1|W,U}(x_1,y_1|w,u)=Q_{X_1|W,U}(x_1|w,u)Q_{Y_1|X_1}(y_1|x_1)\label{EQ:compare_chen_factor2}
\end{equation}
for all $(w,u,x_1,y_1)\in\mathcal{W}\times\mathcal{U}\times\mathcal{X}_1\times\mathcal{Y}_1$. Therefore, for every $(w,u,x_1,y_1)\in\mathcal{W}\times\mathcal{U}\times\mathcal{X}_1\times\mathcal{Y}_1$, either $Q_{X_1|W,U}(x_1|w,u)=0$ or $Q_{Y_1|W,U}(y_1|w,u)=Q_{Y_1|X_1}(y_1|x_1)$. In particular, for $(x_1,y_1)=(1,1)$ and any $(w,u)\in\mathcal{W}\times\mathcal{U}$, either
\begin{subequations}
\begin{equation}
Q_{X_1|W,U}(1|w,u)=0\label{EQ:compare_chen_case1}
\end{equation}
or
\begin{equation}
Q_{Y_1|W,U}(1|w,u)=Q_{Y_1|X_1}(1|1)=0.9.\label{EQ:compare_chen_case2}
\end{equation}
\end{subequations}
If \eqref{EQ:compare_chen_case2} is true, then
\begin{align*}
&Q_{Y_1|W,U}(1|w,u)\\
    &\begin{multlined}[b][.47\textwidth]\stackrel{(a)}=Q_{X_1|W,U}(0|w,u)Q_{Y_1|X_1}(1|0)\\+Q_{X_1|W,U}(1|w,u)Q_{Y_1|X_1}(1|1)\end{multlined}\\
    &=0.1\cdot Q_{X_1|W,U}(0|w,u)+0.9\cdot Q_{X_1|W,U}(1|w,u)\\
    &=0.1+0.8\cdot Q_{X_1|W,U}(1|w,u)\numberthis\label{EQ:compare_chen_PX|U}
\end{align*}
where (a) uses the Markov chain $(W,U)-X_1-Y_1$. When combined with \eqref{EQ:compare_chen_case2}, this gives
\begin{equation}
Q_{X_1|W,U}(1|w,u)=1,\label{EQ:compare_chen_PX|U_conclusion}
\end{equation}
Thus, for any $(w,u)\in\mathcal{W}\times\mathcal{U}$ either \eqref{EQ:compare_chen_case1} or \eqref{EQ:compare_chen_PX|U_conclusion} is true, which implies that $X_1$ is a deterministic function of $(W,U)$. 

Having this, we upper bound $\tilde{R}_2^\star$ as follows.
\begin{align*}
\tilde{R}_2^\star&\stackrel{(a)}=\mspace{-15mu}\sup_{\substack{Q_{W}Q_{U,V,X_2|W,X_1}:\\(W,U,V)-(X_1,X_2)-Y_2}}\mspace{-15mu}\min\left\{\begin{array}{l}
     I(W,V;Y_2)+c,\\
     I(V;Y_2|W)-I(U;V|W),\\
     I(U;Y_1|W)+I(W,V;Y_2)\\\mspace{100mu}-I(U;V|W)
\end{array}\right\}\\
         &\stackrel{(b)}=\sup_{\substack{Q_{W}Q_{U,V,X_2|X_1,W}:\\(W,U,V)-(X_1,X_2)-Y_2}}I(V;Y_2|W)-I(U;V|W)\\
         &\stackrel{(c)}=\sup_{\substack{Q_{W}Q_{U,V,X_2|X_1,W}:\\(W,U,V)-(X_1,X_2)-Y_2}}I(V;Y_2|W)-I(U,X_1;V|W)\\
         &\leq\sup_{\substack{Q_{W}Q_{V,X_2|X_1,W}:\\(W,V)-(X_1,X_2)-Y_2}}I(V;Y_2|W)-I(V;X_1|W)\\
         &\stackrel{(d)}\leq\max_{w\in\mathcal{W}}\mspace{-5mu}\sup_{\substack{Q_{V,X_2|X_1,W=w}:\\V_w-(X_1,X_{2,w})-Y_2}}\mspace{-30mu}I(V;Y_2|W\mspace{-3mu}=\mspace{-3mu}w)-I(V;X_1|W\mspace{-3mu}=\mspace{-3mu}w)\\
         &\leq\sup_{\substack{Q_{V,X_2|X_1}:\\V-(X_1,X_2)-Y_2}}I(V;Y_2)-I(V;X_1)\numberthis\label{EQ:maximal_R2_bad_scheme_UB}
\end{align*}
where:\\
(a) uses the structure of $\tilde{\mathcal{O}}_{\mathsf{NS}}$, the independence of $W$ and $X_1$ and the relation $R_{12}=I(W,U;Y_1)=c$;\\
(b) follows by the non-negativity of mutual information;\\ 
(c) is because $X_1$ is determined by $(W,U)$;\\
(d) follows by defining $(V_w,X_{2,w})$ to be a pair of random variables jointly distributed with $X_1\sim\ber\left(\frac{1}{2}\right)$ according to $Q_{X_1}Q_{V,X_2|X_1,W=w}$, where $w\in\mathcal{W}$.

The lower bound on $R_2^\star$ from \eqref{EQ:maximal_R2_good_scheme_LB} is the capacity of the state-dependent channel $W_{Y_2|X_1,X_2}$ with non-causal CSI $X_1^n$ available at both the transmitting and receiving ends. The upper bound on $\tilde{R}_2^\star$ given in \eqref{EQ:maximal_R2_bad_scheme_UB} is the capacity of the corresponding GP channel, i.e., with non-causal transmitter CSI only. Thus, to show that $\tilde{R}_2^\star<R_2^\star$ it suffices to choose $W_{Y_2|X_1,X_2}$ for which the GP capacity is strictly less than the capacity with full CSI. A simple example for which these capacities are different is the binary dirty-paper (BDP) channel. Specifically, let $W_{Y_2|X_1,X_2}$ be defined by 
\begin{equation}
Y_2=X_2\oplus X_1 \oplus Z
\end{equation}
where $\oplus$ denotes modulo 2 addition, $X_1\sim\ber\left(\frac{1}{2}\right)$ plays the role of the channel's state, and the noise $Z\sim\ber(\epsilon)$, with $\epsilon\in\left[0,\frac{1}{2}\right]$ is independent of $(X_1,X_2)$. The input $X_2$ is subject to a constraint $\frac{1}{n}w_H(\mathbf{x}_2)\leq q$, for $q\in\left[0,\frac{1}{2}\right]$, where $w_H:\big\{0,1\big\}^n\to\mathbb{N}\cup\big\{0\big\}$ is the Hamming weight function. For the BDP channel, the GP capacity is \cite{Zamir_BDPC2002,Barron_BDPC2003,Toly_BDPMAC2010}
\begin{align*}
C_{\mathsf{GP}}^{(\mathsf{BDP})}&=\max_{\substack{Q_{V,X_2|X_1}:\\V-(X_1,X_2)-Y_2}}I(V;Y_2)-I(V;Y_1)\\
                            &=\mathsf{uce}\Big\{\big[H_b(q)-H_b(\epsilon)\big]^+\Big\}\numberthis\label{EQ:GP_CSI_BDPC}
\end{align*}
where `$\mathsf{uce}$' is the upper convex envelope operation with respect to $q$ ($\epsilon$ is constant). On the other hand, the capacity of the BDP channel with full CSI is \cite{Zamir_BDPC2002,Barron_BDPC2003,Toly_BDPMAC2010}
\begin{equation}
C_{\mathsf{F-CSI}}^{(\mathsf{BDP})}=\max_{\substack{Q_{V,X_2|X_1}:\\V-(X_1,X_2)-Y_2}} I(V;Y_2|X_1)= H_b(q*\epsilon)-H_b(\epsilon)\label{EQ:full_CSI_BDPC}
\end{equation}
where $q*\epsilon=q(1-\epsilon)+(1-q)\epsilon$. Clearly, $q$ and $\epsilon$ can be chosen such that $C_{\mathsf{GP}}^{(\mathsf{BDP})}<C_{\mathsf{F-CSI}}^{(\mathsf{BDP})}$, which shows that $\mathcal{R}_{\mathsf{NS}}$ and $\tilde{\mathcal{R}}_{\mathsf{NS}}$ are not equal in general.


\section{Converse Proof for \eqref{EQ:region_Gaussian_BC}}\label{APPEN:Gaussian_proof}

To prove the optimality of \eqref{EQ:region_Gaussian_BC}, we show that $\mathcal{C}_\mathsf{S}^{(\mathsf{PD})}\subseteq\mathcal{C}^{(\mathsf{G})}_\mathsf{S}$ ($\mathcal{C}_\mathsf{S}^{(\mathsf{PD})}$ and $\mathcal{C}^{(\mathsf{G})}_\mathsf{S}$ are given by \eqref{EQ:region_PDBC} and \eqref{EQ:region_Gaussian_BC}, respectively). First note that on one hand
\begin{subequations}
\begin{equation}
h(Y_1|W)\stackrel{(a)}\geq h(Y_1|X)=h(Z_1)= \frac{1}{2}\log(2\pi e\mathrm{N}_1)
\end{equation}
where (a) is because $W-X-Y_1$ forms a Markov chain, while on the other hand
\begin{equation}
h(Y_1|W)\leq h(Y_1)\leq \frac{1}{2}\log\big(2\pi e(\mathrm{P}+\mathrm{N}_1)\big).
\end{equation}\label{EQ:Gaussian_proof_Y1|W}%
\end{subequations}
The intermediate value theorem and \eqref{EQ:Gaussian_proof_Y1|W} imply that there is an $\alpha\in[0,1]$ such that
\begin{equation}
h(Y_1|W)=\frac{1}{2}\log\big(2\pi e(\alpha \mathrm{P}+\mathrm{N}_1)\big).\label{EQ:Gaussian_proof_Y1|W_equality}
\end{equation}
Further, for every $w\in\mathcal{W}$, we have
\begin{align*}
h(Y_2|W=w)&=h(Y_1+Z_2|W=w)\\
          &\stackrel{(a)}\geq\frac{1}{2}\log\left(2^{2h(Y_1|W=w)}+2^{2h(Z_2|W=w)}\right)\\
          &\stackrel{(b)}=\frac{1}{2}\log\left(2^{2h(Y_1|W=w)}+2\pi e(\mathrm{N}_2-\mathrm{N}_1)\right)\\
          &\triangleq \lambda(w)
          \numberthis\label{EQ:Gaussian_proof_Y2|W=w}
\end{align*}
where (a) uses the conditional entropy-power inequality (EPI), while (b) follows by the independence of $Z_2$ and $W$. Using \eqref{EQ:Gaussian_proof_Y2|W=w}, we lower bound $h(Y_2|W)$ in terms of $h(Y_1|W)$ as
\begin{align*}
h(Y_2|W)&\stackrel{(a)}\geq\mathbb{E}_W\lambda(W)\\
        &\stackrel{(b)}\geq\frac{1}{2}\log\left(2^{2h(Y_1|W)}+2\pi e(\mathrm{N}_2-\mathrm{N}_1)\right)\\
        &=\frac{1}{2}\log\big(2\pi e(\alpha \mathrm{P}+\mathrm{N}_2)\big)\numberthis\label{EQ:Gaussian_proof_Y2|W}
\end{align*}
where (a) follows from \eqref{EQ:Gaussian_proof_Y2|W=w}, while (b) uses the convexity of the function $x\mapsto\log(2^x+c)$ for $c\in\mathbb{R}_+$ and Jensen's inequality.

We next present upper bounds on the information terms on the RHS of \eqref{EQ:region_PDBC}. For \eqref{EQ:region_PDBC1}, we have
\begin{align*}
&I(X;Y_1|W)-I(X;Y_2|W)\\
        &\stackrel{(a)}=h(Y_1|W)-h(Y_1|X)-h(Y_2|W)+h(Y_2|X)\\
        &\stackrel{(b)}\leq \frac{1}{2}\log\left(1+\frac{\alpha \mathrm{P}}{\mathrm{N}_1}\right)-\frac{1}{2}\log\left(1+\frac{\alpha \mathrm{P}}{\mathrm{N}_2}\right)\numberthis\label{EQ:Gaussian_proof_R1}
\end{align*}
where (a) follows since the chain $W-X-(Y_1,Y_2)$ is Markov, while (b) relies on \eqref{EQ:Gaussian_proof_Y1|W_equality}, \eqref{EQ:Gaussian_proof_Y2|W} and on the Gaussian distribution maximizing the differential entropy under a variance constraint. Next, using \eqref{EQ:Gaussian_proof_Y2|W} we bound the RHS of \eqref{EQ:region_PDBC02} as
\begin{align*}
I(W;Y_2)+R_{12}&=h(Y_2)-h(Y_2|W)+R_{12}\\
               &\leq \frac{1}{2}\log\left(1+\frac{\bar{\alpha} \mathrm{P}}{\alpha \mathrm{P}+\mathrm{N}_2}\right)+R_{12}.\numberthis
\end{align*}
By repeating arguments similar to those in the derivation of \eqref{EQ:Gaussian_proof_R1}, we bound the sum of rates $R_1+R_2$ as
\begin{equation}
R_1+R_2\leq \frac{1}{2}\log\left(1+\frac{\mathrm{P}}{\mathrm{N}_1}\right)-\frac{1}{2}\log\left(1+\frac{\alpha \mathrm{P}}{\mathrm{N}_2}\right).
\end{equation}

\begin{figure*}[!t]
\setcounter{equation}{126}
\begin{align*}
    \mathbb{E}_{\mathsf{C}^{(n)}_1\big|\mathsf{C}^{(n)}_{0,2}=\mathcal{C}^{(n)}_{0,2}}\bigg[P^{(\mathsf{C}_n)}_{\mathsf{LE}}(i|1,\mathbf{u}_0,\mathbf{u}_2)\mathds{1}_{\big\{\mathbf{U}_1(1,1,1,i)=\mathbf{u}_1\big\}}\bigg]&=\mathbb{E}_{\mathsf{C}^{(n)}_1\big|\mathsf{C}^{(n)}_{0,2}=\mathcal{C}^{(n)}_{0,2}}\mathbb{P}_1\Big(I=i,\mathbf{U}_1(1,1,1,i)=\mathbf{u}_1\Big|\mathsf{C}^{(n)}_1,\mathsf{C}^{(n)}_{0,2}=\mathcal{C}^{(n)}_{0,2}\Big)\\
    &\leq\mathbb{E}_{\tilde{\mathsf{C}}_n}\mathbb{P}_{\tilde{P}}\Big(I=i,\tilde{\mathbf{U}}_1(\mathbf{u}_0,1,i)=\mathbf{u}_1\Big|W=1,\mathbf{U}_0=\mathbf{u}_0,\mathbf{U}_2=\mathbf{u}_2,\tilde{\mathsf{C}}_n\Big)\numberthis\label{EQ:encoding_error_expectation_UB_aux}
\end{align*}
\hrulefill
\end{figure*}
\section{Proof of Lemma \ref{LEMMA:absolute_continues}}\label{APPEN:lemma_absolute_proof}
\setcounter{equation}{121}

For a any $\mathcal{B}_n\in\mathfrak{B}_n$ and  $(\mathbf{s}_0,\mathbf{s},\mathbf{v})\in\mathcal{S}_0^n\times\mathcal{S}^n\times\mathcal{V}^n$, we have
\begin{align*}
&P^{(\mathcal{B}_n)}(\mathbf{s}_0,\mathbf{s},\mathbf{v})\\
&\begin{multlined}[b][.47\textwidth]=Q_{S_0,S}^n(\mathbf{s}_0,\mathbf{s})\mspace{3mu}2^{-n\tilde{R}}\mspace{-10mu}\sum_{(w,i)\in\mathcal{W}_n\times\mathcal{I}_n}\mspace{-10mu}\hat{P}^{(\mathcal{B}_n)}(i|w,\mathbf{s}_0,\mathbf{s})\\\times Q_{V|U,S_0,S}^n\big(\mathbf{v}\big|\mathbf{u}(\mathbf{s}_0,w,i),\mathbf{s}_0,\mathbf{s}\big)\end{multlined}.\numberthis\label{EQ:soft_proof_P_s0sv}
\end{align*}

Let $(\mathbf{s}_0,\mathbf{s},\mathbf{v})\in\mathcal{S}_0^n\times\mathcal{S}^n\times\mathcal{V}^n$ be a triple such that $Q_{S_0,S,V}^n(\mathbf{s}_0,\mathbf{s},\mathbf{v})=0$. Clearly, if $Q_{S_0,S}^n(\mathbf{s}_0,\mathbf{s})=0$ then \eqref{EQ:soft_proof_P_s0sv} implies that $P^{(\mathcal{B}_n)}(\mathbf{s}_0,\mathbf{s},\mathbf{v})=0$. Thus, we henceforth assume that $Q_{S_0,S}^n(\mathbf{s}_0,\mathbf{s})>0$ and $Q_{V|S_0,S}^n(\mathbf{v}|\mathbf{s}_0,\mathbf{s})=0$. By expanding
\begin{align*}
&Q_{V|S_0,S}^n(\mathbf{v}|\mathbf{s}_0,\mathbf{s})\\
    &=\sum_{\mathbf{u}\in\supp\left(Q^n_{U|S_0=\mathbf{s}_0,S=\mathbf{s}}\right)}\mspace{-20mu}Q_{U|S_0,S}^n(\mathbf{u}|\mathbf{s}_0,\mathbf{s})Q_{V|U,S_0,S}^n(\mathbf{v}|\mathbf{u},\mathbf{s}_0,\mathbf{s})\numberthis
\end{align*}
we have $Q_{V|U,S_0,S}^n(\mathbf{v}|\mathbf{u},\mathbf{s}_0,\mathbf{s})=0$ for every $\mathbf{u}\in\supp\left(Q^n_{U|S_0=\mathbf{s}_0,S=\mathbf{s}}\right)$. Thus, to complete the proof it suffices to show that every $u$-codeword that is transmitted with positive probability is in $\supp\left(Q^n_{U|S_0=\mathbf{s}_0,S=\mathbf{s}}\right)$.

By the construction of the codebook, every $\mathbf{u}\in\mathcal{B}_n$ also satisfies $\mathbf{u}\in\supp\left(Q^n_{U|S_0=\mathbf{s}_0}\right)$. Moreover, a necessary condition for a codeword $\mathbf{u}(\mathbf{s}_0,w,i)$ to be chosen by the encoder with positive probability is $\hat{P}^{(\mathcal{B}_n)}(i|w,\mathbf{s}_0,\mathbf{s})>0$, which by the definition of the likelihood encoder implies that $Q_{S|U,S_0}^n\big(\mathbf{s}\big|\mathbf{u}(\mathbf{s}_0,w,i),\mathbf{s}_0\big)>0$. Combining the above, we have that if a codeword $\mathbf{u}(\mathbf{s}_0,w,i)$ is transmitted with positive probability then
\begin{align*}
&Q_{U|S_0,S}^n\big(\mathbf{u}(\mathbf{s}_0,w,i)\big|\mathbf{s}_0,\mathbf{s}\big)\\
    &=\frac{Q_{S_0,S,U}^n\big(\mathbf{s}_0,\mathbf{s},\mathbf{u}(\mathbf{s}_0,w,i)\big)}{Q_{S_0,S}^n(\mathbf{s}_0,\mathbf{s})}\\
    &=\frac{Q_{S_0}^n(\mathbf{s}_0)Q_{U|S_0}^n\big(\mathbf{u}(\mathbf{s}_0,w,i)\big|\mathbf{s}_0\big)Q_{S|U,S_0}^n\big(\mathbf{s}\big|\mathbf{u}(\mathbf{s}_0,w,i),\mathbf{s}_0\big)}{Q_{S_0,S}^n(\mathbf{s}_0,\mathbf{s})}\\
    &>0.
\end{align*}


\begin{figure*}[!t]
\setcounter{equation}{127}
\begin{subequations}
\begin{equation}
\mathcal{D}_0=\Big\{\big(\mathbf{U}_0(1),\mathbf{U}_1(1,1,1,I),\mathbf{U}_2(1,1),\mathbf{Y}_1,\mathbf{Y}_2\big)\in\mathcal{T}_\delta^{n}(Q_{U_0,U_1,U_2,Y_1,Y_2})\Big\}\label{EQ:analysis_event_LLN}
\end{equation}
\hrulefill
\begin{equation}
\mathcal{D}_1(m_p,m_1,w)=\Big\{\big(\mathbf{U}_0(m_p),\mathbf{U}_1(m_p,m_1,w,I),\mathbf{Y}_1\big)\in\mathcal{T}_\delta^{n}(Q_{U_0,U_1,Y_1})\Big\}\label{EQ:analysis_event1}
\end{equation}
\hrulefill
\begin{equation}
\mathcal{D}_2(m_p,m_{22})=\Big\{\big(\mathbf{U}_0(m_p),\mathbf{U}_2(m_p,m_{22}),\mathbf{Y}_2\big)\in\mathcal{T}_\delta^{n}(Q_{U_0,U_2,Y_2})\Big\}\label{EQ:analysis_event2}
\end{equation}
\label{EQ:decoding_errors}
\end{subequations}
\hrulefill
\end{figure*}
\setcounter{equation}{123}

\section{Error Probability Analysis for Theorem \ref{TM:inner_bound}}\label{APPEN:error_analysis}
Since we evaluate the expected value (over the codebook ensemble) of the error probability and because the code is symmetric with respect to the uniformly distributed tuple $(M_p,M_1,M_{22},M)$, we may assume that $(M_p,M_1,M_{22},W)=(1,1,1,1)$. For any event $\mathcal{A}$ from the $\sigma$-algebra over which $\mathbb{P}$ is defined, denote
$$\mathbb{P}_1\triangleq\mathbb{P}\big(\mathcal{A}\big|M_p=1,M_{11}=1,W_1=1,M_{22}=1,W_2=1\big).$$


\textbf{Encoding Error:} An encoding error occurs if the $u_1$-codeword chosen by the likelihood encoder is not jointly typical with $\big(\mathbf{U}_0(M_p),\mathbf{U}_2(M_p,M_{22})\big)$. Based on the aforementioned symmetry, for any $\delta'\in(0,1)$, we set the event of an encoding error as
\begin{equation}
\mathcal{E}=\Big\{\big(\mathbf{U}_0(1),\mathbf{U}_1(1,1,1,I),\mathbf{U}_2(1,1)\big)\notin\mathcal{T}_{\delta'}^n(Q_{U_0,U_1,U_2})\Big\}\label{EQ:analysis_event0}.
\end{equation}
Abbreviating $\mathcal{T}\triangleq\mathcal{T}_{\delta'}^{n}(Q_{U_0,U_1,U_2})$ and recalling that $\mathsf{C}^{(n)}_{0,2}\triangleq\left\{\mathsf{C}^{(n)}_0,\mathsf{C}^{(n)}_2\right\}$, we have
\begin{align*}
&\mathbb{P}_1(\mathcal{E})\\
    &=\mathbb{E}_{\mathsf{C}_n}\mathbb{P}_1\Big(\big(\mathbf{U}_0(1),\mathbf{U}_1(1,1,1,I),\mathbf{U}_2(1,1)\big)\notin\mathcal{T}\Big|\mathsf{C}_n\Big)\\
    &\begin{multlined}[b][.45\textwidth]=\mathbb{E}_{\mathsf{C}_n}\Bigg[\sum_{i,\mathbf{u}_0,\mathbf{u}_1,\mathbf{u}_2}\mathds{1}_{\big\{\big(\mathbf{U}_0(1),\mathbf{U}_2(1,1)\big)=(\mathbf{u}_0,\mathbf{u}_2)\big\}}\\\times P^{(\mathsf{C}_n)}_{\mathsf{LE}}(i|1,\mathbf{u}_0,\mathbf{u}_2)\mathds{1}_{\big\{\mathbf{U}_1(1,1,1,i)=\mathbf{u}_1\big\}}\mathds{1}_{\big\{(\mathbf{u}_0,\mathbf{u}_1,\mathbf{u}_2)\notin\mathcal{T}\big\}}\Bigg]\end{multlined}\\
    &\stackrel{(a)}=\mathbb{E}_{\mathsf{C}^{(n)}_{0,2}}\vast[\sum_{\substack{i,\mathbf{u}_0,\mathbf{u}_1,\mathbf{u}_2:\\(\mathbf{u}_0,\mathbf{u}_1,\mathbf{u}_2)\notin\mathcal{T}}}\mathds{1}_{\big\{\big(\mathbf{U}_0(1),\mathbf{U}_2(1,1)\big)=(\mathbf{u}_0,\mathbf{u}_2)\big\}}\\&\mspace{50mu}\times\mathbb{E}_{\mathsf{C}^{(n)}_1\big|\mathsf{C}^{(n)}_{0,2}}\bigg[P^{(\mathsf{C}_n)}_{\mathsf{LE}}(i|1,\mathbf{u}_0,\mathbf{u}_2)\mathds{1}_{\big\{\mathbf{U}_1(1,1,1,i)=\mathbf{u}_1\big\}}\bigg]\vast]\\
    &\stackrel{(b)}=\mathbb{E}_{\tilde{\mathsf{C}}_n}\vast[\sum_{\substack{i,\mathbf{u}_0,\mathbf{u}_1,\mathbf{u}_2:\\(\mathbf{u}_0,\mathbf{u}_1,\mathbf{u}_2)\notin\mathcal{T}}}Q_{U_0,U_2}^n(\mathbf{u}_0,\mathbf{u}_2)\tilde{P}^{(\tilde{\mathsf{C}}_n)}(i|1,\mathbf{u}_0,\mathbf{u}_2)\\&\mspace{280mu}\times\mathds{1}_{\big\{\tilde{\mathbf{U}}_1(\mathbf{u}_0,1,i)=\mathbf{u}_1\big\}}\vast]\\
    &\stackrel{(c)}=\mathbb{E}_{\tilde{\mathsf{C}}_n}\mathbb{P}_{Q^n_{U_0,U_2}\times\tilde{P}}\Big(\big(\mathbf{U}_0,\tilde{\mathbf{U}}_1\big(\mathbf{U}_0,1,I\big),\mathbf{U}_2\big)\notin\mathcal{T}\Big|\tilde{\mathsf{C}}_n\Big).\numberthis\label{EQ:encoding_error_expectation_UB}
\end{align*}
In the above derivation (a) applies the law of total expectation in a similar fashion as in  \eqref{EQ:expected_divergence_UB1} (an inner expectation over $\mathsf{C}^{(n)}_1$ conditioned on $\mathsf{C}^{(n)}_{0,2}$, and an outer expectation over the possible values of $\mathsf{C}^{(n)}_{0,2}$), while (c) uses \eqref{EQ:achiev_proof_aux_PMF_code}. To justify step (b), for every $\mathcal{C}_n\in\mathfrak{C}_n$, we define (analogously to \eqref{EQ:cond_PMF_bad})
\begin{equation}
P^{(\mathcal{C}_n)}_{\mathsf{LE}}(i|1,\mathbf{u}_0,\mathbf{u}_2)=0\label{EQ:bad_prob_zero}
\end{equation}
whenever $\mathbf{u}_0\neq\mathbf{u}_0(1)$ or $\mathbf{u}_2\neq\mathbf{u}_2(1,1)$, and note that for every fixed $\mathcal{C}^{(n)}_{0,2}$, we have \eqref{EQ:encoding_error_expectation_UB_aux} on the top of this page, where the last step follows by intersecting the event of interest with $\Big\{\big(\mathbf{u}_0(1),\mathbf{u}_2(1,1)\big)=(\mathbf{u}_0,\mathbf{u}_2)\Big\}$ (otherwise the probability is zero due to \eqref{EQ:bad_prob_zero}) and, once again, using \eqref{EQ:achiev_proof_aux_PMF_code}. Inequality (b) then follows by removing the intersection with the aforementioned event and because $\tilde{\mathsf{C}}_n$ and $\mathsf{C}_n$ are independent.
Since the PMF $Q^n_{U_0,U_2}\tilde{P}_{\tilde{\mathsf{C}}_n,W,I,\tilde{\mathbf{U}}_1|\mathbf{U}_0,\mathbf{U}_2}$ is merely a relabeling of the induced distribution \eqref{EQ:soft_proof_P_PMF_code} in our resolvability setup, Lemma \ref{LEMMA:typicallity} implies that the RHS of \eqref{EQ:encoding_error_expectation_UB} approaches 0 as $n\to\infty$, as long as \eqref{EQ:achiev_rb1}-\eqref{EQ:achiev_rb6} are satisfied. 

\textbf{Decoding Errors:} To account for decoding errors, define the events in \eqref{EQ:decoding_errors} at the top of this page.
\setcounter{equation}{128}

\textbf{Expected Average Error Probability:} By the union bound, the expectation of the average error probability over the codebook ensemble\footnote{We slightly abuse notation in writing $\mathbb{E}P_\mathsf{e}(\mathsf{C}_n)$ because $P_e$ is actually a function of the code $c_n$ rather than the codebook $\mathcal{C}_n$. We favor this notation for its simplicity and remind the reader that $\mathcal{C}_n$ uniquely defines $c_n$.} is bounded as \eqref{EQ:analysis_error_prob_UB} at the top of the next page. Note that $P_0^{[1]}$ is the probability of an encoding error, while $P_0^{[2]}$ and $P_j^{[k]}$, for $k\in[1:4]$, correspond to decoding errors of Decoder $j=1,2$. We proceed with the following steps:

\begin{figure*}[!t]
\begin{align*}
\mathbb{E}P_e(\mathsf{C}_n)&\leq\mathbb{P}_1\left(\mathcal{E}\cup\mathcal{D}_0^c\cup\mathcal{D}_1(1,1,1,I)^c\cup\mathcal{D}_2(1,1)^c\cup\left\{\bigcup_{\substack{(\tilde{m}_p,\tilde{m}_1,\tilde{w})\\\neq(1,1,1)}}\mspace{-15mu}\mathcal{D}_1(\tilde{m}_p,\tilde{m}_1,\tilde{w},I)\right\}\cup\left\{\bigcup_{\substack{(\tilde{m}_p,\tilde{m}_{22})\neq(1,1):\\\tilde{m}_p\in\mathcal{B}_n\big(\hat{m}_{12}(1)\big)}}\mspace{-15mu}\mathcal{D}_2(\tilde{m}_p,\tilde{m}_{22})\right\}\right)\\
&\begin{multlined}[b][.89\textwidth]\leq\mathbb{P}_1\big(\mathcal{E}\big)+\mathbb{P}_1\big(\mathcal{D}_0^c\cap\mathcal{E}^c\big)+\mathbb{P}_1\Big(\mathcal{D}_1(1,1,1,I)^c\cap\mathcal{D}_0\Big)+\mathbb{P}_1\left(\bigcup_{(\tilde{m}_p,\tilde{m}_1,\tilde{w})\neq(1,1,1)}\mathcal{D}_1(\tilde{m}_p,\tilde{m}_1,\tilde{w},I)\right)\\+\mathbb{P}_1\Big(\mathcal{D}_2(1,1)^c\cap\mathcal{D}_0\Big)+\mathbb{P}_1\left(\bigcup_{\substack{(\tilde{m}_p,\tilde{m}_{22})\neq(1,1):\\\tilde{m}_p\in\mathcal{B}_n\big(\hat{m}_{12}(1)\big)}}\mathcal{D}_2(\tilde{m}_p,\tilde{m}_{22})\right)\end{multlined}\\
&\begin{multlined}[b][.89\textwidth]\leq\underbrace{\mathbb{P}_1\big(\mathcal{E}\big)}_{P_0^{[1]}}+\underbrace{\mathbb{P}_1\big(\mathcal{D}_0^c\cap\mathcal{E}^c\big)}_{P_0^{[2]}}+\underbrace{\mathbb{P}_1\Big(\mathcal{D}_1(1,1,1,I)^c\cap\mathcal{D}_0\Big)}_{P_1^{[1]}}+\underbrace{\sum_{\tilde{i}\in\mathcal{I}}P\left(\tilde{i}\right)\mathbb{P}_1\left(\bigcup_{\tilde{m}_p\neq 1}\mathcal{D}_1(\tilde{m}_p,1,1,\tilde{i})\right)}_{P_1^{[2]}}\\
+\underbrace{\mathbb{P}_1\left(\bigcup_{\substack{(\tilde{m}_1,\tilde{w})\neq(1,1),\\\tilde{i}\in\mathcal{I}}}\mathcal{D}_1(1,\tilde{m}_1,\tilde{w},\tilde{i})\right)}_{P_1^{[3]}}+\underbrace{\mathbb{P}_1\left(\bigcup_{\substack{(\tilde{m}_p,\tilde{m}_1,\tilde{w})\neq(1,1,1),\\\tilde{i}\in\mathcal{I}}}\mathcal{D}_1(\tilde{m}_p,\tilde{m}_1,\tilde{w},\tilde{i})\right)}_{P_1^{[4]}}+\underbrace{\mathbb{P}_1\Big(\mathcal{D}_2(1,1)^c\cap\mathcal{D}_0\Big)}_{P_2^{[1]}}\\+\underbrace{\mathbb{P}_1\left(\bigcup_{\substack{\tilde{m}_p\neq 1:\\\tilde{m}_p\in\mathcal{B}_n\big(\hat{m}_{12}(1)\big)}}\mathcal{D}_2(\tilde{m}_p,1)\right)}_{P_2^{[2]}}+\underbrace{\mathbb{P}_1\left(\bigcup_{\tilde{m}_{22}\neq 1}\mathcal{D}_2(1,\tilde{m}_{22})\right)}_{P_2^{[3]}}+\underbrace{\mathbb{P}_1\left(\bigcup_{\substack{(\tilde{m}_p,\tilde{m}_{22})\neq(1,1):\\\tilde{m}_p\in\mathcal{B}_n\big(\hat{m}_{12}(1)\big)}}\mathcal{D}_2(\tilde{m}_p,\tilde{m}_{22})\right)}_{P_2^{[4]}}.\end{multlined}\numberthis\label{EQ:analysis_error_prob_UB}
\end{align*}
\hrulefill
\end{figure*}

\begin{enumerate}

\item The encoding error analysis shows that $P_0^{[1]}\to 0$ as $n\to\infty$ if \eqref{EQ:achiev_rb1}-\eqref{EQ:achiev_rb6}.

\item The Conditional Typicality Lemma \cite[Section 2.5]{ElGamal2011} implies that $P_0^{[2]}\to 0$ as $n$ grows. More precisely, there exists a function $\beta(n,\delta,\delta')$ with $\lim_{n\to\infty}\beta(n,\delta,\delta')=0$ for any $0<\delta'<\delta$, such that $P_0^{[2]}\leq \beta(n,\delta,\delta')$. Although the exact exponent of decay is of no consequence for the asymptotic analysis in this work, the interested reader may refer to, e.g., \cite[Theorem 3.16]{Kramer_Lecture_Notes} for the precise expressions. \\

\item The definitions in \eqref{EQ:decoding_errors} clearly give $P_j^{[1]}=0$, for $j=1,2$ and every $n\in\mathbb{N}$.\\


\item
For $P_1^{[3]}$, we have
\begin{align*}
P_1^{[3]}&\stackrel{(a)}\leq\sum_{\substack{(\tilde{m}_1,\tilde{w})\neq (1,1),\\\tilde{i}\in\mathcal{I}}}2^{-n\big(I(U_1;Y_1|U_0)-\tau^{[3]}_1(\delta)\big)}\\
         &\leq2^{n(R_1+\tilde{R}+R')}2^{-n\big(I(U_1;Y_1|U_0)-\tau^{[3]}_1(\delta)\big)}\\
         &=2^{n\big(R_1+\tilde{R}+R'-I(U_1;Y_1|U_0)+\tau^{[3]}_1(\delta)\big)}
\end{align*}
where (a) follows since for any $(\tilde{m}_1,\tilde{w})\neq (1,1)$ and $\tilde{i}\in\mathcal{I}$, $\mathbf{U}_1(1,\tilde{m}_1,\tilde{w},\tilde{i})$ is independent of $\mathbf{Y}_1$ while both of them are drawn conditioned on $\mathbf{U}_0(1)$. Moreover, $\tau^{[3]}_1(\delta)\to 0$ as $\delta\to 0$. Hence, for the probability $P_1^{[3]}$ to vanish as $n\to\infty$, we take:
\begin{equation}
     R_1+\tilde{R}+R'<I(U_1;Y_1|U_0)-\tau^{[3]}_1(\delta).\label{EQ:analysis_RB1}
\end{equation}


\item
For $P_1^{[4]}$, consider
\begin{align*}
P_1^{[4]}&\stackrel{(a)}\leq\sum_{\substack{(\tilde{m}_p,\tilde{m}_1,\tilde{w})\neq (1,1,1),\\\tilde{i}\in\mathcal{I}}}2^{-n\big(I(U_0,U_1;Y_1)-\tau^{[4]}_1(\delta)\big)}\\
         &\leq2^{n(R_p+R_1+\tilde{R}+R')}2^{-n\big(I(U_0,U_1;Y_1)-\tau^{[4]}_1(\delta)\big)}\\
         &=2^{n\big(R_p+R_1+\tilde{R}+R'-I(U_0,U_1;Y_1)+\tau^{[4]}_1(\delta)\big)}
\end{align*}
where (a) follows since for any $(\tilde{m}_p,\tilde{m}_1,\tilde{w})\neq(1,1,1)$ and $\tilde{i}\in\mathcal{I}$, $\mathbf{U}_0(\tilde{m}_p)$ and $\mathbf{U}_1(\tilde{m}_p,\tilde{m}_1,\tilde{w},\tilde{i})$ are correlated with one another but independent of $\mathbf{Y}_1$. As before, $\tau^{[4]}_1(\delta)\to 0$ as $\delta\to 0$, and we have that $P_1^{[4]}\to 0$ as $n\to\infty$ if
\begin{equation}
     R_p+R_1+\tilde{R}+R'<I(U_0,U_1;Y_1)-\tau^{[4]}_1(\delta).\label{EQ:analysis_RB2}
\end{equation}

\item
Similar steps as in the upper bound of $P_1^{[3]}$ show that the rate bound that ensures that $P_1^{[2]}\to0$ as $n\to\infty$ is redundant. This is since for every $\tilde{m}_p\neq 1$ and $\tilde{i}\in\mathcal{I}$, the codewords $\mathbf{U}_0(\tilde{m}_p)$ and $\mathbf{U}_1(\tilde{m}_p,1,1,\tilde{i})$ are independent of $\mathbf{Y}_1$. Hence, the condition
\begin{equation}
R_p<I(U_0,U_1;Y_1)-\tau^{[2]}_1(\delta)\label{EQ:analysis_RB3}
\end{equation}
where $\lim_{\delta\to 0}\tau^{[2]}_1(\delta)=0$ suffices for $P_1^{[2]}$ to vanish. However, up to the vanishing terms, the RHS of \eqref{EQ:analysis_RB3} coincides with the RHS of \eqref{EQ:analysis_RB2}, while the left-hand side (LHS) of \eqref{EQ:analysis_RB3} is with respect to $R_p$ only. Clearly, \eqref{EQ:analysis_RB2} is the dominating constraint.\\

\item
By similar arguments, we find that $P_2^{[j]}$, for $j=2,3,4$, vanish with $n$ if
\begin{align}
R_{22}&<I(U_2;Y_2|U_0)-\tau^{[3]}_2(\delta)\label{EQ:analysis_RB4}\\
R_p+R_{22}-R_{12}&<I(U_0,U_2;Y_2)-\tau^{[4]}_2(\delta)\label{EQ:analysis_RB5}
\end{align}
where $\tau^{[3]}_2(\delta),\tau^{[4]}_2(\delta)\to0$ as $\delta\to 0$.

\end{enumerate}

\par Summarizing the above results, by setting
\begin{equation}
\tau_{\delta}\triangleq\max\left\{\tau^{[k]}_j(\delta)\right\}_{\substack{j=1,2,\\k=3,4}}
\end{equation}
we find that the RHS of \eqref{EQ:analysis_error_prob_UB} decays as
$n\to\infty$ for any $0<\delta'<\delta$ if the conditions in \eqref{EQ:achiev_rb} are met.



\section{Proof of the Markov Relation in \eqref{EQ:SD_converse_Markov} and \eqref{EQ:PD_converse_Markov}}\label{APPEN:Markov_Proof}

We prove that \eqref{EQ:SD_converse_Markov} and \eqref{EQ:PD_converse_Markov} form Markov chains by using the notions of d-separation and fd-separation in functional dependence graphs (FDGs), for which we use the formulation from \cite{Kramer_FDG2003}. Throughout this appendix all probabilities are taken with respect to the PMF $P^{(c_n)}$ that is induced by $c_n$ and given in \eqref{EQ:induced_PMF}. For brevity, we omit the superscript and write $P$ instead of $P^{(c_n)}$.

\subsection{Proof of \eqref{EQ:SD_converse_Markov}}\label{APPEN:SD_Markov_Proof}


\begin{figure}[!t]
\begin{center}
\begin{psfrags}
    \psfragscanon
    \psfrag{A}[][][0.8]{\ \ \ $M_0$}
    \psfrag{Y}[][][0.8]{\ \ \ \ $M_2$}
    \psfrag{B}[][][0.8]{\ \ \ \ \ $M_1$}
    \psfrag{C}[][][0.8]{\ $X^{t-1}$}
    \psfrag{D}[][][0.8]{$X_t$}
    \psfrag{E}[][][0.8]{\ \ \ \ \ \ \ \ $X_{t+1}^n$}
    \psfrag{F}[][][0.8]{\ \ \ $Y_1^{t-1}$}
    \psfrag{G}[][][0.8]{\ $Y_2^{t-1}$}
    \psfrag{H}[][][0.8]{\ \ \ $Y_{1,t}$}
    \psfrag{I}[][][0.8]{\ $Y_{2,t}$}
    \psfrag{J}[][][0.8]{\ \ \ \ $Y_{1,t+1}^n$}
    \psfrag{K}[][][0.8]{\ $Y_{2,t+1}^n$}
    \psfrag{X}[][][0.8]{\ }
\subfloat[]{\includegraphics[scale=0.58]{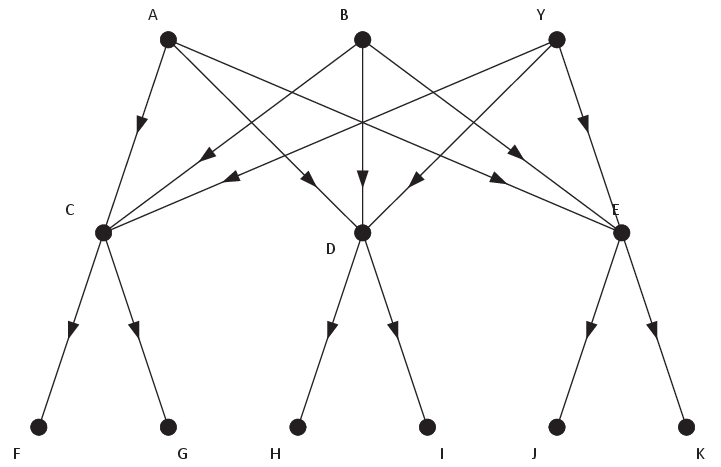}}\\
\subfloat[]{\includegraphics[scale=0.58]{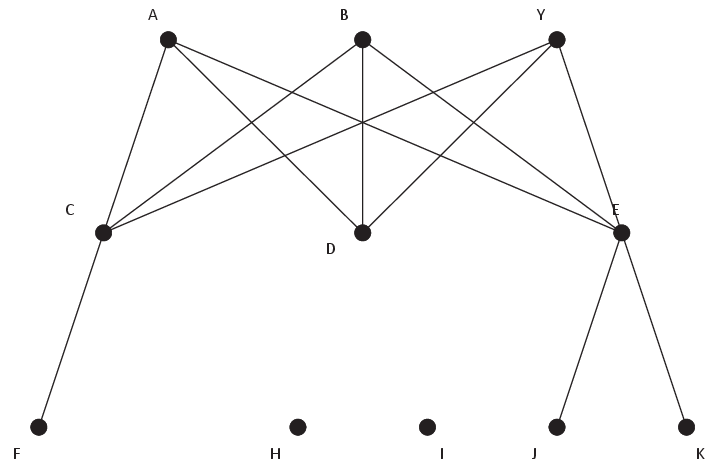}}
\caption{(a) The FDG that stems from \eqref{EQ:SD_Markov_sufficient_temp}: \eqref{EQ:SD_Markov_sufficient2} follows since $\mathcal{C}=\big\{X_t\big\}$ d-separates $\mathcal{A}=\big\{Y_{2,t}\big\}$ from $\mathcal{B}=\big\{M_0,M_2,Y_1^n,Y_{2,t+1}^n\big\}$. (b) The undirected graph obtained from the FDG after the manipulations described in Definition \cite[Definition 1]{Kramer_FDG2003}. Both FDGs omit the dependence of the channel outputs on the noise.} \label{FIG:FDG}
\psfragscanoff
\end{psfrags}
\end{center}
\end{figure}

By the definitions of the auxiliaries $W$ and $V$, it suffices to show that
\begin{equation}
(M_0,M_2,M_{12},Y_1^{t-1},Y_{2,t+1}^n,Y_{1,t})-X_t-Y_{2,t}\label{EQ:SD_Markov_sufficient}
\end{equation}
forms a Markov chain for every $t\in[1:n]$. In fact, we prove the stronger relation
\begin{equation}
(M_0,M_2,Y_1^n,Y_{2,t+1}^n)-X_t-Y_{2,t}\label{EQ:SD_Markov_sufficient2}
\end{equation}
from which \eqref{EQ:SD_Markov_sufficient} follows because $M_{12}$ is a function of $Y_1^n$. Since the channel is SD, memoryless and without feedback, for every $(m_0,m_1,m_2)\in\mathcal{M}^{(n)}_0\times\mathcal{M}^{(n)}_1\times\mathcal{M}^{(n)}_2$, $(x^n,y_1^n,y_2^n)\in\mathcal{X}^n\times\mathcal{Y}_1^n\times\mathcal{Y}_2^n$ and $t\in[1:n]$, we have
\begin{align*}
&P(m_0,m_1,m_2,x^n,y_1^n,y_2^n)\\
    &\begin{multlined}[b][.41\textwidth]=P(m_0)P(m_1)P(m_2)P(x^n|m_0,m_1,m_2)\\\times P\big(y_1^{t-1}\big|x^{t-1}\big)P\big(y_2^{t-1}\big|x^{t-1}\big) P(y_{1,t}|x_t)\\\times P(y_{2,t}|x_t) P\big(y_{1,t+1}^n\big|x_{t+1}^n\big)P\big(y_{2,t+1}^n\big|x_{t+1}^n\big).\end{multlined}\numberthis\label{EQ:SD_Markov_sufficient_temp}
\end{align*}

Fig. \ref{FIG:FDG}(a) shows the FDG induced by \eqref{EQ:SD_Markov_sufficient_temp}. The structure of FDGs allows one to establish the conditional statistical independence of sets of random variables by using d-separation. The Markov relation in \eqref{EQ:SD_Markov_sufficient2} follows by setting $\mathcal{A}=\big\{Y_{2,t}\big\}$, $\mathcal{B}=\big\{M_0,M_2,Y_1^n,Y_{2,t+1}^n\big\}$ and $\mathcal{C}=\big\{X_t\big\}$, and noting that $\mathcal{C}$ d-separates $\mathcal{A}$ from $\mathcal{B}$ by applying the manipulations described in \cite[Definition 1]{Kramer_FDG2003}.

\subsection{Proof of \eqref{EQ:PD_converse_Markov}}\label{APPEN:PD_Markov_Proof}


\begin{figure}[!t]
\begin{center}
\begin{psfrags}
    \psfragscanon
    \psfrag{A}[][][0.8]{\ \ \ $M_0$}
    \psfrag{Y}[][][0.8]{\ \ \ \ $M_2$}
    \psfrag{B}[][][0.8]{\ \ \ \ \ $M_1$}
    \psfrag{C}[][][0.8]{$\mspace{-3mu}X^{t-1}$}
    \psfrag{D}[][][0.8]{\ \ \ \ \ \ \ \ \ \ \ \ $X_t$}
    \psfrag{E}[][][0.8]{\ \ \ \ \ \ \ \ \ $X_{t+1}^n$}
    \psfrag{F}[][][0.8]{\ $Y_1^{t-1}$}
    \psfrag{G}[][][0.8]{\ $Y_2^{t-1}$}
    \psfrag{H}[][][0.8]{\ \ \ \ $Y_{1,t}$}
    \psfrag{I}[][][0.8]{\ \ \ \ $Y_{2,t}$}
    \psfrag{J}[][][0.8]{\ \ \ \ \ \ \ $Y_{1,t+1}^n$}
    \psfrag{K}[][][0.8]{\ \ \ \ \ \ \ $Y_{2,t+1}^n$}
    \psfrag{X}[][][0.8]{\ }
\subfloat[]{\includegraphics[scale=0.58]{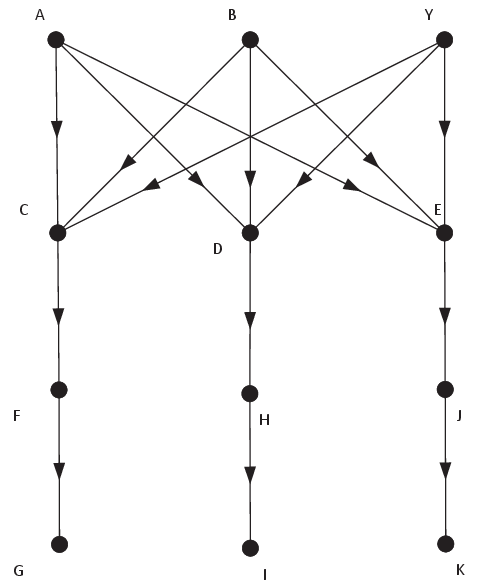}}\\
\subfloat[]{\includegraphics[scale=0.58]{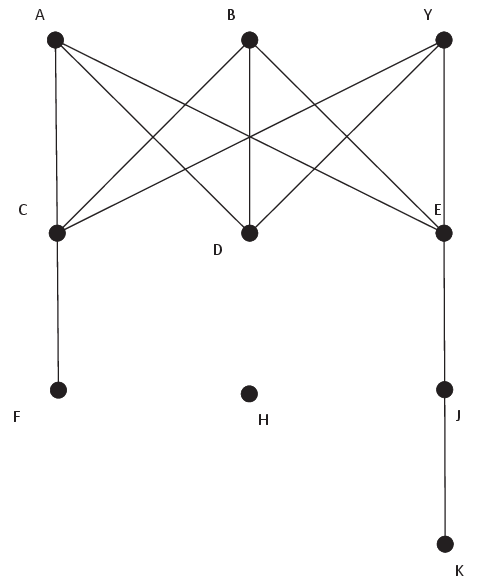}}\\
\subfloat[]{\includegraphics[scale=0.58]{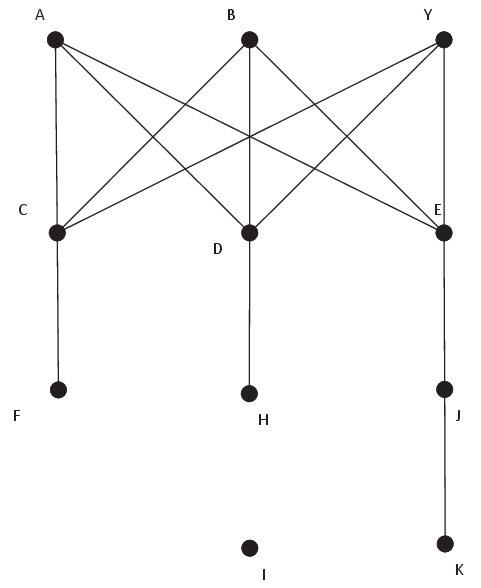}}
\caption{(a) The FDG that stems from \eqref{EQ:PD_Markov_sufficient_temp}: \eqref{EQ:PD_Markov_sufficient} follows since $\mathcal{C}_j$ d-separates $\mathcal{A}_j$ from $\mathcal{B}_j$, for $j=1,2$. (b) The undirected graph that corresponds to $\mathcal{A}_1$, $\mathcal{B}_1$ and $\mathcal{C}_1$. (c) The undirected graph that corresponds to $\mathcal{A}_2$, $\mathcal{B}_2$ and $\mathcal{C}_2$. The FDGs omit the dependence of the channel outputs on the noise.}\label{FIG:PD_FDG}
\psfragscanoff
\end{psfrags}
\end{center}
\end{figure}

To prove \eqref{EQ:PD_converse_Markov}, is suffices to show that Markov relations
\begin{subequations}
\begin{gather}
(M_0,M_2,Y_1^{t-1},Y_{2,t+1}^n)-X_t-Y_{1,t}\label{EQ:PD_Markov_sufficient1}\\
(M_0,M_2,Y_1^{t-1},Y_{2,t+1}^n,X_t)-Y_{1,t}-Y_{2,t}\label{EQ:PD_Markov_sufficient2}
\end{gather}\label{EQ:PD_Markov_sufficient}%
\end{subequations}
hold for every $t\in[1:n]$. By the PD property of the channel, and because it is memoryless and without feedback, for every $(m_0,m_1,m_2)\in\mathcal{M}^{(n)}_0\times\mathcal{M}^{(n)}_1\times\mathcal{M}^{(n)}_2$, $(x^n,y_1^n,y_2^n)\in\mathcal{X}^n\times\mathcal{Y}_1^n\times\mathcal{Y}_2^n$ and $t\in[1:n]$, we have
\begin{align*}
&P(m_0,m_1,m_2,x^n,y_1^n,y_2^n)\\
&\begin{multlined}[b][.41\textwidth]=P(m_0)P(m_1)P(m_2)P(x^n|m_0,m_1,m_2)\\\times P\big(y_1^{t-1}\big|x^{t-1}\big)P\big(y_2^{t-1}\big|y_1^{t-1}\big)
P(y_{1,t}|x_t)\\\times P(y_{2,t}|y_{1,t})P\big(y_{1,t+1}^n\big|x_{t+1}^n\big)P\big(y_{2,t+1}^n\big|y_{1,t+1}^n\big)\end{multlined}.\numberthis\label{EQ:PD_Markov_sufficient_temp}
\end{align*}

The FDG induced by \eqref{EQ:PD_Markov_sufficient_temp} is shown in Fig. \ref{FIG:PD_FDG}(a). Set $\mathcal{A}_1=\big\{Y_{1,t}\big\}$, $\mathcal{B}_1=\big\{M_0,M_2,Y_1^{i-1},Y_{2,t+1}^n\big\}$ and $\mathcal{C}_1=\big\{X_t\big\}$, and $\mathcal{A}_2=\big\{Y_{2,t}\big\}$, $\mathcal{B}_2=\big\{M_0,M_2,Y_1^{i-1},Y_{2,t+1}^n,X_t\big\}$ and $\mathcal{C}_2=\big\{Y_{1,t}\big\}$. The relations in \eqref{EQ:PD_Markov_sufficient} follow by noting that $\mathcal{C}_j$ d-separates $\mathcal{A}_j$ from $\mathcal{B}_j$, for $j=1,2$ by applying the manipulations described in \cite[Definition 1]{Kramer_FDG2003}.

\bibliographystyle{unsrt}
\bibliographystyle{IEEEtran}
\bibliography{ref}

\begin{thebibliography}{10}

\bibitem{Wyner_Wiretap1975}
A.~D. Wyner.
\newblock The wire-tap channel.
\newblock {\em Bell Sys. Techn.}, 54(8):1355--1387, Oct. 1975.

\bibitem{Csiszar_Korner_BCconfidential1978}
I.~Csisz{\'a}r and J.~K{\"o}rner.
\newblock Broadcast channels with confidential messages.
\newblock {\em IEEE Trans. Inf. Theory}, 24(3):339--348, May 1978.

\bibitem{BC_Confidential_Yates2008}
R.~Liu, I.~Maric, P.~Spasojevi{\'c}, and R.~D. Yates.
\newblock {Discrete memoryless interference and broadcast channels with
  confidential messages: Secrecy rate regions}.
\newblock {\em IEEE Trans. Inf. Theory}, 54(6):2493--2507, Jun. 2008.

\bibitem{Semi-det_BC_secrect_two2009}
Y.~Zhao, P.~Xu, Y.~Zhao, W.~Wei, and Y.~Tang.
\newblock Secret communications over semi-deterministic broadcast channels.
\newblock In {\em Fourth Int. Conf. Commun. and Netw. in China (CHINACOM)},
  Xian, China, Aug. 2009.

\bibitem{Semi-det_BC_secrect_one2009}
W.~Kang and N.~Liu.
\newblock The secrecy capacity of the semi-deterministic broadcast channel.
\newblock In {\em Proc. Int. Symp. Inf. Theory}, Seoul, Korea, Jun.-Jul. 2009.

\bibitem{goldfeld_leakage2015}
Z.~Goldfeld, G.~Kramer, and H.~H. Permuter.
\newblock Broadcast channels with privacy leakage constraints.
\newblock {\em Submitted for publication to IEEE Trans. Inf. Theory}, 2015.
\newblock Available on ArXiv at http://arxiv.org/abs/1504.06136.

\bibitem{Ulukus_Cooperative_RBC2011}
E.~Ekrem and S.~Ulukus.
\newblock Secrecy in cooperative relay broadcast channels.
\newblock {\em IEEE Trans. Inf. Theory}, 57(1):137--155, Jan. 2011.

\bibitem{Poor_Gaussian_MIMO_BC_Secrecy2009}
R.~Liu and H.~Poor.
\newblock Secrecy capacity region of a multiple-antenna {Gaussian} broadcast
  channel with confidential messages.
\newblock {\em IEEE Trans. Inf. Theory}, 55(3):1235--1249, Mar. 2009.

\bibitem{Liu_Shamai_MIMOWTC2009}
T.~Liu and S.~Shamai.
\newblock A note on the secrecy capacity of the multiple-antenna wiretap
  channel.
\newblock {\em IEEE Trans. Inf. Theory}, 6(6):2547--2553, Jun. 2009.

\bibitem{Poor_Shamai_Gaussian_MIMO_BC_Secrecy2010}
R.~Liu, T.~Liu, H.~V. Poor, and S.~Shamai.
\newblock Multiple-input multiple-output {Gaussian} broadcast channels with
  confidential messages.
\newblock {\em IEEE Trans. Inf. Theory}, 56(9):4215--4227, Sep. 2010.

\bibitem{Khitsi_MIMOWTC2010}
A.~Khisti and G.~W. Wornell.
\newblock Secure transmission with multiple antennas - part {II}: The {MIMOME}
  channel.
\newblock {\em IEEE Trans. Inf. Theory}, 56(11):5515--5532, Nov. 2010.

\bibitem{Ulukus_Gaussian_Wiretap2011}
E.~Ekrem and S.~Ulukus.
\newblock The secrecy capacity region of the {Gaussian MIMO} multi-receiver
  wiretap channel.
\newblock {\em IEEE Trans. Inf. Theory}, 57(4):2083--2114, Apr. 2011.

\bibitem{Hassibi_MINOWTC2011}
F.~Oggier and B.~Hassibi.
\newblock The secrecy capacity of the {MIMO} wiretap channel.
\newblock {\em IEEE Trans. Inf. Theory}, 57(8):4961--4972, Aug. 2011.

\bibitem{Ulukus_External_Eve2009}
E.~Ekrem and S.~Ulukus.
\newblock Secrecy capacity of a class of broadcast channels with an
  eavesdropper.
\newblock {\em EURASIP Journal on Wireless Commun. and Netw.}, 2009(1):1--29,
  Mar. 2009.

\bibitem{Bagherikaram_Gaussin_External_Eve2009}
G.~Bagherikaram, A.~Motahari, and A.~Khandani.
\newblock Secrecy capacity region of {Gaussian} broadcast channel.
\newblock In {\em 43rd Annual Conf. on Inf. Sci. and Sys. (CISS) 2009}, pages
  152--157, Baltimore, MD, US, Mar. 2009.

\bibitem{Piantanida_External_Eve2014}
M.~Benammar and P.~Piantanida.
\newblock Secrecy capacity region of some classes of wiretap broadcast
  channels.
\newblock {\em IEEE Trans. Inf. Theory}, 61(10):5564--5582, Oct. 2015.

\bibitem{Maurer_Strong_Secrecy_Chapter1994}
U.~Maurer.
\newblock {\em Communications and Cryptography: Two Sides of One Tapestry},
  chapter The Strong Secret Key Rate of Discrete Random Triples, pages
  271--285.
\newblock Springer US, Norwell, MA, USA, 1994.

\bibitem{Maurer_Wolf_Strong_Secrecy2000}
U.~Maurer and S.~Wolf.
\newblock Information-theoretic key agreement: From weak to strong secrecy for
  free.
\newblock In {\em Lecture Notes in Computer Science}, pages 351--368, 2000.

\bibitem{Bloch_Barros_Secrecy_Book2011}
M.~Bloch and J.~Barros.
\newblock {\em Physical-Layer Security: From Information Theory to Security
  Engineering}.
\newblock Cambridge Univ. Press, Cambridge, UK, Oct. 2011.

\bibitem{Csiszar_Strong_Secrecy1996}
I.~Csisz{\'a}r.
\newblock Almost independence and secrecy capacity.
\newblock {\em Prob. Inf. Trans.}, 32(1):40--47, Jan.-Mar. 1996.

\bibitem{Hayashi_Secrecy_Resolvability2006}
M.~Hayashi.
\newblock General nonasymptotic and asymptotic formulas in channel
  resolvability and identification capacity and their application to the
  wiretap channels.
\newblock {\em IEEE Trans. Inf. Theory}, 52(4):1562--1575, Apr. 2006.

\bibitem{Wyner_Common_Information1975}
A.~D. Wyner.
\newblock The common information of two dependent random variables.
\newblock {\em IEEE Trans. Inf. Theory}, 21(2):163--179, Mar. 1975.

\bibitem{Han_Verdu_Resolvability1993}
T.~Han and S.~Verd{\'u}.
\newblock Approximation theory of output statistics.
\newblock {\em IEEE Trans. Inf. Theory}, 39(3):752--772, May 1993.

\bibitem{Kramer_resolvability2013}
J.~Hou and G.~Kramer.
\newblock Informational divergence approximations to product distributions.
\newblock In {\em 13th Canadian Workshop Inf. Theory}, Toronto, Ontario,
  Canada, Jun. 2013.

\bibitem{Cuff_Synthesis2013}
P.~W. Cuff.
\newblock Distributed channel synthesis.
\newblock {\em IEEE. Trans. Inf. Theory}, 59(11):7071--7096, Nov. 2013.

\bibitem{Cuff_Distortion_Secrecy2014}
C.~Schieler and P.~Cuff.
\newblock Rate-distortion theory for secrecy systems.
\newblock {\em IEEE Trans. on Inf. Theory}, 66(12):7584--7605, Dec. 2014.

\bibitem{Cuff_Henchman_Secrecy2014}
C.~Schieler and P.~Cuff.
\newblock The henchman problem: Measuring secrecy by the minimum distortion in
  a list.
\newblock {\em Submitted to IEEE Trans. on Inf. Theory}, 2014.
\newblock Available on ArXiv at http://arxiv.org/abs/1410.2881.

\bibitem{Song_Cuff_Secrecy2014}
E.~Song, P.~Cuff, and V.~Poor.
\newblock A rate-distortion based secrecy system with side information at the
  decoders.
\newblock In {\em Proc. 52nd Annu. Allerton Conf. Commun., Control and
  Comput.}, Monticell, Illinois, United States, Sep. 2014.

\bibitem{Cuff_Secrecy_Coordination2014}
S.~Satpathy and P.~Cuff.
\newblock Secure coordination with a two-sided helper.
\newblock In {\em Proc. Int. Symp. Inf. Theory (ISIT-2014)}, Honolulu, Hawaii,
  US, Jun.-Jul. 2014.

\bibitem{Bloch_Resolvability_Secrecy2013}
M.~Bloch and N.~Laneman.
\newblock Strong secrecy from channel resolvability.
\newblock {\em IEEE Trans. Inf. Theory}, 59(12):8077--8098, Dec. 2013.

\bibitem{Kramer_EffectiveSecrecy2014}
J.~Hou and G.~Kramer.
\newblock Effective secrecy: {Reliability}, confusion and stelth.
\newblock In {\em Proc. Int. Symp. Inf. Theory}, Honolulu, HI, USA, Jun.-Jul.
  2014.

\bibitem{Han_WTC_Cost2014}
T.~S. Han, H.~Endo, and M.~Sasaki.
\newblock Reliability and secrecy functions of the wiretap channel under cost
  constraint.
\newblock {\em IEEE Trans. Inf. Theory}, 60(11):6819--6843, Nov. 2014.

\bibitem{Cuff_Song_Likelihood2014}
E.~Song, P.~Cuff, and V.~Poor.
\newblock The likelihood encoder for lossy compression.
\newblock {\em IEEE Trans. Inf. Theory}, 62(4):1836--1849, Apr. 2016.

\bibitem{CovThom06}
T.~M. Cover and J.~A. Thomas.
\newblock {\em Elements of Information Theory}.
\newblock Wiley, New-York, 2nd edition, 2006.

\bibitem{Goldfeld_BC_Cooperation2014}
Z.~Goldfeld, H.~H. Permuter, and G.~Kramer.
\newblock Duality of a source coding problem and the semi-deterministic
  broadcast channel with rate-limited cooperation.
\newblock {\em IEEE Trans. Inf. Theory}, 65(5):2285--2307, May 2016.

\bibitem{vanderMeulen_Blackwell1975}
E.~C. van~der Meulen.
\newblock Random coding theorems for the general discrete memoryless broadcast
  channel.
\newblock {\em IEEE Trans. Inf. Theory}, IT-21(2):180--190, May 1975.

\bibitem{Gelfand_Blackwell1977}
S.~I. Gelfand.
\newblock Capacity of one broadcast channel.
\newblock {\em Probl. Pered. Inf. (Problems of Inf. Transm.)}, 13(3):106108,
  Jul./Sep. 1977.

\bibitem{Massey_Applied}
J.~L. Massey.
\newblock {\em Applied Digital Information Theory}.
\newblock ETH Zurich, Zurich, Switzerland, 1980-1998.

\bibitem{Perturbation2012}
A.~Gohari and V.~Anantharam.
\newblock Evaluation of {Marton's} inner bound for the general broadcast
  channel.
\newblock {\em IEEE Trans. Inf. Theory}, 58(2):608--619, Feb. 2012.

\bibitem{Eggleston_Convexity1958}
H.~G. Eggleston.
\newblock {\em Convexity}.
\newblock Cambridge University Press, Cambridge, England York, 6th edition
  edition, 1958.

\bibitem{Liang_Veeravalli_RBC2007}
Y.~Liang and V.~V. Veeravalli.
\newblock Cooperative relay broadcast channels.
\newblock {\em IEEE Trans. Inf. Theory}, 53(3):900--928, Mar. 2007.

\bibitem{Liang_Kramer_RBC2007}
Y.~Liang and G.~Kramer.
\newblock Rate regions for relay broadcast channels.
\newblock {\em IEEE Trans. Inf. Theory}, 53(10):3517--3535, Oct. 2007.

\bibitem{Dikstein_PDBC_Cooperation2014}
L.~Dikstein, H.~H. Permuter, and Y.~Steinberg.
\newblock On state dependent broadcast channels with cooperation.
\newblock {\em IEEE Trans. Inf. Theory}, 62(5):2308--2323, May 2016.

\bibitem{Zamir_BDPC2002}
R.~Zamir, S.~Shamai, and U.~Erez.
\newblock Nested linear/lattice codes for structured multiterminal binning.
\newblock {\em IEEE Trans. Inf. Theory}, 48(6):1205--1276, Jun. 2002.

\bibitem{Barron_BDPC2003}
R.~J. Barron, B.~Chen, and G.~W. Wornell.
\newblock The duality between information embedding and source coding with side
  information and some applications.
\newblock {\em IEEE Trans. Inf. Theory}, 49(5):1159--1180, May 2003.

\bibitem{Toly_BDPMAC2010}
A.~Khina, T.~Philosof, U.~Erez, and R.~Zamir.
\newblock Binary dirty {MAC} with common state information.
\newblock In {\em Proc. 26-th Convention of Electrical and Electronics
  Engineers (IEEEI-2010)}, Eilat, Israel, Nov. 2010.

\bibitem{GP_SemideterministicBC1980}
S.~I. Gelfand and M.~S. Pinsker.
\newblock Capacity of a broadcast channel with one deterministic component.
\newblock {\em Prob. Pered. Inf. (Problems of Inf. Transm.)}, 16(1):17--25,
  Jan.-Mar. 1980.

\bibitem{DaboraServetto06BC}
R.~Dabora and S.~D. Servetto.
\newblock Broadcast channels with cooperating decoders.
\newblock {\em IEEE Trans. Inf. Theory}, 52:5438--5454, 2006.

\bibitem{ElGamal2011}
A.~{El Gamal} and Y.-H. Kim.
\newblock {\em Network Information Theory}.
\newblock Cambridge University Press, 2011.

\bibitem{Goldfeld_WTCII_semantic2015}
Z.~Goldfeld, P.~Cuff, and H.~H. Permuter.
\newblock Semantic-security capacity for wiretap channels of type {II}.
\newblock {\em IEEE Trans. Inf. Theory}, 62(7):1--17, Jul. 2016.

\bibitem{FMEIT_newsletter2015}
I.~B. Gattegno, Z.~Goldfeld, and H.~H. Permuter.
\newblock {Fourier-Motzkin} elimination software for information theoretic
  inequalities.
\newblock {\em IEEE Inf. Theory Society Newsletter}, 65(3):25--28, Sep. 2015.

\bibitem{Kramer_telescopic2011}
G.~Kramer.
\newblock Teaching {IT}: {An} identity for the {Gelfand-Pinsker} converse.
\newblock {\em IEEE Inf. Theory Society Newsletter}, 61(4):4--6, Dec. 2011.

\bibitem{Kramer_Lecture_Notes}
G.~Kramer.
\newblock {\em Lecture Notes for Multi-User Information Theory}.
\newblock Ss 2012 edition, 2012.

\bibitem{Kramer_FDG2003}
G.~Kramer.
\newblock Capacity results for the discrete memoryless networks.
\newblock {\em IEEE. Trans. Inf. Theory}, 49(1):4--21, Jan. 2003.

\end{thebibliography}

\begin{IEEEbiographynophoto}{Ziv Goldfeld}
(S'13) received his B.Sc.\@ (summa cum laude) and M.Sc.\@ (summa cum laude) degrees in Electrical and Computer Engineering from the Ben-Gurion University, Israel, in 2012 and 2014, respectively. He is currently a 
student in the direct Ph.D. program for honor students in Electrical and Computer Engineering at that same institution.

Between 2003 and 2006, he served in the intelligence corps of the Israeli Defense Forces.

Ziv is a recipient of several awards, among them are the Dean's List Award, the Basor Fellowship, the Lev-Zion fellowship, IEEEI-2014 best student paper award, a Minerva Short-Term Research Grant (MRG), and a Feder Family Award in the national student contest for outstanding research work in the field of communications technology.
\end{IEEEbiographynophoto}

\begin{IEEEbiographynophoto}{Gerhard Kramer}
(S'91-M'94-SM'08-F'10) received the Dr. sc. techn. (Doktor der technischen Wissenschaften) degree from the Swiss Federal Institute of Technology (ETH), Zurich, in 1998.

From 1998 to 2000, he was with Endora Tech AG, Basel, Switzerland, as a Communications Engineering Consultant. From 2000 to 2008, he was with Bell Labs, Alcatel-Lucent, Murray Hill, NJ, as a Member of Technical Staff. He joined the University of Southern California (USC), Los Angeles, in 2009. Since 2010, he has been a Professor and Head of the Institute for Communications Engineering at the Technical University of Munich (TUM), Munich, Germany.

Dr. Kramer served as the 2013 President of the IEEE Information Theory Society. He has won several awards for his work and teaching, including an Alexander von Humboldt Professorship in 2010 and a Lecturer Award from the Student Association of the TUM Electrical and Computer Engineering Department in 2015. He has been a member of the Bavarian Academy of Sciences and Humanities since 2015.
\end{IEEEbiographynophoto}

\begin{IEEEbiographynophoto}{Haim H. Permuter}
(M'08-SM'13) received his B.Sc.\@ (summa cum laude) and M.Sc.\@ (summa cum laude) degrees in Electrical and Computer Engineering from the Ben-Gurion University, Israel, in 1997 and 2003, respectively, and the Ph.D. degree in Electrical Engineering from Stanford University, California in 2008.

Between 1997 and 2004, he was an officer at a research and development unit of the Israeli Defense Forces. Since 2009 he is with the department of Electrical and Computer Engineering at Ben-Gurion University where he is currently an associate professor.

Prof. Permuter is a recipient of several awards, among them the Fullbright Fellowship, the Stanford Graduate Fellowship (SGF), Allon Fellowship, and the U.S.-Israel Binational Science Foundation Bergmann Memorial Award. Haim is currently serving on the editorial board of the IEEE Transactions on Information Theory.
\end{IEEEbiographynophoto}

\begin{IEEEbiographynophoto}{Paul Cuff}
(S'08-M'10) received the B.S. degree in electrical engineering from Brigham Young University, Provo, UT, in 2004 and the M.S. and Ph.D. degrees in electrical engineering from Stanford University in 2006 and 2009. Since 2009 he has been an Assistant Professor of Electrical Engineering at Princeton University.

As a graduate student, Dr. Cuff was awarded the ISIT 2008 Student Paper Award for his work titled “Communication Requirements for Generating Correlated Random Variables” and was a recipient of the National Defense Science and Engineering Graduate Fellowship and the Numerical Technologies Fellowship. As faculty, he received the NSF Career Award in 2014 and the AFOSR Young Investigator Program Award in 2015.
\end{IEEEbiographynophoto}

\end{document}